\documentclass[sigconf]{acmart}

\usepackage{multirow}
\usepackage{longtable}
\usepackage{xspace}

\usepackage{enumitem}
 
\usepackage{amssymb}
 
\usepackage{pifont}

\usepackage{wrapfig,acmart-taps}

\usepackage{graphicx,graphbox}

\usepackage{enumitem}

\usepackage{tabularx} 

\usepackage{listings} 

  {\list{}{\selectfont\itshape\leftmargin=0.25in\rightmargin=0.25in}\item[]}%
  {\endlist}

\AtBeginDocument{%
  \providecommand\BibTeX{{%
    \normalfont B\kern-0.5em{\scshape i\kern-0.25em b}\kern-0.8em\TeX}}}

 \newcommand{\rev}[1] {{#1}}

\copyrightyear{2025}
\acmYear{2025}
\setcopyright{rightsretained}
\acmConference[CHI '25]{CHI Conference on Human Factors in Computing Systems}{April 26-May 1, 2025}{Yokohama, Japan}
\acmBooktitle{CHI Conference on Human Factors in Computing Systems (CHI '25), April 26-May 1, 2025, Yokohama, Japan}
\acmDOI{10.1145/3706598.3713861}
\acmISBN{979-8-4007-1394-1/25/04}

\begin{document}

\title[Intent Tagging]{Intent Tagging: Exploring Micro-Prompting Interactions for Supporting Granular Human-GenAI Co-Creation Workflows }

\settopmatter{authorsperrow=4}

\author{Frederic Gmeiner}
\affiliation{
  \institution{Microsoft Research}
  \city{Redmond}
  \state{WA}
  \country{USA} 
}
\affiliation{
  \institution{Carnegie Mellon University}
  \city{Pittsburgh}
  \state{PA}
  \country{USA} 
}
\email{gmeiner@cmu.edu}
\authornote{Work done as an intern researcher at Microsoft Research.}

\author{Nicolai Marquardt}
\affiliation{
  \institution{Microsoft Research}
  \city{Redmond}
  \state{WA}
  \country{USA} 
}
\email{nicmarquardt@microsoft.com}

\author{Michael Bentley}
\affiliation{
  \institution{Microsoft}
  \city{Redmond}
  \state{WA}
  \country{USA} 
}
\email{mbentley@microsoft.com}

\author{Hugo Romat}
\affiliation{
  \institution{Microsoft}
  \city{Redmond}
  \state{WA}
  \country{USA} 
}
\email{romathugo@microsoft.com}

\author{Michel Pahud}
\affiliation{
  \institution{Microsoft Research}
  \city{Redmond}
  \state{WA}
  \country{USA} 
}
\email{mpahud@microsoft.com}

\author{David Brown}
\affiliation{
  \institution{Microsoft Research}
  \city{Redmond}
  \state{WA}
  \country{USA} 
}
\email{dabrown@microsoft.com}

\author{Asta Roseway}
\affiliation{
  \institution{Microsoft Research}
  \city{Redmond}
  \state{WA}
  \country{USA} 
}
\email{astar@microsoft.com}

\author{Nikolas Martelaro}
\affiliation{
  \institution{Carnegie Mellon University}
  \city{Pittsburgh}
  \state{PA}
  \country{USA} 
}
\email{nikmart@cmu.edu}

\author{Kenneth Holstein}
\affiliation{
  \institution{Carnegie Mellon University}
  \city{Pittsburgh}
  \state{PA}
  \country{USA} 
}
\email{kjholste@cs.cmu.edu}

\author{Ken Hinckley}
\affiliation{
  \institution{Microsoft Research}
  \city{Redmond}
  \state{WA}
  \country{USA} 
}
\email{kenneth.p.hinckley@gmail.com}

\author{Nathalie Riche}
\affiliation{
  \institution{Microsoft Research}
  \city{Redmond}
  \state{WA}
  \country{USA} 
}
\email{nath@microsoft.com}

\renewcommand{\shortauthors}{Gmeiner et al.}

\begin{abstract}
Despite Generative AI (GenAI) systems' potential for enhancing content creation, users often struggle to effectively integrate GenAI into their creative workflows. Core challenges include misalignment of AI-generated content with user intentions (intent elicitation and alignment), user uncertainty around how to best communicate their intents to the AI system (prompt formulation), and insufficient flexibility of AI systems to support diverse creative workflows (workflow flexibility). Motivated by these challenges, we created IntentTagger: a system for slide creation based on the notion of Intent Tags—small, atomic conceptual units that encapsulate user intent—for exploring granular and non-linear micro-prompting interactions for Human-GenAI co-creation workflows. Our user study with 12 participants provides insights into the value of flexibly expressing intent across varying levels of ambiguity, meta-intent elicitation, and the benefits and challenges of intent tag-driven workflows. We conclude by discussing the broader implications of our findings and design considerations for GenAI-supported content creation workflows.
\end{abstract}

 


\begin{teaserfigure}
  \includegraphics[width=0.95\textwidth]{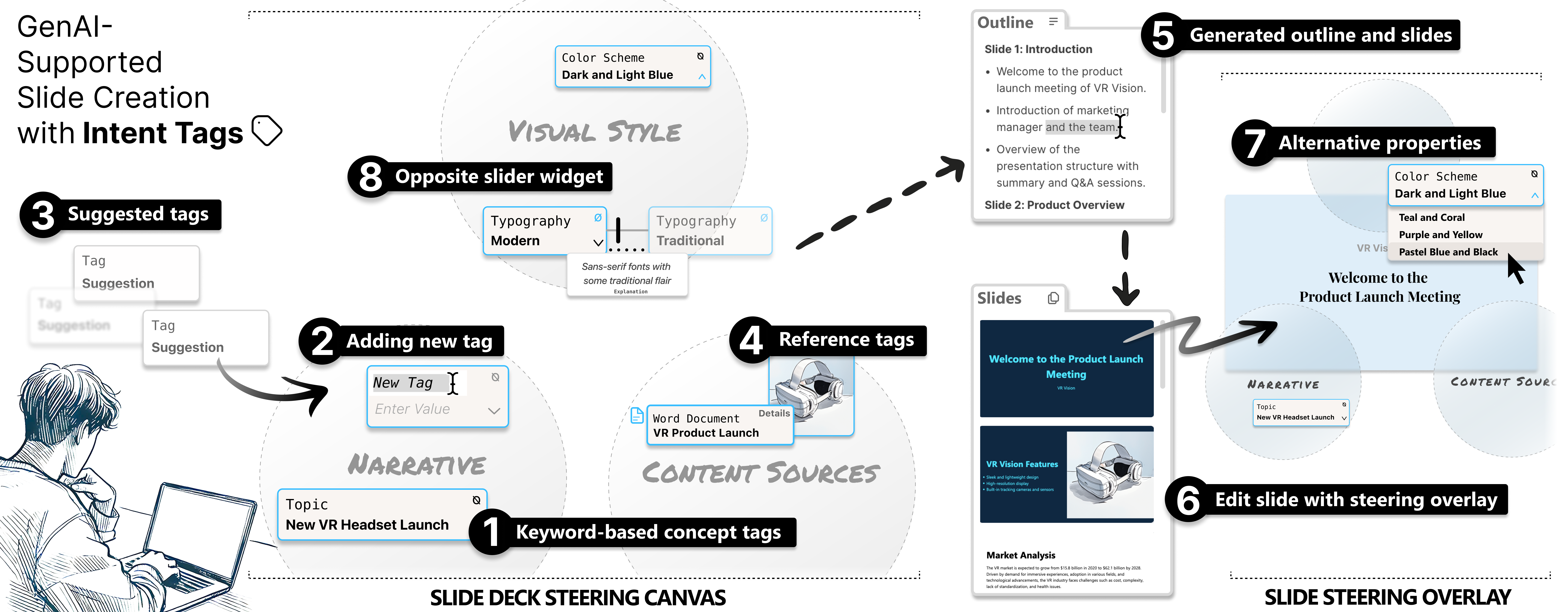}
\caption{\textit{IntentTagger}, an intent tag-based LLM-driven slide creation system: 
Users can steer slide generation with (1) keyword-based concept tags by (2) adding them directly, (3) choosing from suggestions,
or (4) adding existing content.
(5) A user-editable outline and slide deck are generated grounded in all selected tags.
(6) For fine-grained changes, the user can change individual slides with the same (scoped) intent tagging mechanism.
Each tag allows exploring alternative properties in a dynamically generated (7) drop-down list, and 
tag values can be fine-tuned through the (8) opposite slider widget. 
}
  \Description{This figure illustrates the interface and workflow of IntentTagger, a system for LLM-driven slide creation using intent tags. The image is divided into two main sections: the Slide Deck Steering Canvas on the left, and the Slide Steering Overlay on the right. 1. Keyword-based Concept Tags (Bottom Left): Users can steer slide generation by entering specific topics. In this example, the user is adding the tag "New VR Headset Launch" to the "Narrative" group. 2. Adding New Tags (Center Left): A user is shown typing in a new tag under "Narrative," labeled as "New Tag." 3. Suggested Tags (Top Left): The system provides suggested tags for the user to choose from, such as "Tag Suggestion." 4. Reference Tags (Bottom Center): Content sources, such as documents or images, can be added as reference tags. Here, a Word document titled "VR Product Launch" is added. 5. Generated Outline and Slides (Top Right): An automatically generated outline and slide preview are shown. The outline contains a list of slide titles, while the slide preview displays an introductory slide and a "VR Vision Features" slide. 6. Edit Slide with Steering Overlay (Middle Right): A slide from the deck is selected for editing. The steering overlay allows the user to make scoped changes to this slide. 7. Alternative Properties (Top Center Right): A dynamically generated drop-down menu shows alternative properties for tags, allowing the user to select options such as different color schemes for the slides (e.g., "Dark and Light Blue"). 8. Opposite Slider Widget (Top Center): The "Typography" tag is being adjusted with a slider, offering a choice between modern and traditional fonts with descriptive tooltips showing the visual difference.}
  \label{fig:teaser}
\end{teaserfigure}

\maketitle

\section{Introduction}

Generative AI (GenAI) models have become increasingly powerful in content creation tasks, and are rapidly being integrated into a range of professional applications, including coding IDEs \cite{github_github_2024, chen_evaluating_2021}, text processors \cite{grammarly_grammarly_2024, brown_language_2020}, image and video editors \cite{adobe_adobe_2024, saharia_photorealistic_2022, ho_imagen_2022}, and office suites \cite{microsoft_powerpoint_2024b}. These developments offer tremendous promise for AI as a co-creation tool steered by people, but significant challenges remain:

\begin{enumerate}
    \item It is hard to \textbf{align AI-generated content with user intentions} \textit{(intent elicitation and alignment)} \cite{chen_evaluating_2021, terry_ai_2023}, often due to the difficulty in balancing the flexibility of free-form text prompts with guided interfaces;
    \item Users often struggle with \textbf{formulating effective prompts and understanding an AI system’s capabilities}, leading to trial-and-error interactions \textit{(prompt formulation)} \cite{zamfirescu-pereira_why_2023, subramonyam_bridging_2024, mahdavi_goloujeh_is_2024};
    \item The iterative nature of content creation, which requires continuous reflection and refinement, further complicates intent elicitation and prompt formulation since \textbf{users’ needs and intents aren’t often completely clear upfront} \cite{BuxtonSketchingUserExperiences2007, DesignReflectiveConversation1992, TerryCreativeNeedsUIDesign2002, shneiderman_creativity_2007}; 
    \item Many \textbf{GenAI systems impose workflows} that force users to adapt their commonly used creative processes or generate content with each refinement, further complicating human-AI collaboration \textit{(workflow flexibility)} \cite{sarkar_exploring_2023, tankelevitch_metacognitive_2023}.
\end{enumerate}

In this work, we explore potential solutions to these challenges around \textit{intent elicitation and alignment}, \textit{prompt formulation}, and \textit{workflow flexibility} through the lens of GenAI-supported rich content creation tasks, such as slide deck creation.
Slide deck creation involves complex decisions around content structure, visual style, media content, slide sequencing, and narrative flow, all influenced by implicit factors like audience or presentation duration.
Often, slide decks reference external documents, evolve over time,  and may be duplicated and changed for different audiences and purposes.

Technically, GenAI multi-modal foundation models offer powerful capabilities to support such complex content creation tasks, for example, by transforming documents or synthesizing various content sources into a single slide deck \cite{costa_smartedu_2023, Fu2022DOC2PPT, winters_automatically_2019a}. 
However, most human-GenAI interaction challenges are amplified in the context of such complex content creation scenarios---the diverse and non-linear workflows in slide deck creation, coupled with the difficulty of expressing requirements upfront, make it challenging for users to effectively integrate GenAI into their specific creation process.

\begin{table*}
    \centering
    \begin{tabular}{lcccc}
    \toprule
         \textit{Intent Elicitation Interface} &  

         \begin{tabular}[c]{@{}l@{}}
            \textbf{Open Prompt}  \\ \textbf{Format}
         \end{tabular} & 

          \begin{tabular}[c]{@{}l@{}}
            \textbf{System }  \\ \textbf{Guidance}
         \end{tabular} &
         
         \begin{tabular}[c]{@{}l@{}}
                \textbf{GUI-based Option} \\               
                \textbf{Manipulation} 
        \end{tabular} &  
         
         \begin{tabular}[c]{@{}l@{}}
            \textbf{Continuous Option}\\ \textbf{Representation}
         \end{tabular} \\
         \midrule
         \textbf{\textit{GUI-based / Wizard Dialogues}}&  &   \checkmark&  \checkmark &   \checkmark \\
         \textbf{\textit{Text Prompting}}&  \checkmark&  &  &   \\
         \textbf{\textit{Chat Dialogues}}&  &  \checkmark&  &   \\
         \textbf{\textit{Intent Tagging}}&  \checkmark&  \checkmark&  \checkmark&  \checkmark \\
         \bottomrule
    \end{tabular}
    \caption{Outlining the trade-offs between four intent elicitation methods (GUI-based/Wizard Dialogues, Text Prompting, Chat Dialogues, and Intent Tagging); comparing the factors Open Prompt Format (allowing users to express intent freely), System Guidance (guiding users through predefined options), GUI-based Option Manipulation (adjusting single parameters via GUI elements), and Continuous Option Representation (persistent, single mutable objects).}
    \label{tab:ie_comparision}
    \Description{The table outlines the trade-offs between four intent elicitation methods: GUI-based/Wizard Dialogues, Text Prompting, Chat Dialogues, and Intent Tagging. It compares these methods across four criteria: Open Prompt Format (allowing users to express intent freely), System Guidance (guiding users through predefined options), GUI-based Option Manipulation (adjusting single parameters via GUI elements), and Continuous Option Representation (persistent, single mutable objects). Each method is marked with checkmarks to indicate the criteria it supports. For example, Text Prompting supports Open Prompt Format, while GUI-based/Wizard Dialogues supports both System Guidance and GUI-based Option Manipulation.}
\end{table*}

To address these challenges and empower users in rich AI-assisted content creation tasks, we propose \textbf{\textit{intent tags}} as a granular and flexible technique for GenAI-supported workflows via graphical micro-prompting. 
Intent tags represent atomic conceptual units that encapsulate a single aspect of a user's intent. Intent tags help users steer content generation in non-linear workflows---the user instantiates multiple intent tags and edits each individually through adaptive UI elements that allow for granular control of content generation and transparent intent elicitation.
To support a creative design process where users rapidly explore multiple alternatives, iterate, and reflect on the outcomes, our system proactively surfaces AI-generated suggestions through intent tags. These suggestions appear in an unobtrusive manner, allowing people to see the parameter spectrum ~\cite{TerryCreativeNeedsUIDesign2002} of possible outcomes that they might not have considered or been aware of and use them to iteratively refine the system's outputs as desired. 

To explore further possibilities of intent tagging for human-GenAI co-creation workflows for rich content creation, we created \textbf{IntentTagger}: a GenAI-driven system allowing users to iteratively create and modify slide deck presentations using intent tags on a 2D canvas interface. Through intent tags, users steer the generation of single slides or the entire deck by defining keywords \textit{(concept tags)} or including content from other documents or images \textit{(reference tags)}. 
Under the hood, IntentTagger utilizes an LLM for slide generation. IntentTagger also dynamically generates adaptive UI elements such as context-related \textit{tag suggestions}, dynamic \textit{drop-down lists} with \textit{slide preview tooltips}, and interactive \textit{slider widgets} to help users fine-tune tag expressions.    

We explore the benefits and limitations of intent tagging through a \textbf{lab user study} with 12 participants. 
The study comprised \rev{comparative} closed-ended and \rev{semi-open-ended} slide-deck-related tasks that participants completed using our prototype system and, in some comparative tasks, using GenAI features from an existing commercial slide authoring tool. 
Our findings indicate that users felt more in control and satisfied with intent tag-based interactions than with existing chat-based and design gallery-based generative AI systems for slide deck creation.
Participants especially appreciated the \textit{\textbf{support of non-linear and iterative workflows}}, the ability to \textbf{\textit{express their intent in flexible ways}} with \textbf{\textit{varying levels of ambiguity}}, and the \textbf{\textit{integrated system suggestions}} as a valuable and non-distractive aid for \textbf{\textit{helping them think through the task}} and \textbf{\textit{figure out what they want}} while working on the task.

Building upon these study findings, we illustrate how intent tagging could be generally utilized for facilitating human-GenAI co-creation interactions across diverse rich content creation tasks beyond slide deck creation. 
We sketch several \textbf{user experience (UX) interface scenarios} for applications such as web blogging, video creation, and 3D scene creation.
Based on our prototype system, the study's findings, and the alternative UX scenarios, we discuss potential design considerations and future work for intent tag-based interactive generative AI systems.

In sum, this paper makes three main contributions:
\begin{enumerate}
    \item \textit{Intent tagging}: graphical micro-prompting interactions to support granular and non-linear co-creation workflows with generative AI systems in the context of slide deck creation and other rich content creation tasks;

    \item \textit{User study insights on intent tagging for human-AI interaction} in steerability, content creation workflows, and "meta-intent elicitation" by utilizing a GenAI-driven slide creation system called IntentTagger;

    \item \textit{Design considerations for intent tagging} in GenAI-assisted slide authoring applications in particular, and potentially as a new interaction technique for AI-assisted creative design tasks in general. 

\end{enumerate}

Our present work focuses on using intent tags to support the creative design activity of slide deck authoring, which we establish in depth throughout this paper. However, intent tags appear to offer a straightforward design pattern that researchers could explore further in other AI-assisted workflows and scenarios, thereby offering designers a new interaction technique for HCI+AI applications.

\section{Related Work}
\subsection{Intent Elicitation in Human-AI Interaction}

In the field of Human-Computer Interaction (HCI), the concept of Human-AI interaction \cite{amershi_guidelines_2019} and AI as co-creators \cite{davis_enactive_2015, oh_lead_2018} highlight both the opportunities and challenges presented by GenAI models. 
Opportunities lie in automating repetitive and tedious tasks, such as data transformation or formatting, and augmenting creative processes, such as code, text, or image creation \cite{chen_evaluating_2021,brown_language_2020,ramesh_hierarchical_2022, saharia_photorealistic_2022}, while challenges include aligning AI outputs with user intent and users' uncertainties about effective prompts and AI system’s capabilities \cite{chen_evaluating_2021,zamfirescu-pereira_why_2023}.

A core challenge in designing effective interactive GenAI-driven systems is ensuring that AI-generated content matches user intentions. 
This involves two main aspects: the user communicating their intent to the system and the system interpreting the user's intent or guiding them in gradually disclosing it (intent elicitation). 
    
Traditionally, \textbf{wizard interfaces} have emerged to guide users through predefined option GUI dialogues to elicit their intent step-by-step for complex tasks \cite{cooper_face_2014}. 
However, natural language processing (NLP) advancements have introduced text-prompting-based GenAI models, such as large-language foundation models (LLMs), that allow users to input their intent as \textbf{free-form text prompts} \cite{liu_pretrain_2021} or guide users through a \textbf{dialogue} to elicit their intents \cite{park_user_2023,sahijwani_adaptive_2022, cai_predicting_2020, qu_user_2019}. 
    
However, all such approaches present opposing trade-offs (see Table \ref{tab:ie_comparision}): 
GUI-based menus or wizards can intuitively guide user decisions and reduce ambiguity while allowing for continuous representation of options through graphical manipulatable elements, but they only offer a finite set of options that reflect pre-anticipated use case scenarios by software makers. 
Free-form text prompting allows users to communicate their intent more flexibly using natural language, but the lack of scaffolding and guidance requires users to know what they want (or need) and how to formulate it as an effective prompt.
While easy for simple tasks, communicating intents can be challenging in complex content creation tasks when users might not be aware of all necessary options upfront and are required to iteratively figure these out (intent exploration). 
As a compromise, natural language chat dialogues can provide system-guided option exploration and scaffolding. 
However, such systems can only probe on a limited set of options in a reasonable time frame or number of conversational turns, and they often lack continuous option representation and graphical manipulation. 

On the other hand, effectively aligning AI-generated content with user intentions also requires awareness of one's own intentions and understanding of the possible parameter spectrum \cite{TerryCreativeNeedsUIDesign2002}.
\rev{Recent work in AI-mediated intent elicitation and sensemaking has explored various strategies and GenAI-driven tools to assist users in more open-ended and iterative ways to elicit and discover their (creative) intent. 
For example, systems like \textit{Luminate} \cite{suh_luminate_2024}, \textit{Selenite} \cite{liu_selenite_2024}, \textit{Sensecape} \cite{suh_sensecape_2023} and \textit{Graphologue} \cite{jiang_graphologue_2023} utilize dynamic LLM-driven GUIs to support users in sensemaking and intent elicitation (e.g., exploring a design space, available options, or topics) by generating and visually clustering semantically related concepts or criteria to help them explore relevant dimensions and parameters matching their tacit intent and needs. 
Similarly, \textit{SymbolFinder} \cite{petridis_symbolfinder_2021} offers different LLM-driven GUIs to suggest and group related words and images to guide designers in gradually exploring and finding appropriate symbols to illustrate abstract concepts.
Other work by \textit{Kreminski and Chung} \cite{kreminski_intent_2024} uses language models and micro-interactions to elicit users' creative intent by formulating and asking open-ended questions about aspects of a game creator's emerging story world.
Similarly, \textit{Germinate} \cite{kreminski_germinate_2020}, a system for game generation, features an interface to help break down the design intents of casual creators by allowing users to specify game design-related parameters, such as game entities or resources through graphical text-based tags that users can freely define. 
Based on these user-defined tags, the system attempts to infer additional appropriate tags that users might incorporate or negate (cf.~\cite{martens_languages_2017}). 
Similarly, Lin et al. \cite{lin_prompts_2023} outlined a design space for user interactions with co-creative systems, highlighting the role of explicit communication mechanisms about creative intent, which might be initiated by either the user or the system (mixed-initiative).}

\rev{From a broader cognitive perspective}, self-awareness \rev{about one's intentions} relates to \textit{metacognition} \cite{flavell_metacognition_1979}, and recent research by \textit{Tankelevitch et al.} \cite{tankelevitch_metacognitive_2023} have proposed a stronger focus on supporting users' metacognitive processes, such as \textit{self-awareness} and \textit{task decomposition}, for mitigating challenges of effective GenAI workflows. 
In a recent design fiction, \textit{Vaithilingam et al.} \cite{vaithilingam_imagining_2024} explore metacognitive support using LLM dialogue-based guided intent exploration inspired by human-human communication patterns such as dynamic grounding, constructive negotiation, and sustainable motivation to support the design process. 

Previous work has explored various \rev{AI-mediated} user intent elicitation approaches and proposed metacognitive user support to improve GenAI-driven tasks. 
Building atop this prior work, we aim to derive interaction principles for human-GenAI co-creation that combine the benefits of existing intent elicitation interfaces without their drawbacks while supporting users' intent exploration process through metacognitive support.

\subsection{Systems for Supporting Prompting and Steering GenAI}

Generative AI \textit{foundation} models, such as large language or diffusion models \cite{bommasani_opportunities_2022,rombach_highresolution_2022}, represent a paradigm shift in artificial intelligence, offering task-agnostic pre-training on large-scale data for various downstream applications \cite{schneider_foundation_2022}. 
Although extremely versatile, research suggests three properties that make interacting with GenAI challenging: \textit{input flexibility} (in handling free-form language, images, code, etc.), \textit{generality} (applicability to a wide range of tasks), and \textit{originality} (ability to generate novel content) \cite{schellaert_your_2023}.

Prior HCI research has documented interaction challenges in prompting and steering GenAI systems across domains such as coding, illustration design, or engineering \cite{liu_what_2023,liu_opal_2022, gmeiner_exploring_2023}. 
For example, users frequently struggle to craft text-based input prompts that will achieve desired outcomes, and face difficulties interpreting and repairing erroneous outputs \cite{zamfirescu-pereira_why_2023}.
Research has begun to explore mechanisms and interfaces to better support users in working with text prompt-based GenAI models. 
For example, various works have proposed mechanisms to support users in \textit{prompt engineering} \cite{liu_pretrain_2021} through \textit{"prompt augmentation,"} which automatically modifies and extends a users' input prompt to improve the model's generated output \cite{brade_promptify_2023, shin_autoprompt_2020, betker_improving_2023}.
While such techniques can improve model output quality, users are bound to express their intent in an open-ended text format. 

Other work has explored ways of \textit{combining direct manipulation interfaces with text-based GenAI systems} to enable more structured user inputs and reduce semantic ambiguity. For example, \textit{GhostWriter} \cite{yeh_ghostwriter_2024} offers buttons for predefined prompts for LLM-driven text manipulation tasks. 
While such approaches reduce misalignment and improve feature discoverability by visually exposing model capabilities, they also restrict the available task options to predefined sets and limit the GenAI models' generality.

Recent work has also proposed contextually bounded prompting mechanisms to offer more fine-grained control over Gen-AI-driven content generation. 
\textit{DirectGPT} \cite{masson_directgpt_2024} explores direct manipulation principles for LLMs, such as the continuous representation of manipulatable objects, physical actions to localize the effect of prompts, and reusable prompts in a toolbar.
Similarly, \textit{Cococo} \cite{louie_noviceai_2020} offers users a set of GUI widgets for steering GenAI music generation, including slider elements to nudge music generation in high-level directions.
Other ideas include painting-like interactions, such as how \textit{PromptPaint} \cite{chung_promptpaint_2023} enables users to steer diffusion-based text-to-image generation through paint medium-like interactions and \textit{TaleBrush} \cite{chung_talebrush_2022} supports generative story ideation through line sketching interactions to graphically steer LLM-driven story generation.

Building on this work, we contribute to steering interactions for GenAI by investigating mixed interfaces that blend free text input with dynamic graphical elements.
We propose intent tagging as graphical micro-prompting interactions to flexibly support intent expression and elicitation using GUI widgets to allow exploration across various levels of ambiguity.

\subsection{Non-Linear Content Creation and Iterative Design Workflows}
Creating rich content documents, such as blog posts or slide presentations, involves crafting and integrating text, visuals, and multimedia to convey complex ideas while maintaining coherence and audience engagement.
Guidelines for creating slide presentations often suggest standardized workflows such as crafting outlines before creating slides \cite{reynolds_presentation_2020,zanders_presentation_2018, anholt_dazzle_2006}. 
However, other literature emphasizes that slide-creation workflows are less rigid and are mostly shaped by cultural factors and organizational norms \cite{yates_powerpoint_2007}. 
For example, in organizations, people often start or continue presentations from different starting points, such as from existing documents, other slide decks, or templates. 

Rich content creation processes are also iterative and diverse in nature, typically progressing through cyclically occurring stages \cite{baker_ideas_2010, dorta_signs_2010}, where creators alternate between \textit{‘exploration’} and \textit{‘exploitation’} in a sense-making process to reach a final outcome \cite{pirolli_sensemaking_2005}.
Content creators continuously develop, reflect on, and act on their ideas and plans \cite{kolko_sensemaking_2010, schon_reflective_1983}. 
The interaction between users' perception of the material and the material itself enables creators to develop, reflect on, and question their understanding, leading to new ideas and improved plans \cite{klein_making_2006}. 
Research has underscored the role of reflection in content creation as a necessary moment for creators to situate their ideas and plans in the appropriate context \cite{dove_argument_2016,mols_informing_2016,sharmin_reflect_2011a}.

However, existing challenges around prompt formulation engage users in cognitively demanding trial-and-error processes and hinder reflection related to content creation. 
Furthermore, many of today’s GenAI systems assume a specific workflow or require users to adapt their processes to integrate GenAI effectively \cite{tankelevitch_metacognitive_2023, sarkar_exploring_2023}. 

In conclusion, despite recommended workflows for content creation (such as starting with an outline), content creation processes are typically highly situational, non-linear, and iterative in nature. Therefore, we aim to derive interfaces for GenAI-driven content creation systems to promote non-linear and diverse workflows while supporting reflection on the content creation process.  

\subsection{Systems for Supporting Slide Deck Creation}

Creating slide presentations is a multifaceted task that involves various sub-tasks, and previous research has introduced a range of approaches to support different stages of this process.

 Besides popular slide-by-slide authoring tools like PowerPoint or Keynote, research has suggested interfaces to support users in \textbf{iterative slide creation workflows} by managing multiple presentation versions \cite{drucker_comparing_2006}, prototyping slide decks using markup language \cite{edge_hyperslides_2013} or authoring slide presentations on a 2D canvas interface \cite{lichtschlag_fly_2009} for supporting non-sequential workflows.

Besides manual authoring systems, numerous works have explored ways to \textbf{automatically generate slide presentations} and rich content documents for single topics \cite{winters_automatically_2019a}, or scientific \cite{Fu2022DOC2PPT}, technical \cite{M2009SlidesGen}, educational \cite{costa_smartedu_2023}, or semantically annotated source documents \cite{masao_automatic_1999}. Other systems focus on automatically generating presentation-specific aspects, such as visual-textual layouts \cite{yang_automatic_2016}.
While these systems allow for automatic content transformation and slide creation, they lack methods for users to influence or control the generation process.  

Several works have, therefore, proposed mechanisms to enable users to \textbf{generate presentations in more controllable and workflow-integrated ways}, such as allowing data scientists to steer the generation of slide presentations from Jupyter notebooks by interactively selecting code cells and slide layouts \cite{wang_slide4n_2023} or through a user-controllable text outline \cite{wang_outlinespark_2024}. 
Similarly, \textit{Knowledge-Decks} \cite{christino_knowledgedecks_2022} allows users to generate slides documenting data science knowledge-discovery processes from automatically collected user behavior events from visual analytic tools.

Previous work has explored various authoring interfaces, automatic slide generation from source documents, and user-controllable slide generation in domain-specific applications like data science. Expanding on these approaches, we aim to explore interaction principles for GenAI-supported slide deck creation across diverse content tasks and workflows. Our approach combines GenAI-driven slide generation from multiple input types (e.g., prompts, text documents, images) with \textit{non-predefined} and \textit{non-linear authoring workflows} in mind that let users seamlessly switch between generation and editing modes across outline, individual slide, or entire deck level.

\section{Exploring Design Principles for GenAI-supported Rich Content Creation}

In this section, we lay out the challenges of rich content document creation in the context of slide creation and how these informed the design principles that led us to intent tagging as an interaction paradigm for enabling novel human-GenAI co-creation workflows.

Slide deck creation is a complex task that requires balancing multiple facets, such as narrative coherence, visual style consistency, and the integration of diverse content sources \cite{yates_powerpoint_2007, reynolds_presentation_2020, anholt_dazzle_2006}. 
These elements must not only align individually but also work together harmoniously to deliver an impactful presentation. The challenges arise from the need to manage and sequence multimedia content while simultaneously crafting a narrative that flows logically across slides. Achieving this requires careful consideration of storytelling, visual design, and the inclusion of relevant data or references, often making the task overwhelming, especially under time constraints or when dealing with open-ended creative briefs.

One of the core difficulties in slide deck creation is the need to manage these distinct yet interdependent elements—\textit{narrative}, \textit{visual style}, and \textit{content sources}—without losing sight of the overall presentation goals. 

Traditional tools often compartmentalize these tasks, forcing users to switch between different modes or tabs, which can disrupt the creative process. Linear workflows or predefined templates, while useful in some cases, often fail to accommodate the nuanced adjustments needed to refine a slide deck to a professional standard.

Moreover, as discussed in the previous section, the integration of Generative AI (GenAI) into slide deck creation introduces both opportunities and challenges. Current GenAI approaches tend to offer either (semi-)automated solutions with little ability for users to steer the content generation, design galleries with limited pre-defined option sets, or chat-based interactions that can feel too linear, unstructured, and limited for exploring alternatives.

\begin{figure*}[ht!]
  \centering
  \includegraphics[width=\linewidth]{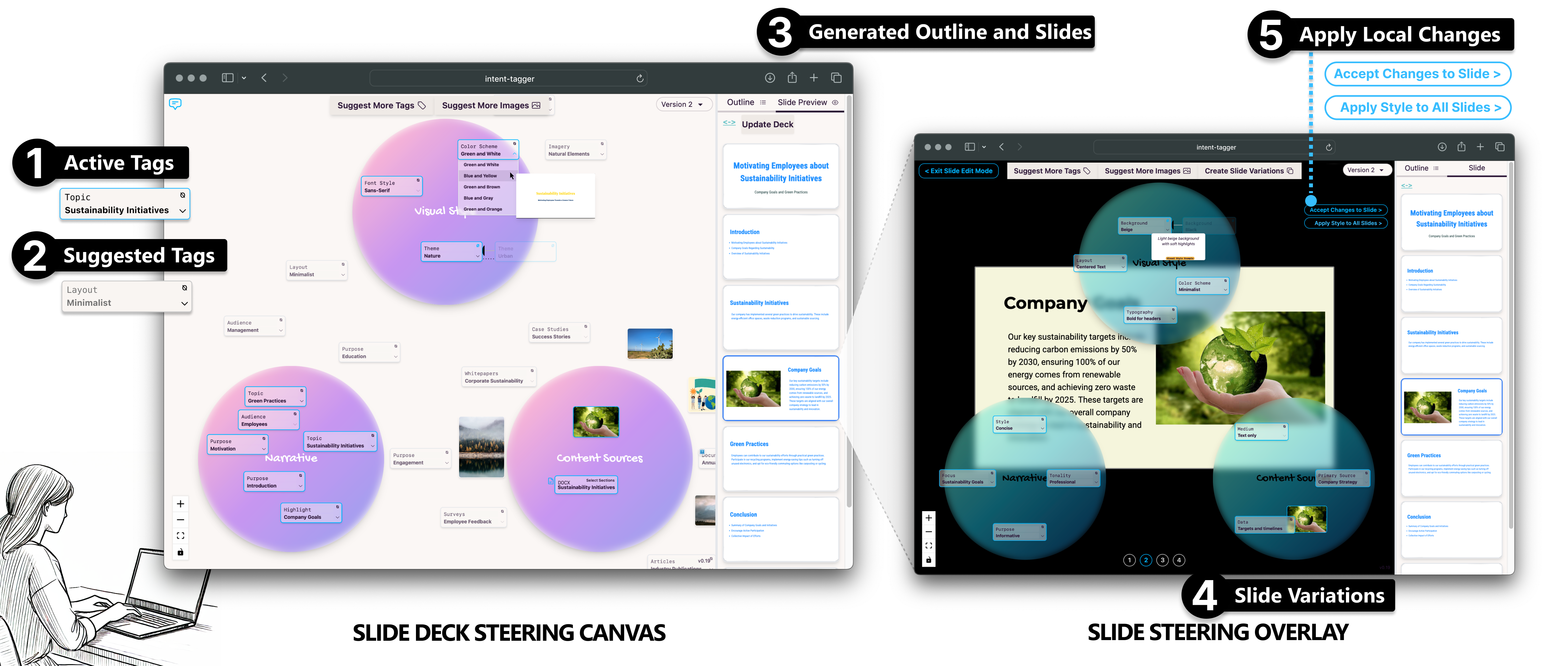}
  \caption{Screenshots of the \textit{IntentTagger} system: (Left window) A user creates a slide presentation using the Slide Deck Steering Canvas with (1) active tags in the three circular tag groups surrounded by (2) inactive system-suggested tags. (3) The generated outline and slides are displayed in the tabbed sidebar. (Right window) The user makes (scoped) adjustments to a single slide via the Slide Steering Overlay and (4) explores generated slide variations. (5) Then, local slide changes can be accepted or applied to the entire deck.}
  \Description{This figure shows the interface of the IntentTagger system with two main sections: the Slide Deck Steering Canvas on the left and the Slide Steering Overlay on the right. (1) Active Tags (Bottom Left): Users have selected active tags, such as "Sustainability Initiatives" under the "Narrative" group. (2) Suggested Tags: The system suggests additional tags like "Layout: Minimalist." (3) Generated Outline and Slides: The sidebar shows the generated outline and slide previews. (4) Slide Variations: The user explores slide variations for individual slides in the Slide Steering Overlay, such as a "Company" slide showing sustainability targets. (5) Apply Local Changes: Users can apply changes to individual slides or the entire deck. The figure demonstrates how users organize tags and make fine-tuned slide adjustments using intent tags and overlay features.}
  \label{fig:system-ui}
\end{figure*}

\subsection{Design Principles}

Based on these requirements, we developed the following interaction design principles for GenAI-supported slide creation:

\begin{itemize}[font=\bfseries,
  align=left]
    \item[DP1] \textbf{Enabling flexible, non-linear, and iterative workflows } \\
        To cater to users' diverse slide creation needs and individual working styles, users should be empowered to start and refine their presentations in a variety of ways. Whether beginning with an outline, building on an existing deck, crafting a single slide, or incorporating content from other documents, users should have the flexibility to approach their tasks from any angle. Additionally, the interface should allow seamless transitions between different content views—such as slide, deck, or outline—enabling GenAI-assisted iterative enhancements at any stage of the creation process.  
        
    \item[DP2] \textbf{Accommodating diverse steering input types} \\ 
    Slide creation often requires the integration of multiple sources of information, ranging from explicit, well-defined data (e.g., an annual sales report) to more abstract, nuanced considerations (e.g., audience-specific messaging). To ensure that GenAI-generated outputs align closely with user intentions, the interface should facilitate the input of a wide range of content types, including natural language descriptions, existing documents, images, and other relevant media. By enabling users to provide rich, varied inputs, the system can better interpret and fulfill the specific needs of each presentation.   
    
\vspace{\baselineskip} 
    
     \item[DP3] \textbf{Supporting intent expression across varying levels of abstraction} \\
    In many slide creation scenarios, users may have clear, specific ideas about certain aspects of their presentation, while other elements remain less defined or harder to articulate. To accommodate this variability, a GenAI interface should allow users to express their intent at different levels of abstraction—ranging from detailed, precise instructions to more general, high-level directives. This flexibility not only helps users articulate their ideas more effectively but also supports iterative exploration and refinement of their presentation content.     

    \item[DP4] \textbf{Leveraging AI for contextual content and terminology suggestions  } \\
    To assist users in overcoming challenges like the "blank page syndrome" or finding terminology to describe their intent, the interface should harness the associative power of LLMs to provide contextual suggestions and inspiration. By offering alternative ideas, vocabulary, and creative prompts, the system can stimulate users' thinking, helping them to refine their content and explore new directions in an iterative process.

    \item[DP5] \textbf{Providing pre-computed real-time previews} \\
    To allow users to better and quicker anticipate how specific choices will impact generated slides, the tool should provide pre-computed real-time previews to, for example, let users rapidly explore how different fonts or colors would look on a given slide.

\end{itemize}

\section{IntentTagger: An Intent Tagging-based Slide Creation System}

Driven by \rev{the requirements of rich content creation tasks and} the design principles described in the previous section, we explored novel input strategies through graphical micro-prompting interactions---an interaction notion we coined \textit{intent tagging}. 
\rev{In addition to our design principles, a central inspiration for conceptualizing creative intent elicitation through granular and flexible micro prompts that represent users’ intention was the ``mood board'' technique, in which creatives iteratively compile visual and textual elements into a collage to convey an overarching theme or creative direction \cite{garner_problem_2001}. 
Similar to how mood boards enable capturing and expressing designers’ intent through compiling different media with varying levels of abstractions---such as words and pictures ranging from abstract concepts, stylistic references, or specific material details---intent tagging allows users to express their intent openly through collections of intent tags.}

\rev{To explore further the idea of intent tagging,} we created \textit{IntentTagger}, an LLM-driven system allowing users to iteratively create and modify slide deck presentations using intent tags on a 2D canvas interface (see Figure \ref{fig:system-ui}). 
In this section, we demonstrate the utility of \textit{IntentTagger} by first illustrating its functionalities in a use-case scenario. We then follow with a detailed description of the proposed interface mechanisms and their implementation.

\subsection{Example Use Case Scenario}

Lucy is a sustainability manager at a company and wants to prepare a short slide presentation for new employees to inform them about sustainability initiatives at her company. 

Lucy starts IntentTagger and clicks on the \textit{"Create a new presentation from prompt..."} button. 
A modal popup appears, and she types \textit{"a presentation for introducing sustainability initiatives for new employees"} in the prompt field and presses enter. 
The modal disappears, and the three circular steering groups on the steering canvas pulse, then three concept tags appear in the \textit{"Narrative"} circle with the parameters \textit{"Topic: Sustainability Initiatives," "Audience: New Employees"} and \textit{"Purpose: Introduction."} 

\textbf{Tag Suggestions:} To get inspiration for further presentation talking points, she clicks \textit{"Suggest More Tags,"} and shortly after, new tags appear outside of the circular tag groups. Lucy finds some of the suggested tags useful and drags these into the \textit{"Narrative"} group to include them in the generation, such as \textit{"Focus: Eco-friendly Practices," "Highlight: Company Goals," "Section: Key Initiatives,"} and \textit{"Objective: Corporate Responsibility."}

\textbf{Including External Content:}
Lucy also wants to include sections from an existing Word document detailing the company's policies. With the mouse cursor, she drags the document from her file browser onto the IntentTagger app, and a new reference tag appears in the \textit{"Content Sources"} group. She clicks on the tag's \textit{"Select Sections"} button, which opens up the tag's Text Selection Widget showing the document's content. In the widget, she highlights two document sections, clicks \textit{"Include Section in Presentation"} from the widget's context menu, and closes the widget. 

\textbf{Generating an Outline:}
Next, she wants to generate a first draft of the presentation's outline and presses \textit{"update outline from intent tags"} in the outline view panel. The tag groups start pulsing, and then the new outline appears in the outline text editor. 

\textbf{Manually Crafting Concept Tags:}
Lucy is happy with the structure overall but finds it too long (20 sections). Instead of manually editing the outline, she adds a new concept tag to the \textit{"Narrative"} group and enters \textit{"Number of slides"} into the tag's upper text field and \textit{"10"} into the lower and clicks the \textit{"update outline from intent tags"} button again. The newly generated outline is now shorter and only comprises ten sections. 

\textbf{Generating Slides:}
Next, Lucy wants to generate a first draft of the slide deck and clicks the \textit{"Update slides from intent tags"} button. Shortly after, the generated slides appear in the slide preview side panel. After evaluating the slides, Lucy realizes that adding some images and adjusting the visual style could enhance the presentation's appeal. 

\textbf{Including Suggested Images:}
She clicks the \textit{"suggest more images"} button on the tag steering board, and shortly after, new Reference Image Tags appear outside the \textit{"Content Sources"} tag group with eco-friendly-themed stock photos. Lucy drags some of the image tags into the \textit{"Content Sources"} group to add them to the presentation.   

\textbf{Adjusting the Visual Style through Tags' Alternatives Drop-down:}
To make visual adjustments to the slides, she drags in some of the previously suggested tags, such as \textit{"Typography: Sans-Serif," "ColorPalette: Green and Blue,"} and \textit{"Theme: Nature"} onto the \textit{"Visual Style"} tag group. To get a better sense of typography alternatives, she clicks on the tags' drop-down, and a list of alternative tag values appears, such as \textit{"Serif," "Monospace,"} and \textit{"Handwritten."} While hovering over each option with the mouse, Lucy sees a preview thumbnail of that font style applied to the current slide. Lucy chooses \textit{"Monospace."} Now, Lucy clicks \textit{"Update slides from intent tags"} again, and shortly after, she sees the revised slides reflecting the intended new visual style, including the images. 

\textbf{Adjusting a Single Slide with Tag Steering Overlay:}
Lucy decides to refine the slide deck further. She clicks on the third slide in the \textit{Slide Preview} panel to open the slide edit view and review its content in detail. 
The slide contains a text paragraph and an image, but she wants to show a list of bullet points instead. 
To modify the slide, she clicks on the intent tag icon next to the slide, and the three steering tag groups appear as an overlay containing tags representing the slide's \textit{Narrative}, \textit{Visual Style}, and \textit{Content Sources}. 
Lucy selects the \textit{"Text Format"} tag inside the \textit{Narrative} group and clicks on its drop-down widget. 
A list of alternative tag values appears, including \textit{"Bullet Points," "List,"} and \textit{"Table."} Lucy chooses \textit{"Bullet Points."} 
To adjust the background color, Lucy creates a new tag in the \textit{Visual Style} group and types in \textit{"Background"} and \textit{"pastel color."} 
Now, she clicks on the \textit{"Explore slide variations"} button, and the system generates several alternative versions of the slide, each with slightly different text variations and pastel background colors.  
Lucy clicks through these variations and selects one that better aligns with her presentation’s tone.

\textbf{Applying a Slide's Style to All Slides:}
To maintain consistency across the entire presentation, Lucy uses the same visual style and selects \textit{"Apply to all slides."} 

Satisfied with the changes, Lucy saves the presentation, ready to deliver it in the upcoming meeting with her colleagues.

\textbf{Adjusting the Presentation to New Requirements:}
After a couple of months, Lucy needs to adapt the presentation for a different audience—existing employees rather than new hires. She clicks on the \textit{"Audience"} tag in the \textit{Narrative} group and changes it from \textit{"New Employees"} to \textit{"Existing Employees."} The system instantly updates the content and tone of the slides to better fit this new audience. For example, introductory sections are replaced with more in-depth discussions of ongoing sustainability projects and their impact.

\subsection{Interface Features}

Here, we provide a more detailed explanation of IntentTagger's features:

\subsubsection{\textbf{Intent Tags: Concept Tags, Reference Tags and Groups (DP2)}}
Intent tags are collections of single keywords, phrases, or media artifacts describing users' intended outcomes. 
Tags serve as granular and flexible micro prompts on a zoomable 2D canvas. 
Users can steer the generative slide creation system through two types of tags:

\aptLtoX[graphic=no,type=html]{\begin{figure}[H]
    \includegraphics[align=t,width=\columnwidth]{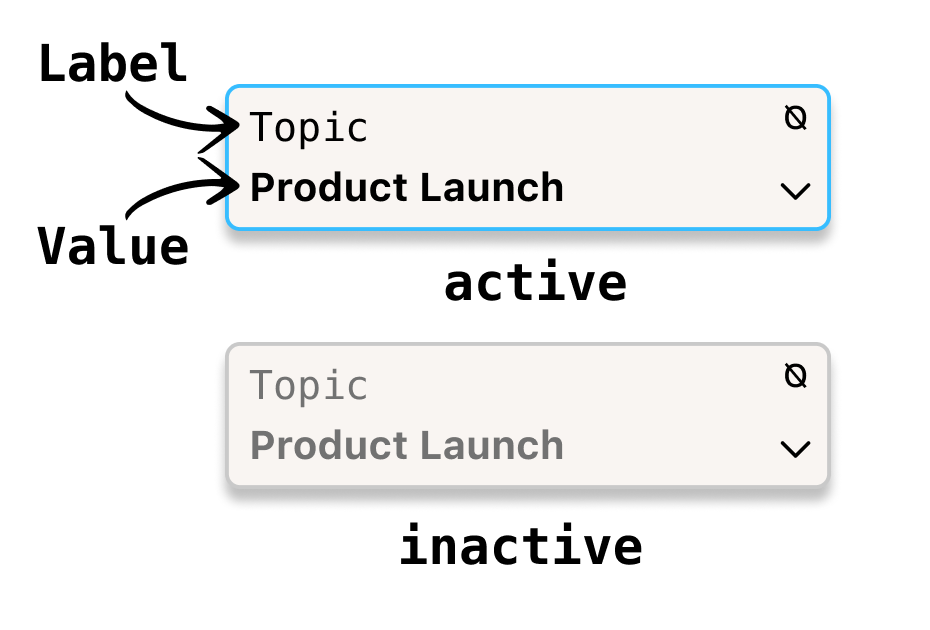}
    \caption{ \ ~ \ ~ \ ~ \ ~ \ }
    \label{fig:concept-tag}
    \Description{The figure shows two examples of concept tags with a label and value format. The first example is an active tag labeled “Topic” with the value “Product Launch,” displayed with a bold border. The second example is an inactive tag with the same label and value, but shown with a lighter, grayed-out appearance, indicating that it is not currently influencing the content generation. Both tags have a drop-down arrow for selecting or modifying the value.}
\end{figure}
          \textbf{(1) Concept Tags:} Concept Tags are two-part text micro prompts of the format [label: value], such as \textit{”Topic: Product launch”} or \textit{”Color scheme: Corporate.”} Tags can be active or inactive to toggle their influence on generated content.
}{
\begin{figure}[H]
  \begin{minipage}[t]{3cm}
    \includegraphics[align=t,width=\columnwidth]{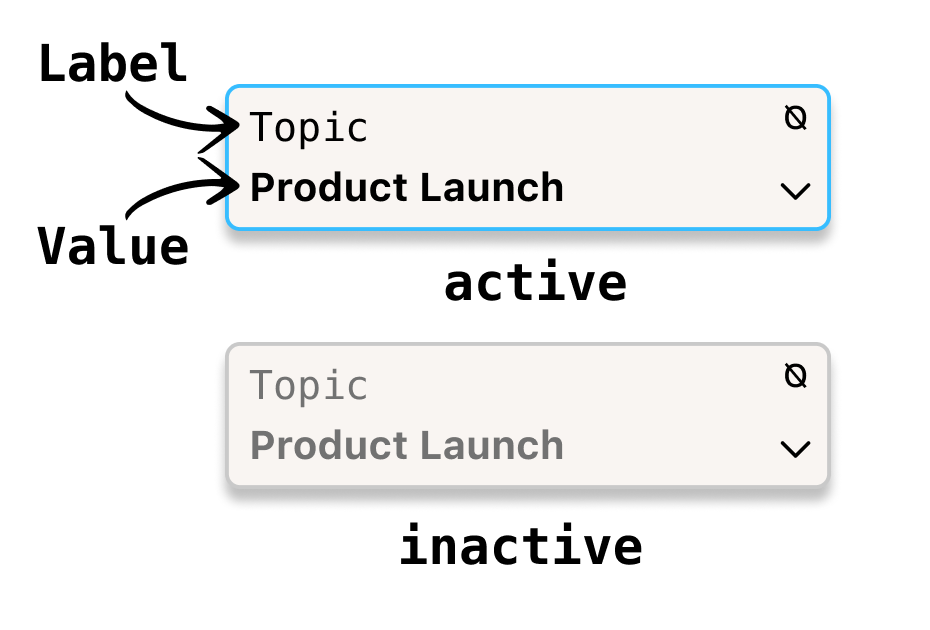}
    \caption{}
    \label{fig:concept-tag}
    \Description{The figure shows two examples of concept tags with a label and value format. The first example is an active tag labeled “Topic” with the value “Product Launch,” displayed with a bold border. The second example is an inactive tag with the same label and value, but shown with a lighter, grayed-out appearance, indicating that it is not currently influencing the content generation. Both tags have a drop-down arrow for selecting or modifying the value.}
  \end{minipage}
   \begin{minipage}[t]{\dimexpr\columnwidth-3.2cm\relax}
          \textbf{(1) Concept Tags:} Concept Tags are two-part text micro prompts of the format [label: value], such as \textit{”Topic: Product launch”} or \textit{”Color scheme: Corporate.”} Tags can be active or inactive to toggle their influence on generated content.
  \end{minipage}
\end{figure}}

\aptLtoX[graphic=no,type=html]{\begin{figure}[H]
%  \begin{minipage}[t]{3cm}
    \includegraphics[align=t,width=\columnwidth]{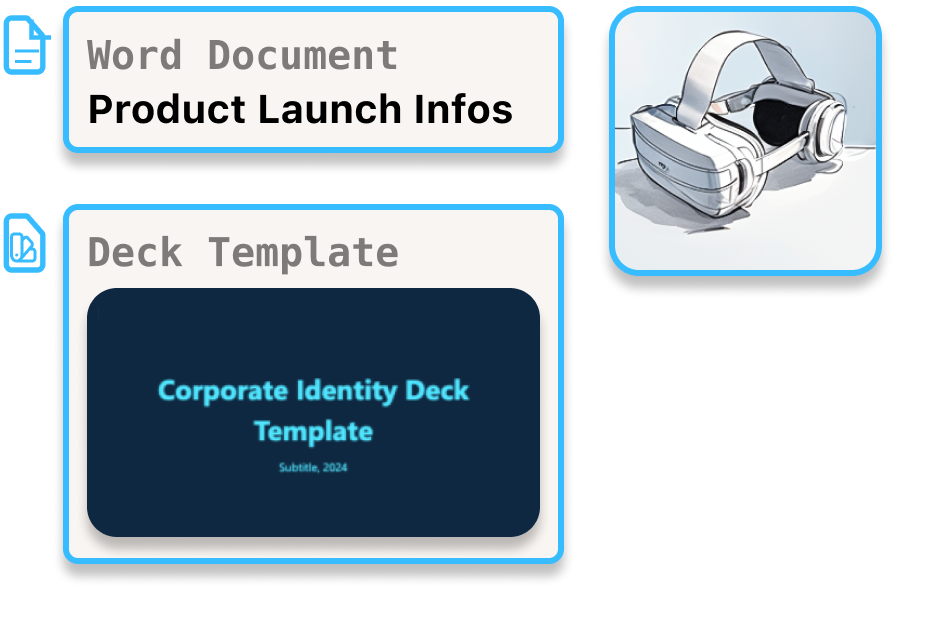}
    \caption{ \ ~ \ ~ \ ~ \ ~ \ }
    \label{fig:reference-tags}
    \Description{The figure shows two examples of reference tags. The first tag represents a Word document titled “Product Launch Infos,” with an icon of a document beside it. The second tag represents a slide deck template titled “Corporate Identity Deck Template,” displayed with a thumbnail of a presentation slide and an image of a 3D headset. Both tags are visually enclosed in boxes, indicating that these external files are used as references for slide generation.}
%  \end{minipage}
%   \begin{minipage}[t]{\dimexpr\columnwidth-3.2cm\relax}
%  \end{minipage}
\end{figure}
\textbf{(2) Reference Tags:} \textit{Reference Tags} allow users to provide external documents such as Word documents, images, or other slide decks as references for slide generation. Users can add these files by drag-and-drop from their file browser.  }{
\begin{figure}[H]
  \begin{minipage}[t]{3cm}
    \includegraphics[align=t,width=\columnwidth]{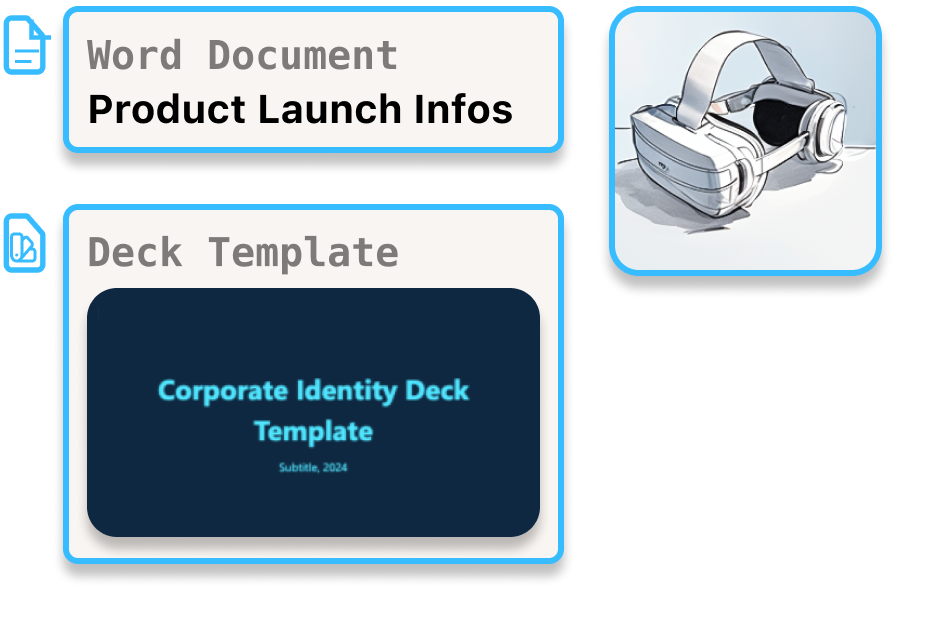}
    \caption{}
    \label{fig:reference-tags}
    \Description{The figure shows two examples of reference tags. The first tag represents a Word document titled “Product Launch Infos,” with an icon of a document beside it. The second tag represents a slide deck template titled “Corporate Identity Deck Template,” displayed with a thumbnail of a presentation slide and an image of a 3D headset. Both tags are visually enclosed in boxes, indicating that these external files are used as references for slide generation.}
  \end{minipage}
   \begin{minipage}[t]{\dimexpr\columnwidth-3.2cm\relax}
          \textbf{(2) Reference Tags:} \textit{Reference Tags} allow users to provide external documents such as Word documents, images, or other slide decks as references for slide generation. Users can add these files by drag-and-drop from their file browser.  
  \end{minipage}
\end{figure}}

\aptLtoX[graphic=no,type=html]{\begin{figure}[H]
%  \begin{minipage}[t]{3cm}
    \includegraphics[align=t,width=\columnwidth]{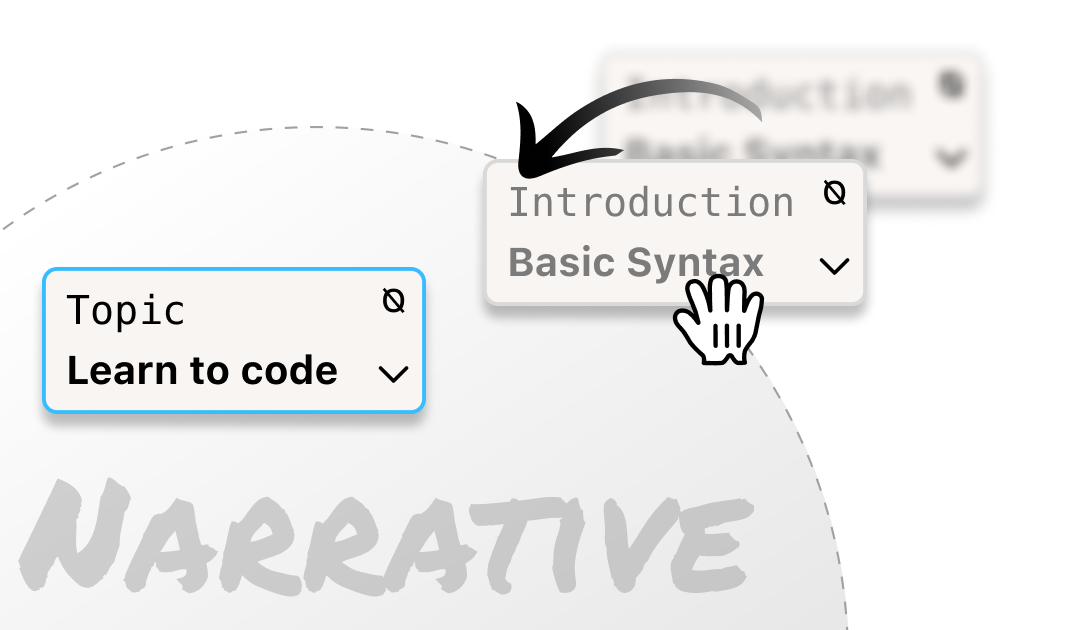}
    \caption{ \ ~ \ ~ \ ~ \ ~ \ }
    \label{fig:tag-groups}
    \Description{The figure displays a circular group labeled “Narrative,” containing an active concept tag with the topic “Learn to code.” Above the active tag, an inactive tag is dragged into the circular group  reading “Introduction” and “Basic Syntax,” with a cursor pointing to “Basic Syntax.” The tags are clustered visually within the circular group, demonstrating how tags can be organized and managed within the interface. Users can drag and drop tags into or out of the group to activate or deactivate them.}
%  \end{minipage}
%   \begin{minipage}[t]{\dimexpr\columnwidth-3.2cm\relax}
%  \end{minipage}
\end{figure}          \textbf{Tag Groups}: All active tags are visually clustered into three circular groups on the interface, each representing different slide creation aspects: \textit{Narrative}, \textit{Visual Style}, and \textit{Content Sources}. Users can drag tags in and out of groups to activate and deactivate them. 
}{\begin{figure}[H]
  \begin{minipage}[t]{3cm}
    \includegraphics[align=t,width=\columnwidth]{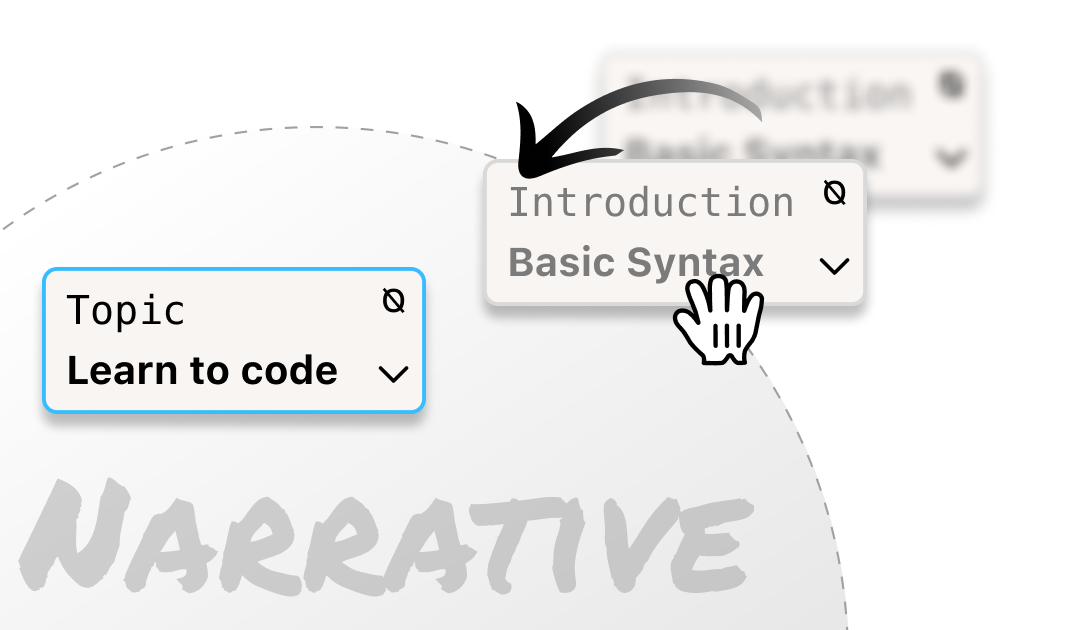}
    \caption{}
    \label{fig:tag-groups}
    \Description{The figure displays a circular group labeled “Narrative,” containing an active concept tag with the topic “Learn to code.” Above the active tag, an inactive tag is dragged into the circular group  reading “Introduction” and “Basic Syntax,” with a cursor pointing to “Basic Syntax.” The tags are clustered visually within the circular group, demonstrating how tags can be organized and managed within the interface. Users can drag and drop tags into or out of the group to activate or deactivate them.}
  \end{minipage}
   \begin{minipage}[t]{\dimexpr\columnwidth-3.2cm\relax}
          \textbf{Tag Groups}: All active tags are visually clustered into three circular groups on the interface, each representing different slide creation aspects: \textit{Narrative}, \textit{Visual Style}, and \textit{Content Sources}. Users can drag tags in and out of groups to activate and deactivate them. 
  \end{minipage}
\end{figure}
}

\subsubsection{\textbf{Deck Steering Canvas and Slide Steering Overlay (DP1, DP3)} }
 Users can generate and modify single slides or entire slide decks using Intent Tags.
The \textit{Deck Steering Canvas} allows users to steer the (re)generation of entire slide decks. For example, specifying the number of total slides or changing the overall tone of the deck's text from brief to more verbose. 
Alternatively, the \textit{Slide Steering Overlay} allows users to steer the (re)generation and explore alternatives of single slides.

\subsubsection{\textbf{Outline Editor (DP1)}}

Parallel to slides, the system also generates a presentation outline, which can be optionally edited using the outline editor. This editor also allows participants to start a new presentation by first drafting an outline or letting the system generate an outline from provided intent tags.  

\subsubsection{\textbf{Editing Intent Tags, Tag Widgets, and System Suggestions (DP3, DP4, DP5)}}

Intent tags allow users to formulate their intent in several ways, such as manually creating and editing tags or choosing from options generated through LLM-driven dynamic adaptive UI mechanisms:

\aptLtoX[graphic=no,type=html]{\begin{figure}[H]
%  \begin{minipage}[t]{3cm}
    \includegraphics[align=t,width=\columnwidth]{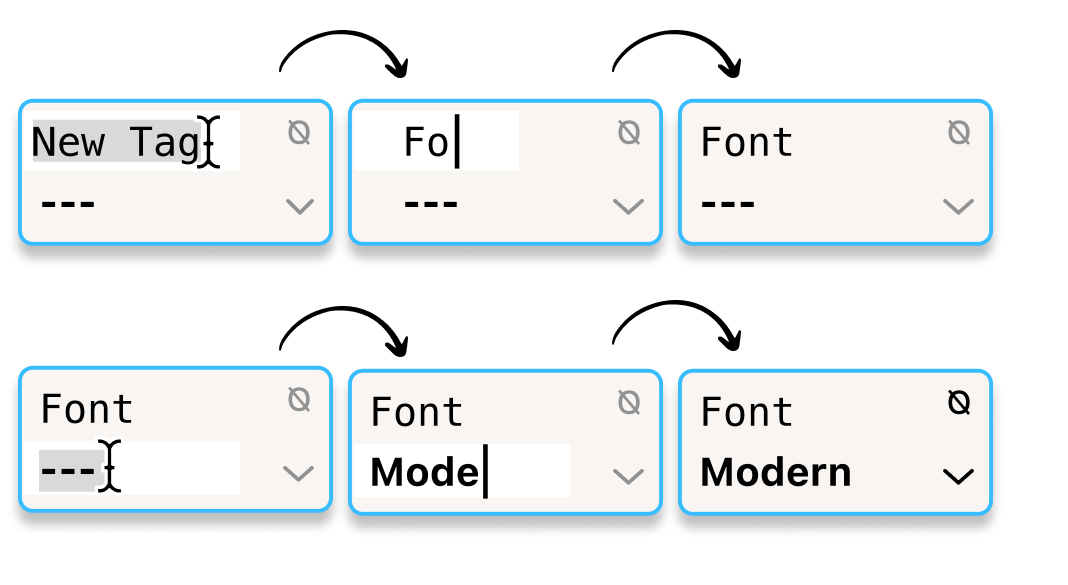}%
    \caption{ \ ~ \ ~ \ ~ \ ~ \ }
    \label{fig:crafting-tags}
    \Description{The figure shows the process of manually creating tags by typing into two text fields. The first example displays an incomplete tag with “New Tag” and an empty text field. Below it, examples show typed tags such as “Font: Modern” and “Font: Mode,” with the latter still being typed. Arrows indicate the progression of typing from incomplete to completed tags. This illustrates how users can manually define tag properties by entering custom labels and values.}
%  \end{minipage}
 %  \begin{minipage}[t]{\dimexpr\columnwidth-3.2cm\relax} 
%  \end{minipage}
\end{figure}          \textbf{Manually Crafting Tags}
          Users can manually create and modify tags by freely typing words or phrases into its two text fields, such as "Font: Modern" or "Tonality: Engaging."
}{\begin{figure}[H]
  \begin{minipage}[t]{3cm}
    \includegraphics[align=t,width=\columnwidth]{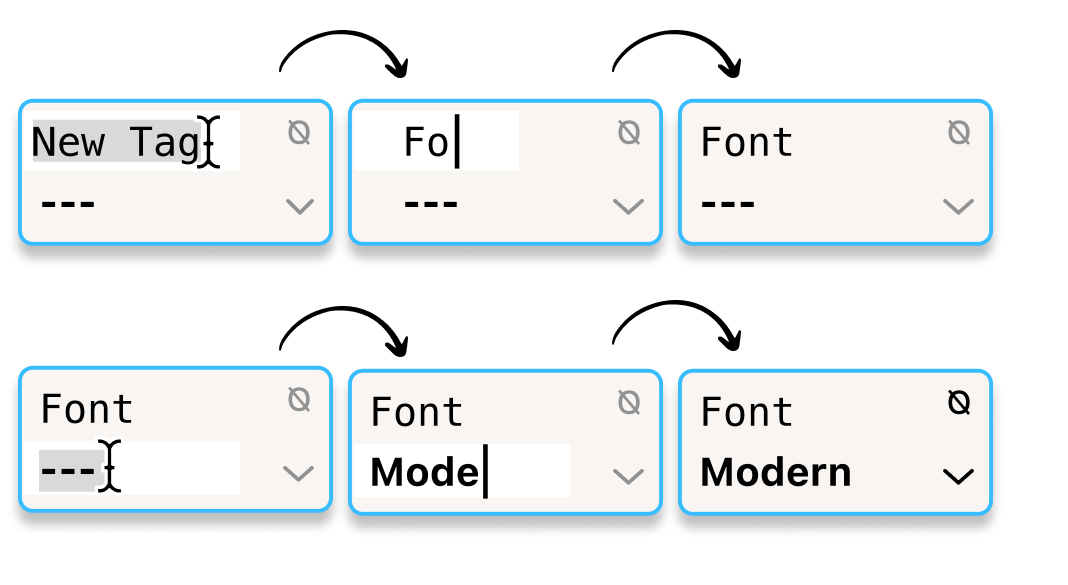}%
    \caption{}
    \label{fig:crafting-tags}
    \Description{The figure shows the process of manually creating tags by typing into two text fields. The first example displays an incomplete tag with “New Tag” and an empty text field. Below it, examples show typed tags such as “Font: Modern” and “Font: Mode,” with the latter still being typed. Arrows indicate the progression of typing from incomplete to completed tags. This illustrates how users can manually define tag properties by entering custom labels and values.}
  \end{minipage}
   \begin{minipage}[t]{\dimexpr\columnwidth-3.2cm\relax}
   
          \textbf{Manually Crafting Tags}
          Users can manually create and modify tags by freely typing words or phrases into its two text fields, such as "Font: Modern" or "Tonality: Engaging."
  \end{minipage}
\end{figure}}

\aptLtoX[graphic=no,type=html]{
\begin{figure}[H]
%  \begin{minipage}[t]{3cm}
    \includegraphics[align=t,width=\columnwidth]{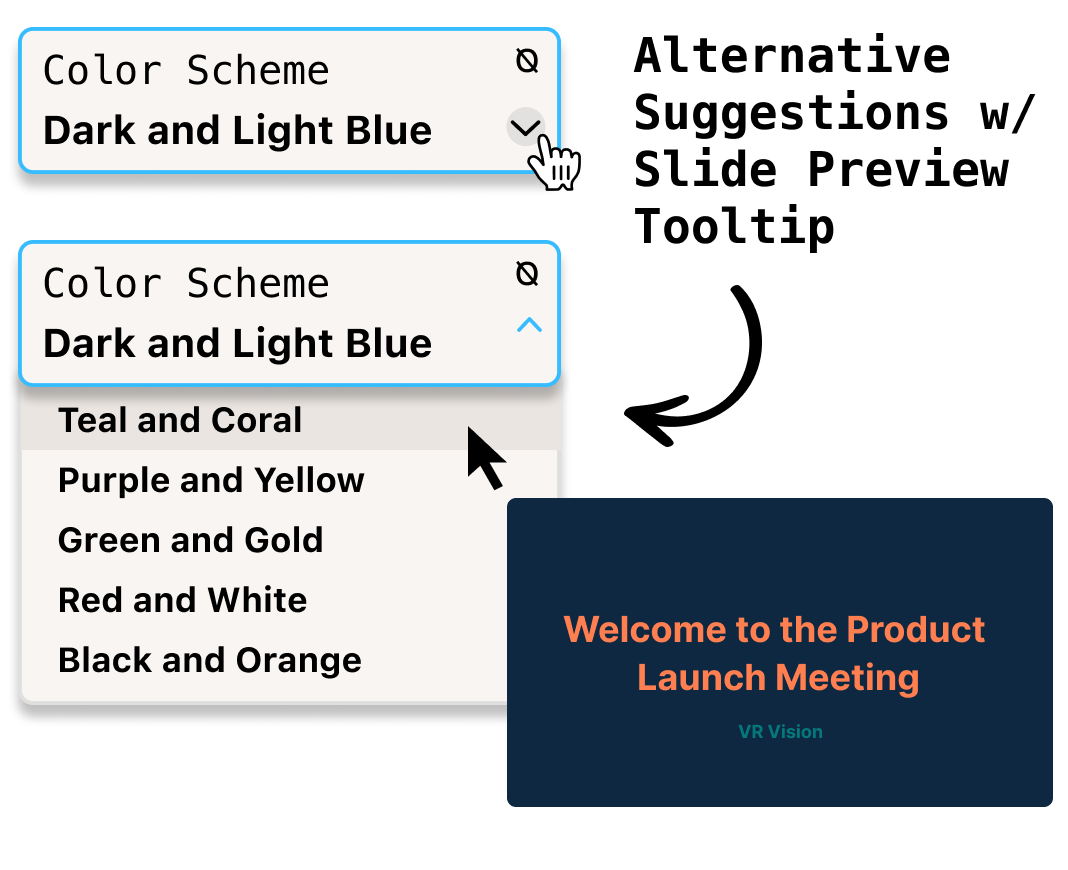}%
    \caption{ \ ~ \ ~ \ ~ \ ~ \ }
    \label{fig:drop-down}
    \Description{The figure shows a drop-down list for the “Color Scheme” tag, currently set to “Dark and Light Blue.” Below it are alternative color scheme options such as “Teal and Coral,” “Purple and Yellow,” and “Green and Gold.” A slide preview tooltip appears next to the list, showing a thumbnail of a slide with the selected “Dark and Light Blue” scheme. This demonstrates how users can explore alternative properties for tags and view pre-generated visual previews in hover tooltips.}
  %\end{minipage}
   %\begin{minipage}[t]{\dimexpr\columnwidth-3.2cm\relax}
%  \end{minipage}
\end{figure}\textbf{Drop-Down List} 
Users can also explore alternative properties for each concept tag by browsing its dynamically generated drop-down list, which also provides pre-generated visual previews in a hover tooltip. }{\begin{figure}[H]
  \begin{minipage}[t]{3cm}
    \includegraphics[align=t,width=\columnwidth]{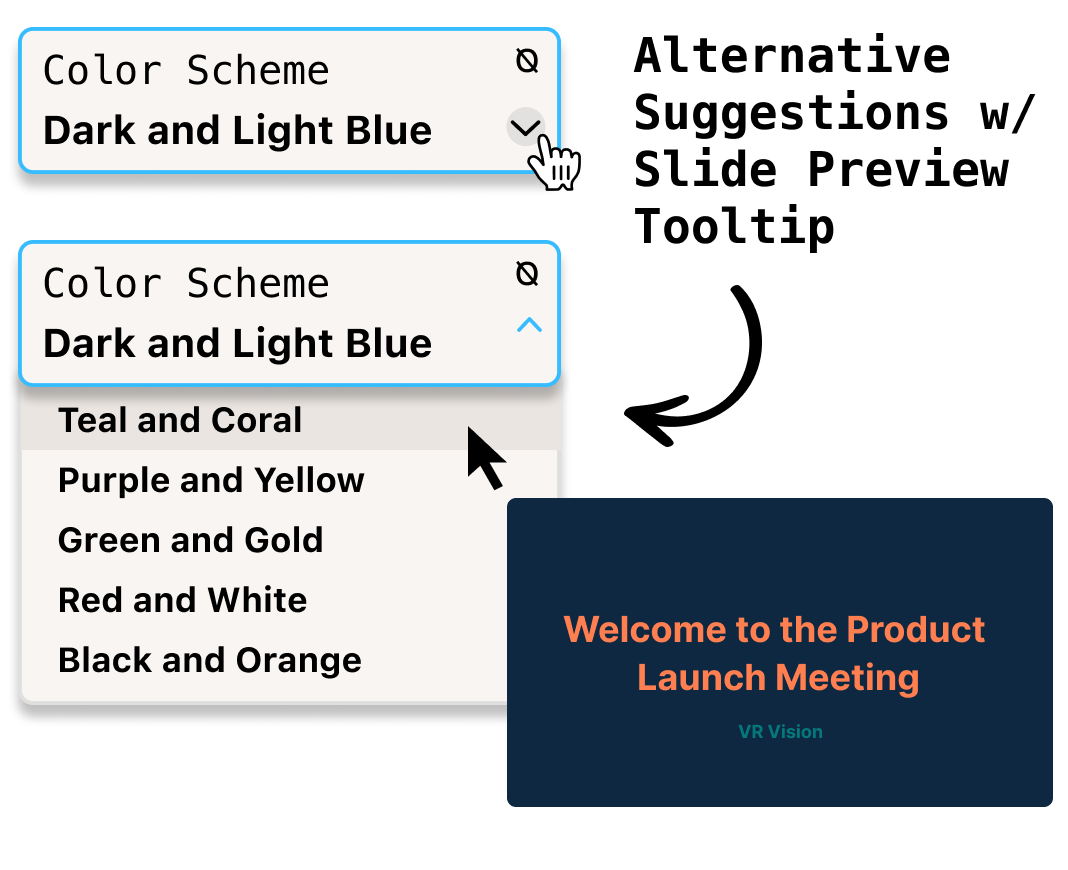}%
    \caption{}
    \label{fig:drop-down}
    \Description{The figure shows a drop-down list for the “Color Scheme” tag, currently set to “Dark and Light Blue.” Below it are alternative color scheme options such as “Teal and Coral,” “Purple and Yellow,” and “Green and Gold.” A slide preview tooltip appears next to the list, showing a thumbnail of a slide with the selected “Dark and Light Blue” scheme. This demonstrates how users can explore alternative properties for tags and view pre-generated visual previews in hover tooltips.}
  \end{minipage}
   \begin{minipage}[t]{\dimexpr\columnwidth-3.2cm\relax}
   
          \textbf{Drop-Down List} 
Users can also explore alternative properties for each concept tag by browsing its dynamically generated drop-down list, which also provides pre-generated visual previews in a hover tooltip. 
  \end{minipage}
\end{figure}}

\aptLtoX[graphic=no,type=html]{\begin{figure}[H]
%  \begin{minipage}[t]{3cm}
    \includegraphics[align=t,width=\columnwidth]{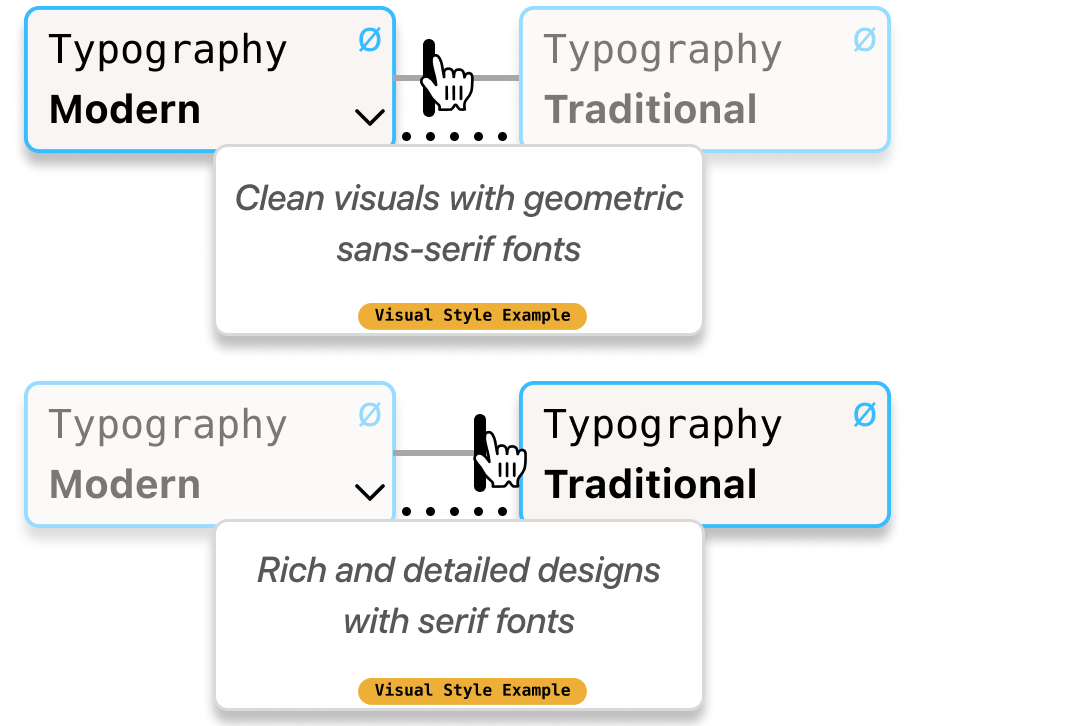}%
    \caption{ \ ~ \ ~ \ ~ \ ~ \ }
    \label{fig:slider}
    \Description{The figure shows a typography tag with a slider widget. The current selection is “Typography: Modern,” and the opposite value is “Typography: Traditional.” The slider allows users to adjust between the two extremes, with a five-step continuous range in between. Below each option are descriptive tooltips: “Clean visuals with geometric sans-serif fonts” for “Modern” and “Rich and detailed designs with serif fonts” for “Traditional.” The slider enables fine-tuning of tag values with visual examples illustrating the effect of each step.}
%  \end{minipage}
%   \begin{minipage}[t]{\dimexpr\columnwidth-3.2cm\relax}
%  \end{minipage}
\end{figure}          \textbf{Opposite Slider Widget}
Users can fine-tune tag values with more granularity by using a slider widget. For this, the system automatically generates an opposite value for each tag with a five-step continuous range slider in between, including descriptive examples illustrating each step's value.
}{\begin{figure}[H]
  \begin{minipage}[t]{3cm}
    \includegraphics[align=t,width=\columnwidth]{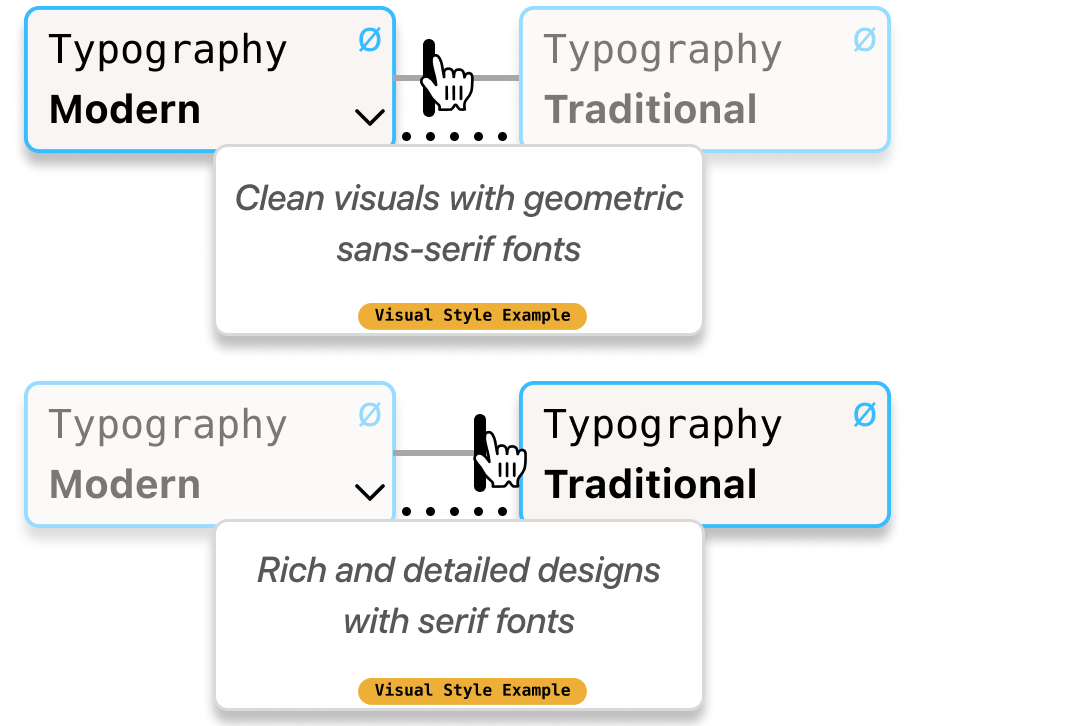}%
    \caption{}
    \label{fig:slider}
    \Description{The figure shows a typography tag with a slider widget. The current selection is “Typography: Modern,” and the opposite value is “Typography: Traditional.” The slider allows users to adjust between the two extremes, with a five-step continuous range in between. Below each option are descriptive tooltips: “Clean visuals with geometric sans-serif fonts” for “Modern” and “Rich and detailed designs with serif fonts” for “Traditional.” The slider enables fine-tuning of tag values with visual examples illustrating the effect of each step.}
  \end{minipage}
   \begin{minipage}[t]{\dimexpr\columnwidth-3.2cm\relax}
          \textbf{Opposite Slider Widget}
Users can fine-tune tag values with more granularity by using a slider widget. For this, the system automatically generates an opposite value for each tag with a five-step continuous range slider in between, including descriptive examples illustrating each step's value.
  \end{minipage}
\end{figure}}

\aptLtoX[graphic=no,type=html]{\begin{figure}[h]
%  \begin{minipage}[t]{3cm}
    \includegraphics[align=t,width=\columnwidth]{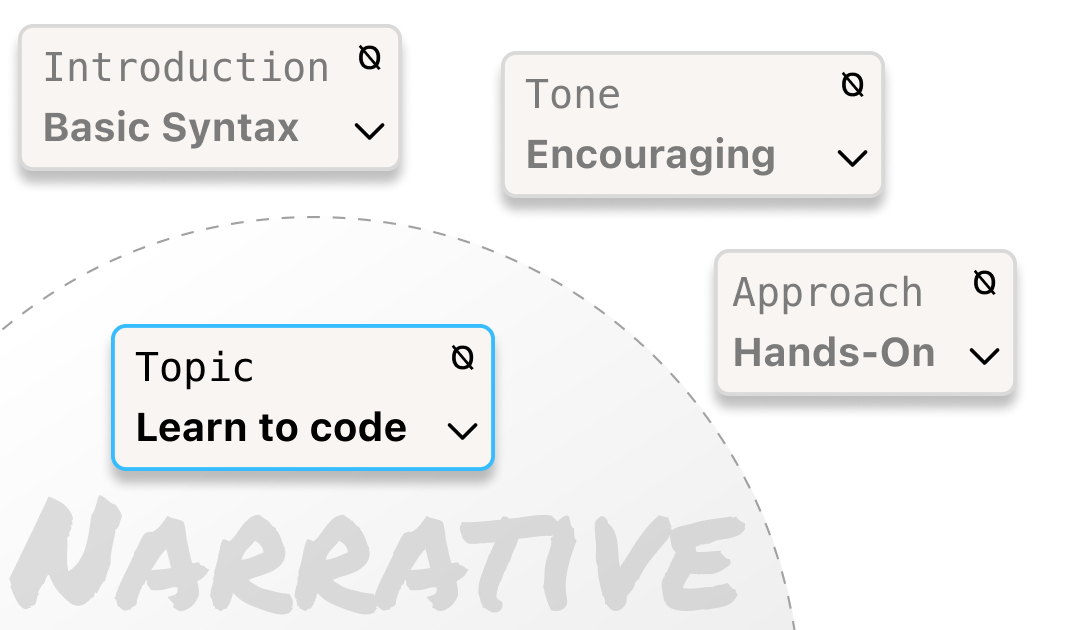}
    \caption{ \ ~ \ ~ \ ~ \ ~ \ }
    \label{fig:tag-suggestions}
    \Description{The figure shows several suggested tags within the “Narrative” tag group. Tags such as “Introduction: Basic Syntax,” “Tone: Encouraging,” “Approach: Hands-On,” and “Topic: Learn to code” are displayed. The suggested tags appear outside the circular “Narrative” group, illustrating that users can drag these suggestions into the group to include them in the content generation process. This allows users to request new tag suggestions for talking points, visual styles, or content sources based on existing tags.}
%  \end{minipage}
%   \begin{minipage}[t]{\dimexpr\columnwidth-3.2cm\relax}
 %  \end{minipage}
\end{figure}         \textbf{Tag Suggestions}
For each tag group, users can request new tag suggestions, such as additional talking points, visual styles, or content sources, from the system based on already specified tags. Generally, all suggested tags first appear outside the circular groups, and users can drag suggested tags into a group to include them in the generation.
}{\begin{figure}[h]
  \begin{minipage}[t]{3cm}
    \includegraphics[align=t,width=\columnwidth]{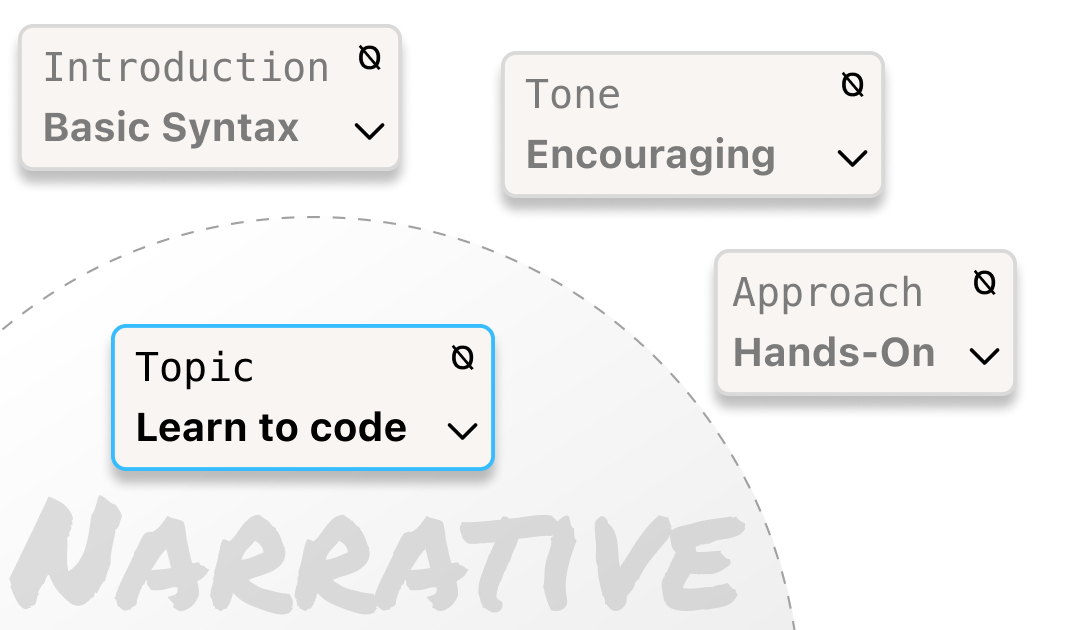}
    \caption{}
    \label{fig:tag-suggestions}
    \Description{The figure shows several suggested tags within the “Narrative” tag group. Tags such as “Introduction: Basic Syntax,” “Tone: Encouraging,” “Approach: Hands-On,” and “Topic: Learn to code” are displayed. The suggested tags appear outside the circular “Narrative” group, illustrating that users can drag these suggestions into the group to include them in the content generation process. This allows users to request new tag suggestions for talking points, visual styles, or content sources based on existing tags.}
  \end{minipage}
   \begin{minipage}[t]{\dimexpr\columnwidth-3.2cm\relax}
          \textbf{Tag Suggestions}
For each tag group, users can request new tag suggestions, such as additional talking points, visual styles, or content sources, from the system based on already specified tags. Generally, all suggested tags first appear outside the circular groups, and users can drag suggested tags into a group to include them in the generation.
  \end{minipage}
\end{figure}}

\aptLtoX[graphic=no,type=html]{\begin{figure}[h]
%  \begin{minipage}[t]{3cm}
    \includegraphics[align=t,width=\columnwidth]{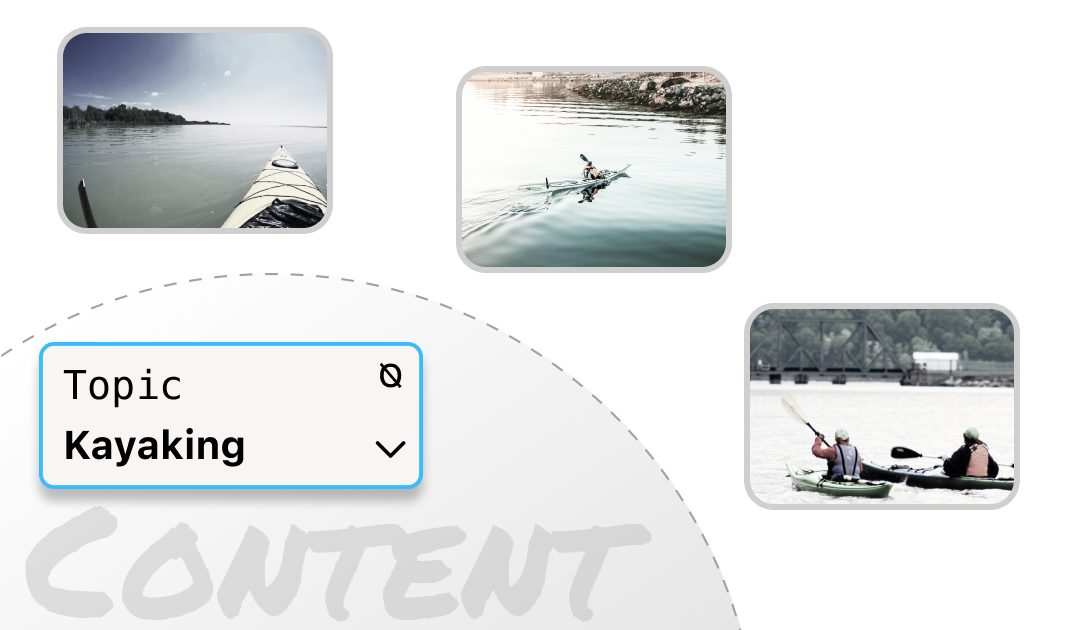}
    \caption{ \ ~ \ ~ \ ~ \ ~ \ }
    \label{fig:image-suggestions}
    \Description{The figure shows image suggestions related to the tag “Kayaking” within the “Content” group. Several images of kayaking and water scenes are displayed around the tag, indicating that the system has retrieved these from an online image search. Users can request system-generated image tags, and each time different images are returned based on the active tags on the board.}
%  \end{minipage}
%   \begin{minipage}[t]{\dimexpr\columnwidth-3.2cm\relax}
%  \end{minipage}
\end{figure}          \textbf{Image Suggestions}
In addition to tags, users can also request system suggestions for new image tags. When requested, each time the system returns different images from a background online image search related to the board's active tags.  
}{\begin{figure}[h]
  \begin{minipage}[t]{3cm}
    \includegraphics[align=t,width=\columnwidth]{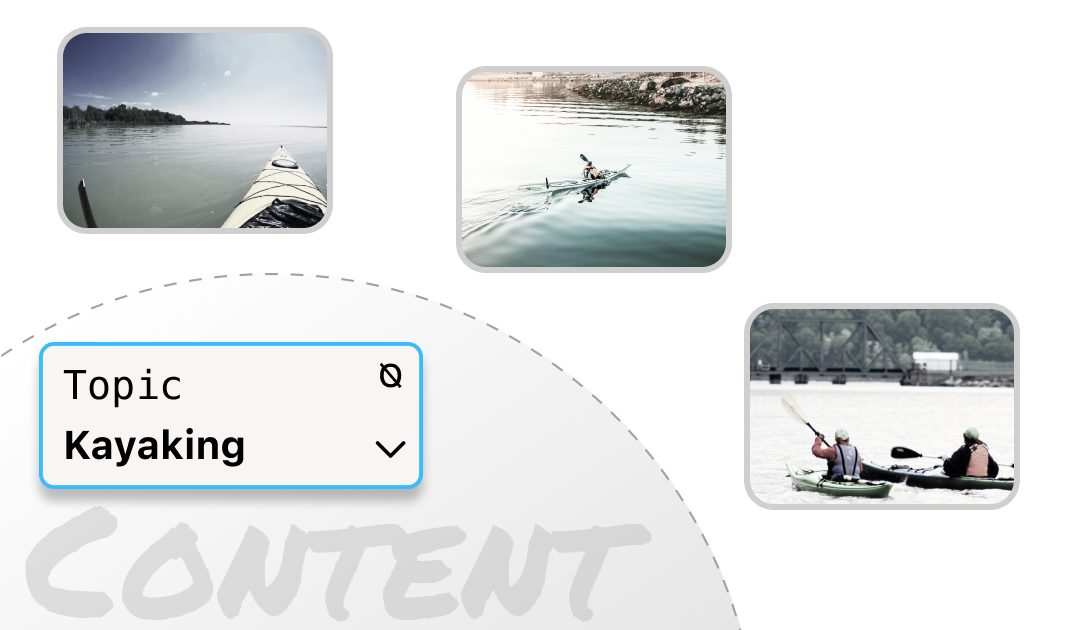}
    \caption{}
    \label{fig:image-suggestions}
    \Description{The figure shows image suggestions related to the tag “Kayaking” within the “Content” group. Several images of kayaking and water scenes are displayed around the tag, indicating that the system has retrieved these from an online image search. Users can request system-generated image tags, and each time different images are returned based on the active tags on the board.}
  \end{minipage}
   \begin{minipage}[t]{\dimexpr\columnwidth-3.2cm\relax}
          \textbf{Image Suggestions}
In addition to tags, users can also request system suggestions for new image tags. When requested, each time the system returns different images from a background online image search related to the board's active tags.  
  \end{minipage}
\end{figure}}

\begin{figure*}[h!]
  \centering
  \includegraphics[width=\linewidth]{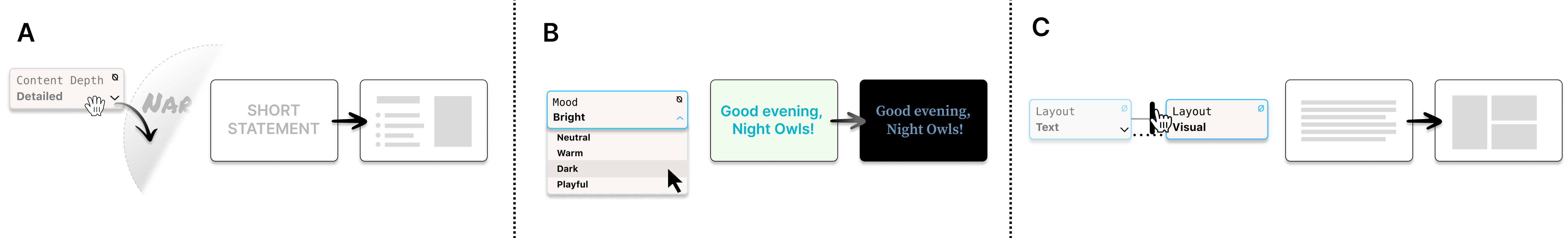}
  \caption{Examples of steering GenAI slide generation with Intent Tags: (A) Influencing the detail level of the displayed information, (B) steering the "mood" from "bright" to "dark," or (C) gradually blending the layout between text- and image-heavy.    }
  \Description{The figure shows three examples of how Intent Tags can be used to steer slide generation. (A) A tag labeled “Content Depth: Detailed” is used to influence the amount of information shown, transitioning from a short statement to a more detailed layout. (B) The “Mood” tag is being adjusted from “Bright” to “Dark,” changing the background color and text style from light to dark. (C) The “Layout” tag is set to adjust between “Text” and “Visual,” gradually blending the layout from a text-heavy format to a more image-focused layout. These examples illustrate how users can fine-tune the content, mood, and layout of slides using intent tags.}
  \label{fig:tag-impact}
\end{figure*}

\subsubsection{\textbf{Steering Slide Generation}}

To summarize the different ways of steering slide generation with Intent Tags, Figure \ref{fig:tag-impact} illustrates how including a tag suggestion modifies the content of a slide, using the drop-down list changes a slide's mood, and using the opposite slider widget blends between a text- and image-heavy layout.

\subsubsection{\textbf{Previews of Slides and Slider Values (DP5)}}
IntentTagger integrates several preview features, allowing users to explore alternative tag values without committing to long waiting times: The tag's \textit{Drop-down} list shows pre-generated previews as tooltips of how the current slide would look like with that tag value applied (Figure \ref{fig:drop-down}). 
The \textit{Opposite Slider} widget shows generated explanations describing the resulting effect if applied to the generation (Figure \ref{fig:slider}). All tag previews and explanations get asynchronously pre-generated in the background, allowing users to explore these options in real time.

\subsubsection{\textbf{Referencing External Documents (DP2)}}
Besides \textit{Concept Tags}, users can also include external files for the generative slide creation as \textit{Reference Tags}. 

\aptLtoX[graphic=no,type=html]{\begin{figure}[H]
%  \begin{minipage}[t]{3cm}
    \includegraphics[align=t,width=\columnwidth]{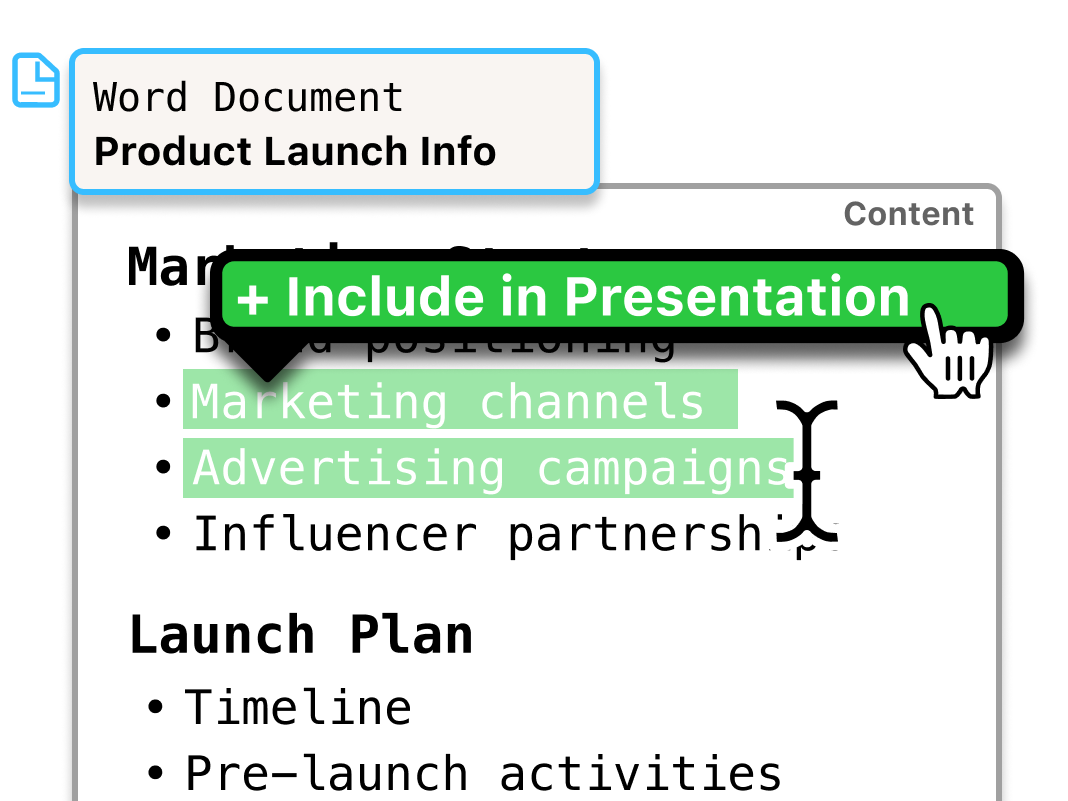}
    \caption{ \ ~ \ ~ \ ~ \ ~ \ }
    \label{fig:word-tag}
    \Description{The figure shows a Word document titled “Product Launch Info” being used as a content source for the presentation. The content from the document is displayed, with the user hovering over a section labeled “Marketing channels,” revealing a green button labeled “+ Include in Presentation.” This allows users to selectively include specific parts of the document, such as “Launch Plan” and “Timeline,” into the generated presentation by using a text selection pop-up.}
%  \end{minipage}
%   \begin{minipage}[t]{\dimexpr\columnwidth-3.2cm\relax}
%  \end{minipage}
\end{figure}          \textbf{Including Content From Word Documents}
Users may drag in Word documents as content sources for the generated presentation. Optionally, users can select sections through a text selection pop-up to include only specific parts in the generated 
 presentation. 
}{\begin{figure}[H]
  \begin{minipage}[t]{3cm}
    \includegraphics[align=t,width=\columnwidth]{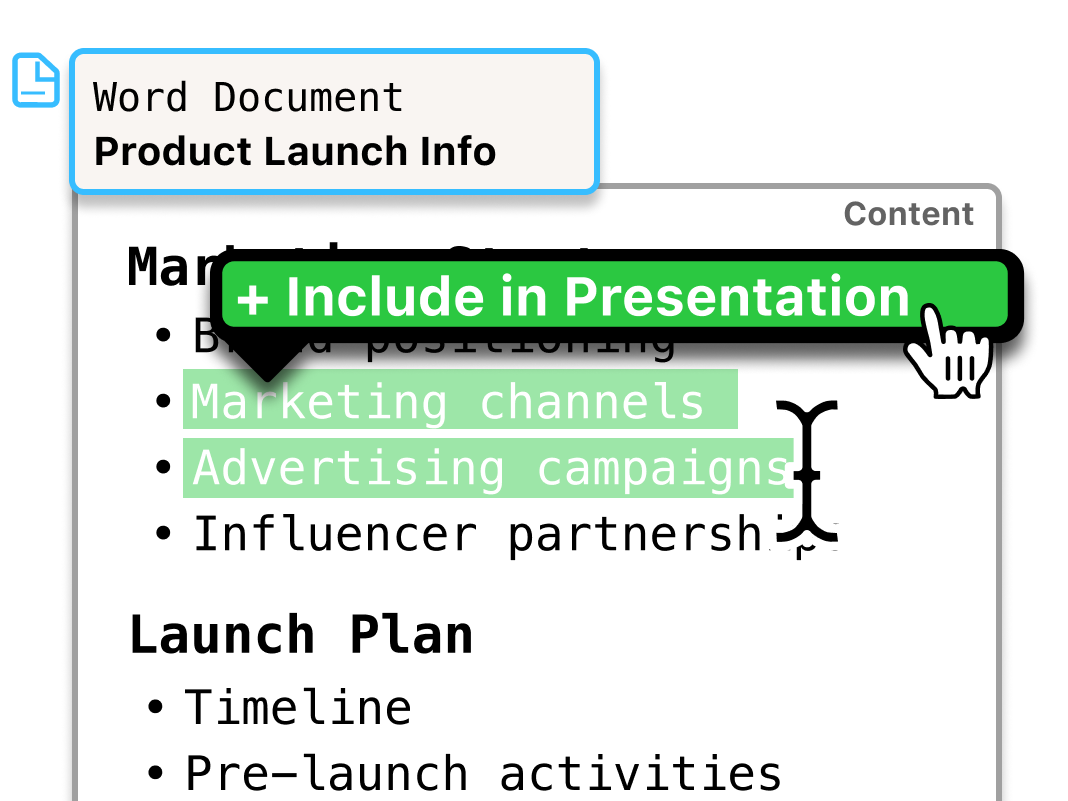}
    \caption{}
    \label{fig:word-tag}
    \Description{The figure shows a Word document titled “Product Launch Info” being used as a content source for the presentation. The content from the document is displayed, with the user hovering over a section labeled “Marketing channels,” revealing a green button labeled “+ Include in Presentation.” This allows users to selectively include specific parts of the document, such as “Launch Plan” and “Timeline,” into the generated presentation by using a text selection pop-up.}
  \end{minipage}
   \begin{minipage}[t]{\dimexpr\columnwidth-3.2cm\relax}
          \textbf{Including Content From Word Documents}
Users may drag in Word documents as content sources for the generated presentation. Optionally, users can select sections through a text selection pop-up to include only specific parts in the generated 
 presentation. 
  \end{minipage}
\end{figure}}

\aptLtoX[graphic=no,type=html]{\begin{figure}[H]
%  \begin{minipage}[t]{3cm}
    \includegraphics[align=t,width=\columnwidth]{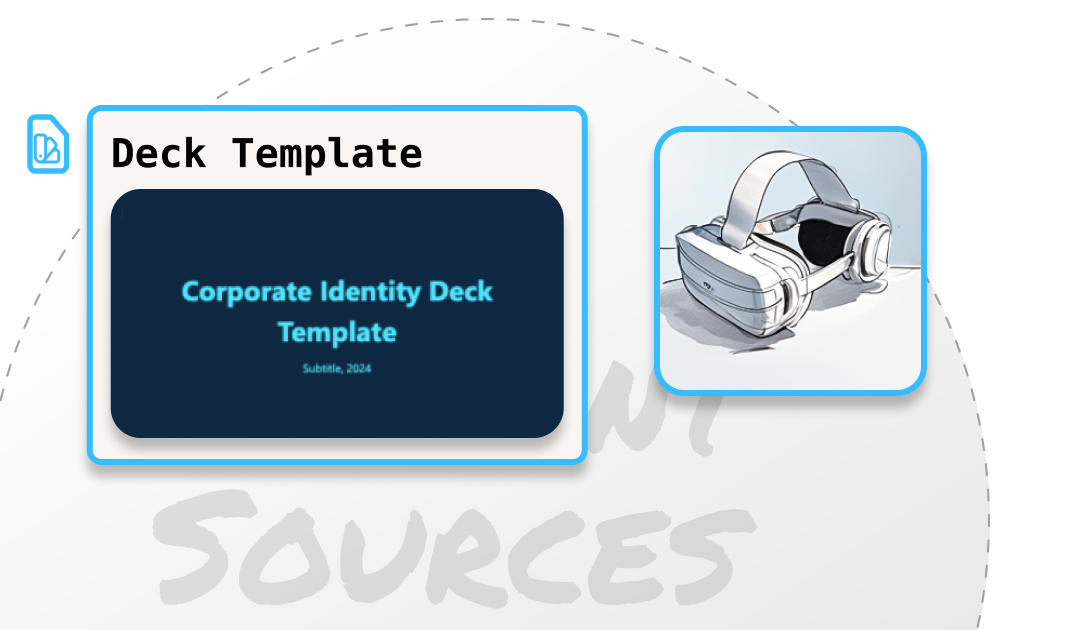}
    \caption{ \ ~ \ ~ \ ~ \ ~ \ }
    \label{fig:include-images}
    \Description{The figure shows a “Deck Template” tag titled “Corporate Identity Deck Template” alongside an image of a 3D headset. These elements are within the “Content Sources” group, illustrating how users can drag external images or existing slide deck templates onto the canvas to include them in the presentation. The templates can serve as visual references or be used to structure the presentation.}
%  \end{minipage}
%   \begin{minipage}[t]{\dimexpr\columnwidth-3.2cm\relax}
%  \end{minipage}
\end{figure}          \textbf{Including Images and Slide Deck Templates }
Users can drag external images onto the canvas to include these in the presentation. 
Users also have the option to reference other existing slide decks as visual templates. 
}{\begin{figure}[H]
  \begin{minipage}[t]{3cm}
    \includegraphics[align=t,width=\columnwidth]{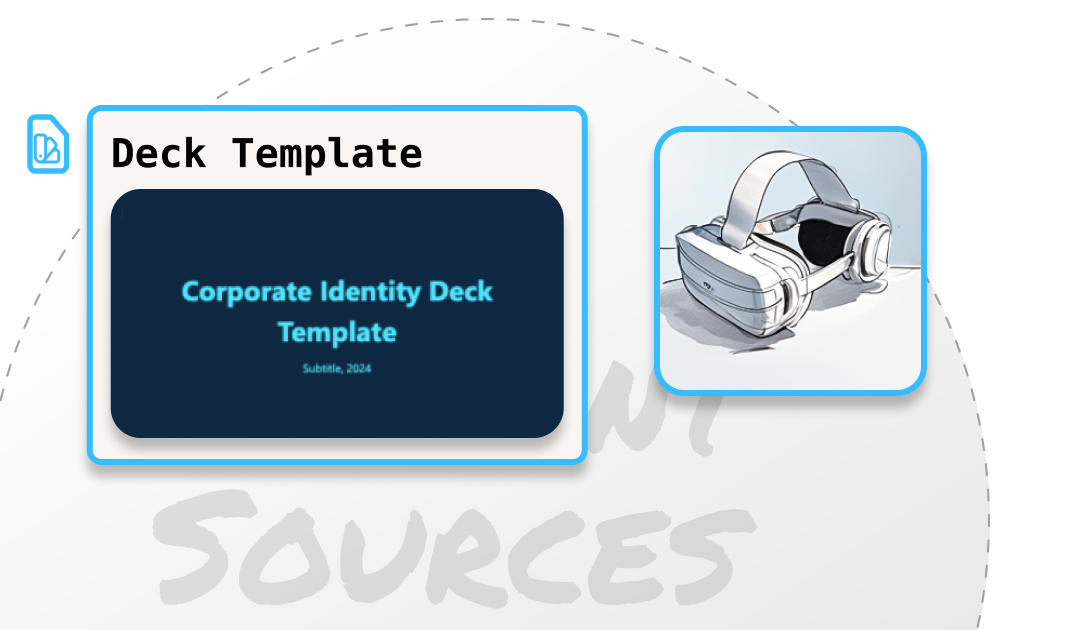}
    \caption{}
    \label{fig:include-images}
    \Description{The figure shows a “Deck Template” tag titled “Corporate Identity Deck Template” alongside an image of a 3D headset. These elements are within the “Content Sources” group, illustrating how users can drag external images or existing slide deck templates onto the canvas to include them in the presentation. The templates can serve as visual references or be used to structure the presentation.}
  \end{minipage}
   \begin{minipage}[t]{\dimexpr\columnwidth-3.2cm\relax}
          \textbf{Including Images and Slide Deck Templates }
Users can drag external images onto the canvas to include these in the presentation. 
Users also have the option to reference other existing slide decks as visual templates. 
  \end{minipage}
\end{figure}}

\subsubsection{\textbf{Tag Grounding Acts (DP4)}}
While Intent Tagging presents granular and flexible mechanisms for users to communicate their intentions to the generative AI slide creation system, this principle also works backward in so-called \textit{Tag Grounding Acts}, where the system creates intent tags from user inputs such as longer text prompts or slides.

\aptLtoX[graphic=no,type=html]{\begin{figure}[H]
%  \begin{minipage}[t]{3cm}
    \includegraphics[align=t,width=\columnwidth]{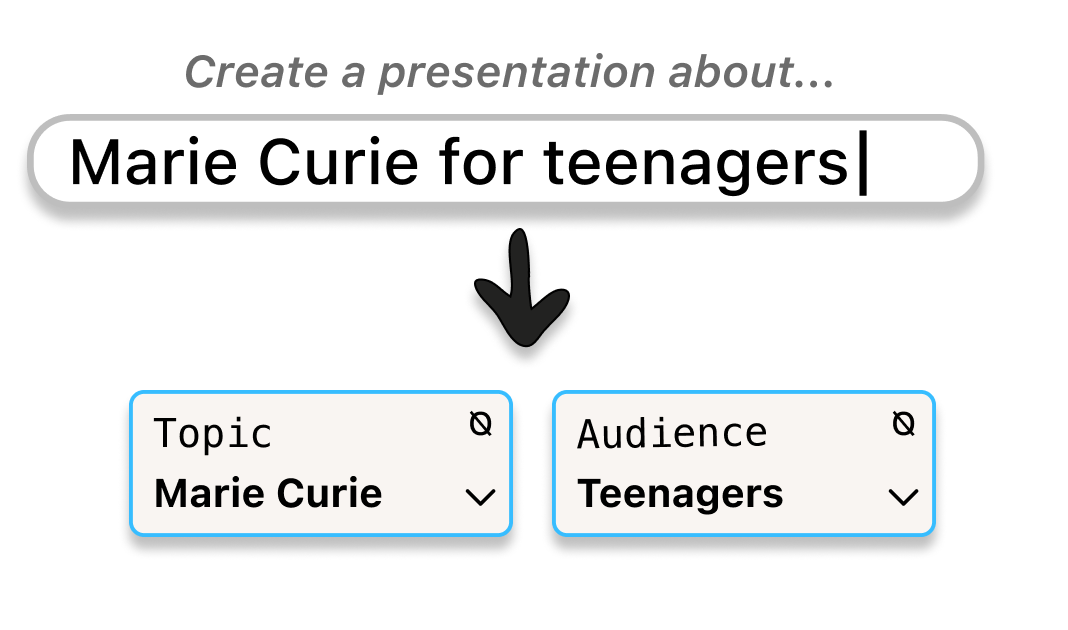}
    \caption{ \ ~ \ ~ \ ~ \ ~ \ }
    \label{fig:text-grounding}
    \Description{The figure shows a text input field where a user types the prompt “Marie Curie for teenagers,” which is automatically parsed into two Intent Tags: “Topic: Marie Curie” and “Audience: Teenagers.” This demonstrates how the system can take a longer text prompt and decompose it into individual tags, allowing users to start a new presentation based on their initial instructions.}
%  \end{minipage}
%  \begin{minipage}[t]{\dimexpr\columnwidth-3.2cm\relax}
%  \end{minipage}
\end{figure}          \textbf{Grounding from Text Prompt}
This feature allows users to start a new presentation by providing instructions via a longer conventional text prompt, which the system will then automatically decompose into individual Intent Tags.
}{\begin{figure}[H]
  \begin{minipage}[t]{3cm}
    \includegraphics[align=t,width=\columnwidth]{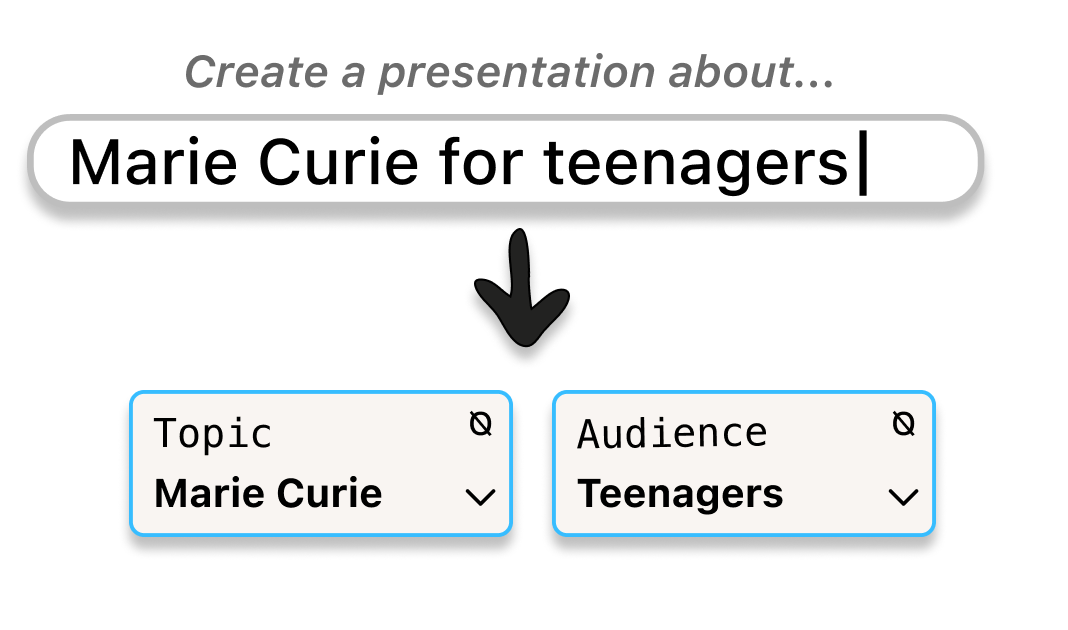}
    \caption{}
    \label{fig:text-grounding}
    \Description{The figure shows a text input field where a user types the prompt “Marie Curie for teenagers,” which is automatically parsed into two Intent Tags: “Topic: Marie Curie” and “Audience: Teenagers.” This demonstrates how the system can take a longer text prompt and decompose it into individual tags, allowing users to start a new presentation based on their initial instructions.}
  \end{minipage}
   \begin{minipage}[t]{\dimexpr\columnwidth-3.2cm\relax}
          \textbf{Grounding from Text Prompt}
This feature allows users to start a new presentation by providing instructions via a longer conventional text prompt, which the system will then automatically decompose into individual Intent Tags.
  \end{minipage}
\end{figure}}

\aptLtoX[graphic=no,type=html]{\begin{figure}[H]
%  \begin{minipage}[t]{3cm}
    \includegraphics[align=t,width=\columnwidth]{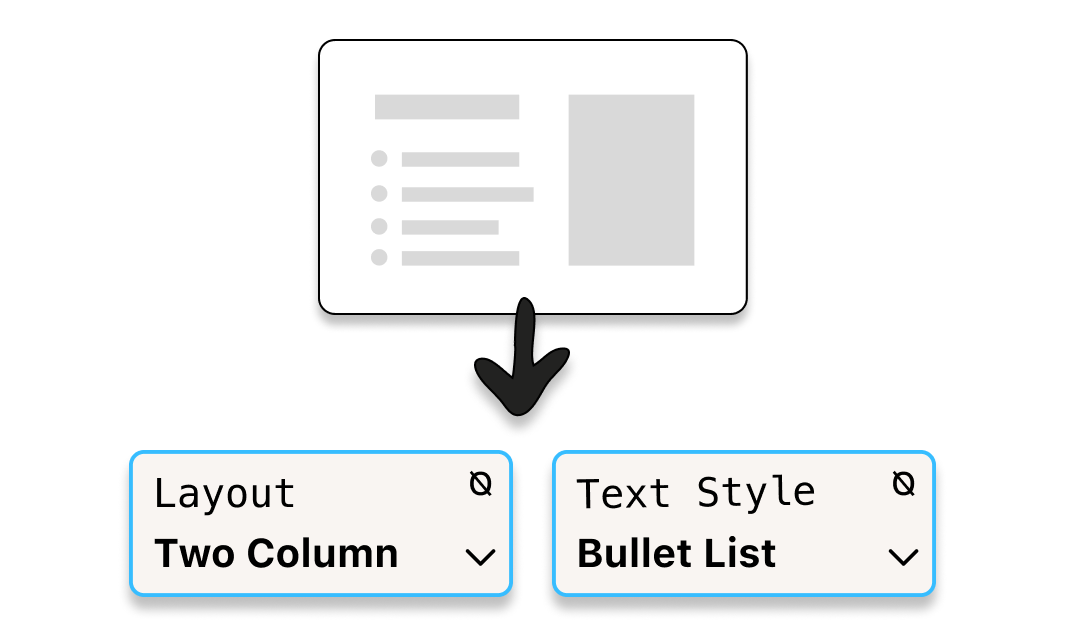}
    \caption{ \ ~ \ ~ \ ~ \ ~ \ }
    \label{fig:slide-grounding}
    \Description{The figure shows how the system analyzes a slide’s layout and text style when a user invokes the Slide Steering Overlay. The current layout is “Two Column,” and the text style is set to “Bullet List.” The system pre-populates these attributes as intent tags, representing the slide’s current content sources, narrative, and visual style, providing a starting point for users to make adjustments through the intent tagging interface.}
%  \end{minipage}
%   \begin{minipage}[t]{\dimexpr\columnwidth-3.2cm\relax}
%  \end{minipage}
\end{figure}          \textbf{Grounding from Slide}
Each time a user invokes the \textit{Slide Steering Overlay} to adjust single slides, the system first analyzes the slide's \textit{content sources}, \textit{narrative}, and \textit{visual style} and then pre-populates the interface with intent tags representing the slide's current state. This provides an easy starting point for users to make adjustments through Intent Tags. 
}{\begin{figure}[H]
  \begin{minipage}[t]{3cm}
    \includegraphics[align=t,width=\columnwidth]{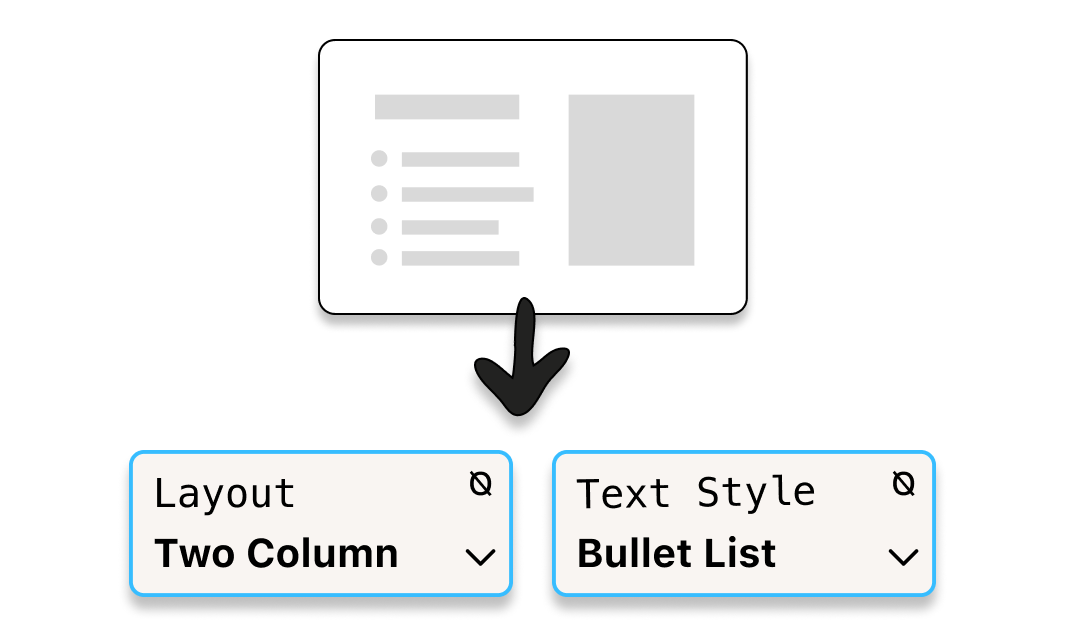}
    \caption{}
    \label{fig:slide-grounding}
    \Description{The figure shows how the system analyzes a slide’s layout and text style when a user invokes the Slide Steering Overlay. The current layout is “Two Column,” and the text style is set to “Bullet List.” The system pre-populates these attributes as intent tags, representing the slide’s current content sources, narrative, and visual style, providing a starting point for users to make adjustments through the intent tagging interface.}
  \end{minipage}
   \begin{minipage}[t]{\dimexpr\columnwidth-3.2cm\relax}
          \textbf{Grounding from Slide}
Each time a user invokes the \textit{Slide Steering Overlay} to adjust single slides, the system first analyzes the slide's \textit{content sources}, \textit{narrative}, and \textit{visual style} and then pre-populates the interface with intent tags representing the slide's current state. This provides an easy starting point for users to make adjustments through Intent Tags. 
  \end{minipage}
\end{figure}}

\subsection{Implementation Details}

IntentTagger is implemented in TypeScript using ReactJS \cite{react.js_reactjs_2024} with ReactFlow \cite{reactflow_react_2024} for the tag canvas interface and Blocknote \cite{blocknote_blocknote_2024} for the text editor features. 
Mommoth.js \cite{williamson_mwilliamson_2024} is used for importing Word documents, and slide rendering is based on a modified fork of spectacle.js \cite{formidablelabs_formidablelabs_2024}. 
For generating slides and adaptive UI, we use the official OpenAI API \cite{openai_openai_2024} to execute prompts using “gpt-4o” and the Bing search API \cite{microsoft_bing_2024} for image suggestions.
\rev{To suggest new tags, we prompt GPT to generate a list of \textit{[label:value]} pairs for each tag group to augment the existing set of active intent tags.
For outlines, GPT is instructed to generate a presentation outline in markdown format from a list of a user's active concept and reference tags (including images if present).
To generate slides, we then prompt GPT to return a template-based JSON slide deck representation by providing the outline, active intent tags, and deck references (if specified).}
For generating the drop-down values, slide previews, and slider explanations, the system asynchronously requests these from GPT in the background for each new or modified tag on the canvas.
\rev{Please see the Appendix for detailed prompts.}

\section{User Study}
To better understand the possible benefits and limitations of intent tagging-based interactions, we conducted a lab user study aimed at providing insights into these research questions:
\begin{itemize}[font=\bfseries,
  align=left]
    \item[RQ1] \textit{What are the key differences between chat-based and intent tag-based interactions with GenAI?}
    \item[RQ2] \textit{How do people work with intent tags?}
    \item[RQ3] \textit{What are users’ perceived benefits and challenges for GenAI-driven slide creation with intent tags?}
   
 \end{itemize}

\rev{For RQ1, we decided to compare intent tag-based interactions with chat-based and design gallery-based approaches (text prompting and choosing from a set of options) since these represent the currently most common forms of interacting with GenAI for content creation in commercial systems, such as in OpenAI's ChatGPT/DALL-E \cite{betker_improving_2023}, Adobe's Firefly \cite{adobe_adobe_2024} or Microsoft's PowerPoint Copilot \cite{microsoft_powerpoint_2024b} and Designer \cite{microsoft_powerpoint_2024a} features.}

\subsection{Participants}
We recruited 12 participants\textit{ (8 self-identified as females, 3 males, and 1 non-binary, age M=31.6 years (SD = 7.65 ))} with professional slide presentation creation experience via email lists at a large software company. The participant pool consisted of individuals with diverse job titles, and the majority of selected participants used PowerPoint at least multiple times per month in their jobs (see Table \ref{tab:participants} in the Appendix). All participants had previous experience using the \textit{Designer} and \textit{Copilot} features in PowerPoint, ensuring existing familiarity with generative AI functionalities for slide deck creation. Participants signed an IRB-approved consent form and were compensated with a \$50 gift card after study completion. 

\subsection{Procedure and Tasks} \label{sec:study-procedure}
The lab study was structured into four phases: 

\textbf{1) On-Boarding (20 min):} At the beginning of the session, after a general study introduction, participants watched a video tutorial demonstrating IntentTagger's core functionalities with a step-by-step example. Following the video, participants were asked to complete two five-minute guided hands-on structured tasks (editing an existing slide and making adjustments to an entire deck) to familiarize themselves with the tools' interface and operation.

\textbf{2) Comparative Tasks ( 2 x 10 min): }
After the onboarding phase, participants completed two comparative tasks, each lasting \rev{10} minutes, to evaluate their ability to create presentations using IntentTagger and a baseline system. 
\rev{In both tasks, participants were asked to create a 6-slide presentation aimed at educating teenagers about the inventions and scientific discoveries of a historical figure (\textit{“The Discoveries of Marie Curie”} or \textit{“The Inventions of Nikola Tesla”}).
Participants alternated between using IntentTagger and Microsoft PowerPoint \cite{microsoft_powerpoint_2024} across the two tasks in randomized order. 
For PowerPoint, participants were restricted to using only the integrated Copilot feature \cite{microsoft_powerpoint_2024b} (chatbot interface with optional document upload) and the Designer feature \cite{microsoft_powerpoint_2024a} (design gallery for slide layouts) for slide generation and modification. 
As a starting point, participants were provided with a Word document containing the relevant Wikipedia article in both tasks.} 
Participants were required to think aloud during the tasks, and after completing each task, they filled out a survey with attitudinal 6-point Likert scale questions. 
To mitigate order effects, the order of systems and presentation topics was randomized across participants. 

\textbf{3) \rev{Semi-structured} Task (10 min):}
In the third phase, participants were tasked with creating a 7-slide presentation from scratch using IntentTagger. 
They chose a topic related to a hobby they enjoy, aiming to convince others of its value and explain how to get started. 
Participants had ten minutes to complete the task, using only the prototype system without directly editing text or images on the canvas. 
They were required to think aloud as they worked, and before starting, they briefly described their topic and intentions for the presentation. 
\rev{We created this task to encourage participants to engage more freely with IntentTagger’s features, focusing on content creation and presentation design related to a topic they are knowledgeable about and emotionally connected to.
Since IntentTagger's deck generation time increases per slide, we deliberately limited the number of slides to six and seven per task in phases 2 and 3 to keep the processing time for each cycle within 15 seconds.}

\textbf{4) Exit Interview (20 min): }
In the final phase, participants participated in a semi-structured interview to provide feedback on their experience with IntentTagger, focusing on its overall utility, comparison with tools like PowerPoint’s Copilot, and the effectiveness of intent tagging interactions for slide creation. They shared insights on the tool’s strengths, areas for improvement, and potential integration into their professional workflows.

\subsection{Collected Data, Measures, and Analysis}

Across the study, we collected the following data:
\begin{itemize}
\item  Video, screen, and audio recordings and machine-generated transcripts of the task think-aloud sessions 
\item Audio recordings and machine-generated transcripts of the post-task interviews 
\item Post-task survey data
\item Participant-created presentations and related IntentTagger project files
\end{itemize}

To compare chat-based and intent tag-based interactions (Q1), we analyzed the post-task surveys from phase 2 that probed on participants' perceived ease of use, efficiency, and control over the slide generation process on a 6-point Likert scale. We applied the Wilcoxon signed-rank test to assess statistical significance and calculated 95\% confidence intervals for mean differences via bootstrapping with 10,000 replications using R \cite{rcoreteam_language_2024}. This approach has been suggested for similar data and studies \cite{zhu_assessing_2018, masson_statslator_2023}.

To answer how people work with intent tags (Q2), we conducted a video interaction analysis \cite{baumer_comparing_2011} of the video recordings collected in the semi-structured task \textit{(study phase 3)}. We manually coded participants' interactions with IntentTagger's features in Atlas.ti \cite{atlas.ti_atlasti_2024}, such as their interactions with tags or the moments they triggered the outline and slide generation.  

Finally, to investigate users’ perceived benefits and challenges of intent tags (Q3), we conducted a reflexive thematic analysis \cite{braun_reflecting_2019} of the interview transcripts. We followed an iterative inductive coding process (using Marvin \cite{marvin_marvin_2024}) and generated themes through affinity diagramming using Miro \cite{miro_miro_2024}.

\begin{figure*}[t]
  \centering
  \includegraphics[width=1.025\linewidth]{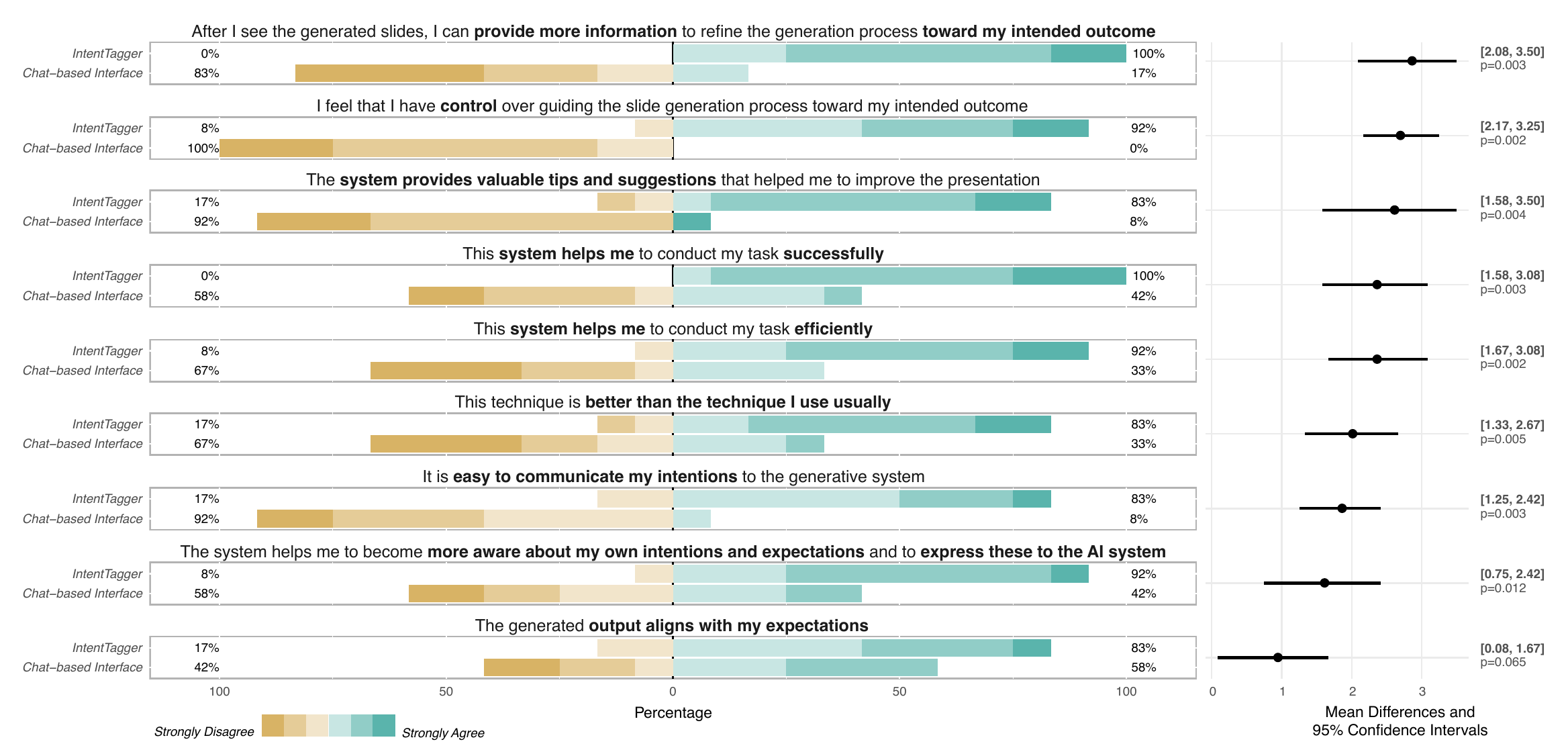}
  \caption{(Left) Participants’ responses when rating the 6-point Likert statements for IntentTagger and Chat-based Interface, \rev{ranked from largest to smallest effects}; (Right) Dots show the mean difference of IntentTagger compared to Chat-based Interface; Bars are the 95\% CIs calculated with the studentized bootstrap method. }
  
  \Description{The figure shows participants’ responses to 6-point Likert scale statements comparing the IntentTagger and Chat-based Interface systems across various dimensions. The left side of the chart displays bar graphs representing the percentage of agreement (from “Strongly Disagree” to “Strongly Agree”) for each statement, with IntentTagger responses in light brown and Chat-based Interface responses in teal. Statements include aspects such as the ability to provide more information, control over the slide generation process, receiving valuable suggestions, task efficiency, and output alignment with expectations. The right side shows the mean difference between the two interfaces, indicated by black dots, with bars representing 95\% confidence intervals using the studentized bootstrap method. The chart highlights significant contrasts in user perceptions of both systems.}
  \label{fig:survey-results}
\end{figure*}

\section{Study Findings}

Participants went through the tasks without abandoning them. 

\subsection{Key differences: Chat vs Intent Tag-based Interactions (RQ1)}

\subsubsection{Observations}

In the two comparative tasks (phase \rev{2}), participants had to create a presentation about Nikola Tesla or Marie Curie on six slides using engaging images and language suitable for teenagers, based on a provided Word document containing information from a Wikipedia article.
Here, we report on the observed key differences between participants' processes during these tasks when working with the chat-based system and IntentTagger. 
\\ \\ 
\textbf{Chat-based System.} In the chat-based system, participants typically began by providing a straightforward task prompt, such as \textit{“Create a new presentation from /Tesla.docx about Nikola Tesla for teenagers on six slides.”} However, the system consistently generated a 20-slide PowerPoint presentation based on all content from the document in a generic way, including generic and vaguely related stock images. When participants tried to refine their prompts, such as specifying the desired number of slides or requesting specific content to be highlighted, the system frequently ignored these requests and generated a new slide deck similar to the previous one. 
As a consequence, participants found themselves repeating their instructions or trying different phrasings to guide the system, often without noticeable changes. The system frequently returned messages like, \textit{“I don't recognize that wording as something I can do,”} offering alternative actions that were not relevant to the task. Consequently, participants spend most of their time trying out different text prompts to get the system to generate a more targeted, shorter presentation; however, in all cases, without success. Occasionally, participants went on to tweak the layout and visual style of single slides using the Designer feature by selecting a layout from a design gallery list. While they managed to adjust the style of single slides in isolation, they could not achieve a consistent look and feel for the entire deck this way.

\textbf{IntentTagger}
With IntentTagger, participants typically began by providing the Word document as a reference tag and then continued by providing further concept tags for key elements like the number of slides, audience, and the focus areas of the content.
In addition, many participants used the reference tag's text selection widget to specify specific sections of the provided Word document to ensure that only relevant information was included in the presentation. After specifying tags matching the task brief, most participants also made use of the image suggestion feature from which they added some images to the content source tag group. Then, after generating slides, they either continued to make refinements on single slides and then updated the deck's style based on the adjusted slide, or they continued refining the entire deck by adding or modifying intent tags on the deck steering canvas. Participants also frequently used the system’s tag suggestions to refine the presentation's content and visual style.

From a usability perspective, some participants struggled initially with understanding how to effectively use the intent tags, particularly in determining how granular or broad their tags should be or to which tag group certain tags would belong. However, they soon became more confident and many participants commented on the ability to better convey their intent with intent tags and tag groups: \textit{"it definitely feels like I get something close to what I had in mind [...] Looks like something efficient, I will be happy to use it in my work."} (P09)

\subsubsection{Questionnaire}

In the comparative post-task surveys, participants rated all nine 6-point statements higher for Intent Tagger (Figure \ref{fig:survey-results}). 
\rev{Participants expressed a \textbf{significant improvement in intent elicitation and alignment with IntentTagger} over the chat-based system regarding the 
\textbf{iterative process} (\textit{“After I see the generated slides, I can provide more information to refine the generation toward my intended outcome”}, $MD = 2.83$), 
\textbf{control over guiding the system} (\textit{“I feel that I have control over guiding the slide generation process toward my intended outcome”}, $MD = 2.67$), 
\textbf{intent communication} (\textit{“It is easy to communicate my intentions to the generative system”}, $MD = 1.83$),  
\textbf{meta intent elicitation} (\textit{“The system helps me to become more aware about my own intentions and expectations during the task and to express these to the AI system”}, $MD = 1.58$) and  \textbf{alignment with their expectations} ($M = 0.92$).
Participants also valued the \textbf{guidance for slide creation} of IntentTagger higher (\textit{“The system provides valuable tips and suggestions that helped me to improve the presentation”}, $MD = 2.58$) 
and felt they were able to \textbf{conduct the task more successfully} ($MD = 2.33$) and \textbf{efficiently} ($MD = 2.33$) than with the chat-based system. 
They also felt that IntentTagger would be \textbf{better than the technique they usually use} ($MD = 2.0$).}

\begin{table*}[t]
\caption{Overview of semi-structured task results (phase 3); reporting the self-chosen topic of the slide deck and task process summary statistics; *Note: During P04’s session, the image suggestions failed to load due to an API outage.}
\begin{tabularx}{\textwidth}{p{0.5cm}p{3cm}p{1cm}p{1cm}p{1cm}p{1cm}p{1cm}p{1cm}p{1cm}p{1cm}}
\hline
\rotatebox{55}{\textbf{ID}}
& \rotatebox{55}{\begin{tabular}[c]{@{}l@{}}\textbf{Slide Deck} \\ \textbf{Topic}\end{tabular}}
& \rotatebox{55}{\begin{tabular}[c]{@{}l@{}}\textbf{Task Time}\\(mm:ss)\end{tabular}}
& \rotatebox{55}{\begin{tabular}[c]{@{}l@{}}\textbf{\# Outlines} \\ \textbf{Generated}\end{tabular}}
& \rotatebox{55}{\begin{tabular}[c]{@{}l@{}}\textbf{\# Decks} \\ \textbf{Generated}\end{tabular}} 
& \rotatebox{55}{\textbf{\# Total Tags }} 
& \rotatebox{55}{\begin{tabular}[c]{@{}l@{}}\textbf{\# \rev{ConceptTag}} \\ \textbf{Sugg Included}\end{tabular}} 
& \rotatebox{55}{\begin{tabular}[c]{@{}l@{}}\textbf{\# ImgTag} \\ \textbf{Sugg Included}\end{tabular}} 
& \rotatebox{55}{\begin{tabular}[c]{@{}l@{}}\textbf{\% Sugg} \\ \textbf{Included}\end{tabular}} 
& \rotatebox{55}{\begin{tabular}[c]{@{}l@{}}\textbf{\# Drop-Dwn} \\ \textbf{Included}\end{tabular}} \\
\hline
P01         & Hiking                                                                 & 11:19                                                                    & 3                                                                      & 4                                                                   & 14                                                            & 3                                                                    & 4                                                                      & 50\%                                                                          & 3                                                                        \\
P02         & Yoga                                                                   & 06:32                                                                    & 3                                                                      & 3                                                                   & 21                                                            & 12                                                                   & 5                                                                      & 81\%                                                                          & 2                                                                        \\
P03         & Kayaking                                                               & 07:01                                                                    & 3                                                                      & 2                                                                   & 25                                                            & 13                                                                   & 5                                                                      & 72\%                                                                          & 1                                                                        \\
P04*        & Homemade Pizza                                                         & 08:55                                                                    & 2                                                                      & 3                                                                   & 20                                                            & 17                                                                   & 0*                                                                      & 85\%                                                                          & 4                                                                        \\
P05         & Growing dahlias                                                        & 08:12                                                                    & 2                                                                      & 5                                                                   & 16                                                            & 6                                                                    & 5                                                                      & 69\%                                                                          & 1                                                                        \\
P06         & Racing road bikes                                          & 08:20                                                                    & 2                                                                      & 1                                                                   & 14                                                            & 4                                                                    & 3                                                                      & 50\%                                                                          & 2                                                                        \\
P07         & Skateboarding                                                          & 07:53                                                                    & 1                                                                      & 3                                                                   & 12                                                            & 8                                                                    & 3                                                                      & 92\%                                                                          & 2                                                                        \\
P08         & Dungeons and dragons                                                   & 06:42                                                                    & 2                                                                      & 1                                                                   & 21                                                            & 9                                                                    & 6                                                                      & 71\%                                                                          & 0                                                                        \\
P09         & Practicing Taekwondo                                                   & 11:43                                                                    & 5                                                                      & 3                                                                   & 11                                                            & 5                                                                    & 3                                                                      & 73\%                                                                          & 2                                                                        \\
P10         & Tennis                                                                 & 09:01                                                                    & 3                                                                      & 4                                                                   & 16                                                            & 8                                                                    & 3                                                                      & 69\%                                                                          & 0                                                                        \\
P11         & Why trombone is cool                                                   & 09:07                                                                    & 3                                                                      & 2                                                                   & 16                                                            & 8                                                                    & 4                                                                      & 75\%                                                                          & 0                                                                        \\
P12         & Paddle boarding                                                        & 09:07                                                                    & 2                                                                      & 2                                                                   & 24                                                            & 17                                                                   & 6                                                                      & 96\%                                                                          & 2                                                                        \\
\hline
\textit{Mean}        &                                                                        & \textit{08:39}                                                                    & \textit{2.6}                                                                    & \textit{2.8 }                                                                & \textit{17.5}                                                          & \textit{9.2}                                                                  & \textit{3.9 }                                                                   & \textit{73.5\%}                                                                        & \textit{1.6}  \\
\bottomrule
\end{tabularx}
\Description{The table presents task statistics for 12 participants (P01 to P12) who created slide decks on self-chosen topics. Columns include Task Time (duration in minutes and seconds), \# Outlines Generated, # Decks Generated, \# Total Tags, \# Tag Suggestions Included, \# Image Suggestions Included, \% Suggestions Included, and \# Drop-Downs Included. Topics range from “Hiking” to “Paddle Boarding.” Mean values are provided at the bottom. Notable rows include P04, where image suggestions failed to load due to an API outage. Participants varied in their use of system-generated suggestions, with a mean of 73.5\% suggestions included and an average task time of 08:39.}\label{tab:task4-results}
\end{table*}

\begin{figure*}[t]
  \centering
  \includegraphics[width=1.15\linewidth]{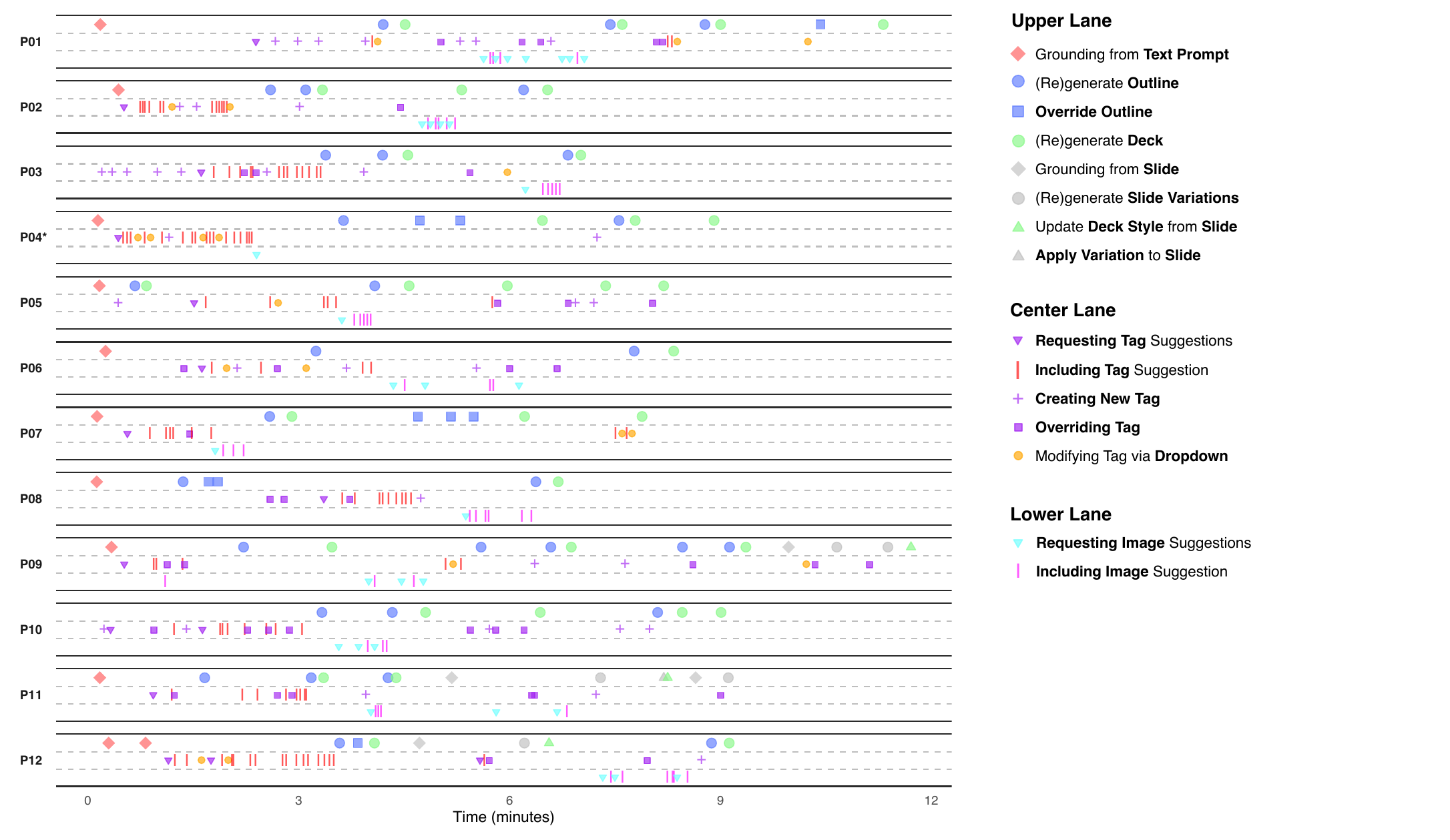}
  \caption{Timeline plots visualizing participants' interaction events of the semi-structured task (phase 3); 
  upper lanes show events related to outline and slide generation; 
  center lanes show interactions with concept tags; lower lanes show interactions with image reference tags;
  \rev{tag modifications via the \textit{Opposite Slider Widget} are omitted since only a few participants occasionally explored slider options but without applying these to the slide generation (see section \ref{Interactions_With_System_Suggested_Tags})  }
  *Note: During P04’s session, the image suggestions failed due to an API outage. }
  \Description{The figure shows timeline plots visualizing participants’ interaction events during the semi-structured task (Phase 3). Each row represents a participant (P01 to P12), with interactions plotted over time (in minutes) on the x-axis. The Upper Lane captures interactions related to slide generation and outlines, including events like Grounding from Text Prompt, (Re)generate Outline, and Apply Variation to Slide, marked by distinct shapes and colors. The Center Lane records interactions with concept tags, such as Requesting Tag Suggestions, Creating New Tag, and Overriding Tag. The Lower Lane shows interactions with image reference tags, such as Requesting Image Suggestions and Including Image Suggestion. P04 is marked with an asterisk indicating that image suggestions failed to load due to an API outage.}
  \label{fig:timelines}
\end{figure*}

\subsection{Observed Content Creation Workflows Using Intent Tags (RQ2)}

During the semi-structured task \textit{(study phase 3)} participants created a presentation from scratch about a personal hobby by using IntentTagger within 7 – 12 minutes (see Table \ref{tab:task4-results}).
Most participants started by using the \textit{Grounding from text prompt} feature to formulate an initial longer text prompt, which the system decomposed into an initial set of intent tags from where the participants continued to work on the Deck Steering Canvas (see Figure \ref{fig:timelines} upper lanes). 
Only two participants (P03 and P10) decided to start by adding concept tags manually.
On average, at the end of their sessions, users had 17.5 active intent tags (concept and image tags combined) and had generated 2.6 outlines and 2.8 slide decks. Of those tags, on average, 9.5 were tags users took from the system's suggested concept tags, and 3.9 were image tags (ranging between 50–96\% of active tags that users chose to include from system suggestions). 
Optionally, we included screenshots in the Appendix to give an impression of the variety of participant-created slide decks and tag steering boards.

\subsubsection{\textbf{Workflows}}
\rev{All participants generated an outline before generating a slide deck (see Figure \ref{fig:timelines} upper lanes). 
While this order was partially driven by IntentTagger’s requirement for an outline to enable slide deck generation, many participants also expressed in the post-task interviews that this order felt intuitive to them (see section \ref{supporting-diverse-presentation-needs}).} 
Almost all users first specified several intent tags for the narrative before generating a first outline after 2-4 minutes. 
However, two users (P05 and P08) generated a first outline before making substantial tag edits and then refined these using tags. 
Most participants did exclusively steer the outline and slide deck generation using intent tags, while three users (P04, P07, P08) made manual edits to the outline at a point.
In addition, most users worked exclusively in the Deck Steering Canvas, but three users (P09, P11, P12) also used the Slide Steering Overlay to adjust single slides and then updated the deck's style from that slide. 
Overall, a common pattern was adding or editing intent tags in the Narrative tag group and then optionally generating a first outline and slide deck. From there, most people continued adjusting the visual style through tags, and lastly, they added image tags to the content sources group.

\subsubsection{\textbf{Prompting With Intent Tags}}

Participants demonstrated a variety of approaches when prompting with intent tags, particularly when dealing with ambiguous or open-ended tags like color or style. For example, one participant specified the header color as \textit{“Gold”} and the color scheme as \textit{“Dark and Moody”} (P11), while another participant used hexadecimal notation \textit{(”\#FF33CC”)} for body color (P05).

\textbf{Users' Initial Uncertainty Around Tag Naming}

Despite the flexibility, some participants experienced uncertainty about how to “correctly” name certain tags, particularly when it came to specifying the number of slides. The task required a seven-slide presentation, and users created tags like \textit{“Number of Slides: 7,” “Slide number: 7 slides,” and “length: 7 slides in total.”} During the think-aloud sessions, participants expressed hesitation about whether they were using the right terminology. However, after seeing the generated slide deck, they gained confidence in the system’s ability to interpret a variety of tag expressions effectively.

\textbf{Users' Initial Uncertainty Around Tag Placing}

Interestingly, while most participants placed the tag for the number of slides within the narrative tag group, there was some variation in how users organized their tags. For instance, one participant placed the slide number tag under visual style (P08), while another placed it in the content sources group (P10). Regardless of where the tag was placed, the system consistently produced a seven-slide presentation. This adaptability underscored the system’s capacity to handle user input flexibly, ensuring that the end result matched the users’ intent even when the tags were categorized differently.

\textbf{Flexible Handling of Longer Prompts}

One participant created an exceptionally long and detailed prompt when using the initial grounding through text prompt, which led to the generation of tags containing chains of comma-separated phrases. While these tags were visually less ideal and cluttered the workspace, the system still managed to process them effectively (P08).

\textbf{User-Driven Tag Hierarchy and Organizational Strategies}

In addition to varying naming conventions, some participants developed a sense of structure and hierarchy by visually organizing the tags within the tag groups. Although the generative system currently does not inherently recognize such tag hierarchies, it was interesting to observe how participants imposed their own sense of order. 
Another strategy for creating hierarchy through tags was observed when P01 explicitly labeled tags to represent an ordered flow, such as \textit{“Topic 1: hiking”} and \textit{“Topic 2: hiking shoes.”}

Overall, the variety of tag naming and placement strategies showed that while some users were initially unsure about the “right” way to prompt the system, they quickly became more comfortable with the process and how to utilize the tag's granularity and various levels of ambiguity.

\subsubsection{\textbf{Interactions With System Suggested Tags}} \label{Interactions_With_System_Suggested_Tags}

Participants actively engaged with system-suggested tags early in their content creation process, with all users requesting tag suggestions within the first three minutes. This was typically followed by an intense phase of tag selection and editing that lasted 1 – 3 minutes (see clusters of red vertical strokes ||| in Figure \ref{fig:timelines}). 
For example, in a presentation about road cycling, the system suggested visual style tags like \textit{“fonts: sporty and dynamic”} and \textit{“backgrounds: scenic roadways,”} which the participant included (P06). Similarly, in a kayaking presentation, the system suggested \textit{“Font: Casual and Readable”} and \textit{“Icons: Outdoor activity symbols,”} demonstrating the system’s ability to offer contextually relevant suggestions that aligned with participants’ content themes (P03).

\textbf{Users leveraged suggestions to guide and refine the narrative structure.}
For narrative content, the system’s suggestions also played a pivotal role in helping participants develop the flow and focus of their presentations. For instance, during a presentation on the \textit{“Introduction and benefits of yoga,”} the system suggested tags such as \textit{“Audience: Beginners”} and \textit{“Benefits: Mental Clarity,”} which the user incorporated into their narrative structure (P02). These suggestions helped guide users in tailoring their presentation content to the intended audience and goals, adding more depth to the narrative while reducing the cognitive load of coming up with all the necessary elements from scratch.

\textbf{Users frequently explored and chose drop-down suggestions.}
We observed several distinct patterns in how participants interacted with and utilized the suggested tags: In some cases, users directly dragged the suggested tag into a relevant group and then changed its value by selecting an option from the alternatives drop-down widget (see sequences of a red vertical stroke followed by an orange circle |\ding{108} in Figure \ref{fig:timelines}, center lane).

For example, when the system suggested \textit{“cycling: cycling for fitness”} for the topic of road riding, one participant dragged it into the narrative group and then refined it by selecting \textit{“cycling for environmental benefit”} from the dropdown (P06, see Figure \ref{fig:timelines} around minute 2). 

\textbf{Users frequently overrode system suggestions.}
Another pattern we frequently observed was that participants often took more exploratory approaches, such as dragging the tag into the group and examining the available dropdown values before manually overriding the tag’s value or label to better suit their vision (see sequences of a red vertical stroke followed by a purple square |\ding{110} in Figure \ref{fig:timelines}, center lanes).

\textbf{Participants requested image suggestions in quick succession.}
Image suggestions triggered by the system followed different workflows, as the system provided five different images each time, prompting some participants to drag and drop the suggested images into the content sources before triggering the system again for more options. 
(see sequences of a cyan triangle followed by pink vertical strokes \ding{116} | in Figure \ref{fig:timelines}, lower lanes)

\textbf{Users explored the slider but not for steering. }
Additionally, while participants occasionally explored slider options to see the opposite ends of a spectrum, this was done more for curiosity than for direct steering slide generation. 
This indicates a potential area for further refinement in how the slider functionality is integrated into the generation process.

\textbf{In the Slide Steering Overlay, users primarily used "Grounding Tags" for adjustments. }
Lastly, those participants who used the Slide Steering Overlay to adjust single slides primarily engaged with the tags that the system had generated through its grounding from slide feature (see sequences of a grey diamond in the upper lanes followed by events in the center lanes in Figure \ref{fig:timelines}, upper lanes). For example, when P09 invoked the \textit{Tag Steering Overlay} for adjusting a single slide, the system created multiple tags representing the slide’s current state, one in the visual style group stating \textit{“Color: Red and Black Text.”} To make adjustments, the user first explored the drop-down widget’s alternative suggestions and then overrode the tag’s value with \textit{“blue, red, and white,”} resulting in a slide variation with red and white text on a blue background.

\subsection{Users’ Perceived Benefits and Challenges for Working with Intent Tags (RQ3)}

We clustered the perceived benefits and challenges for users working with intent tags into three themes: \textit{Expressing intents with tags}, \textit{catering users' presentation needs and workflows}, and \textit{meta-intent elicitation}.

\subsubsection{\textbf{\rev{Expressing Intents with Tags}}} \label{expressing-intents-with-tags}

All participants found intent tags advantageous compared to chat-based prompts, appreciating their conciseness and the ease of specifying inputs without needing to write out long sentences: \textit{"it's quite like fun to be able to sort of click and not have to write out a lot [...] I think the benefit of having these labels [over chat-based] is the conciseness. I don't have that pressure to have to form a sentence."} (P02)

\textbf{Users liked micro-prompting for its granularity and control.}
Many participants also appreciated the granularity and control of intent tags while being able to quickly see which presentation-related aspects they can specify: \textit{"I really liked that granularity where you could specify, like, this is exactly the font style and I want the kind of theme of it. I like that aspect a lot [...] visually seeing these buckets of information where I can specify different tags. I think that is helpful."} (P07)

\textbf{Users valued the structure that tag grouping provided.}
Many participants found the tag groups (e.g., narrative, content source, visual style) intuitive and helpful for sequencing and structuring their process and thinking more explicitly about presentation-related aspects: \textit{"I think that narrative is super cool in terms of thinking about the audience [...] it helped me be more explicit about who my audience is in the deck that I'm creating. And it helped me make the right deck faster."} (P11)

\textbf{Users expected to define their own tag groups.}
However, other users questioned the necessity of pre-defined groups, suggesting the ability to customize or even work without tag groupings, which might better align with their workflow, as their needs can vary between different slide decks: \textit{"But also I would want the ability to just make my own [groups] because I think between different slide decks, it would change"} (P08)

\textbf{Users needed a moment to get used to tags' flexible terminology. }
Some participants experienced initial uncertainty around the terminology of intent tags, particularly when classifying and naming tags or choosing appropriate groups: \textit{"So I found it slightly intimidating at the beginning because there is some kind of arbitrary way of classifying."} (P09)
However, with experimentation and practice, they realized these decisions were not critical as the system does not rely on precise syntax but allows for flexible ways of specifying tags:
\textit{"But after doing it for a while, it seems like it's not that important either. Like, this decision of what to put in the title and this kind of things... As far as I'm putting text, I figure that it does something."} (P09)

\textbf{Users appreciated the preview features. }
Many appreciated the preview feature as it provided a sense of security and confidence when using the system. Knowing that they could see a preview before finalizing their choices reduced the fear of unexpected outcomes but also helped their imagination:
\textit{"I thought that the slider was neat. [...] I do appreciate how I'm not just imagining, like on a scale from one to five, how much detail do you want? I don't know... So, I appreciate that it has a little preview down there. That's neat."} (P05)

\textbf{Users expressed a desire for predefined tags. }
Some participants also noted that starting with a blank tag canvas could be overwhelming, so having a set of predefined tags (such as audience, narrative style, and number of slides) would provide guidance and help users get started: \textit{"Let's say empty tags where there's maybe already kind of like a label... like audience, but there's no audience mentioned yet. But for some people, they don't think about audience before they make a PowerPoint. So, just to be like, hey, who is this for? Is sometimes the most helpful feedback you can give someone. So you can get it here."} (P05)

\textbf{Users expected to rank tags. }
Some participants also expressed the need for a feature that allows them to rank or prioritize tags based on their importance to the presentation. One participant (P03) mentioned how they tried to emphasize certain points by grouping similar tags together, such as audience and objective, and incorporating system-suggested tags. They also noted that some tags, like colors or themes, were less important to them.

\subsubsection{\textbf{\rev{Supporting Diverse Presentation Needs and Workflows}}} \label{supporting-diverse-presentation-needs}

In the interviews, participants also reflected on their general professional slide-creation experience and how IntentTagger could enhance or alleviate their process. Many described the challenges of traditional tools where the manual slide-by-slide process of creating a deck often feels tedious and uninspiring: \textit{"This was a really fun way to create a deck. I think it would appeal to most people that work in PowerPoint, because I don't think most people in PowerPoint creating decks just don't enjoy it unless they do."} (P02)

\textbf{Users valued the system for overcoming ‘blank page’ barrier.}
Many participants found the system particularly helpful in overcoming the initial challenges of starting a presentation, which they identified as one of the hardest parts of any creative project. The ability to quickly generate a basic draft allowed them to move past the initial “blank page” barrier, making it easier to start thinking about additional content and refinements. This was especially valuable for those who often start presentations from scratch, as the system provided a visual outline that lowered the barrier to getting started, allowing them to focus more on editing and refining the content: \textit{"... I feel like the system got the basics going, which then prompted me to think of additional things to add on top."} (P05)

\textbf{Participants appreciated IntentTagger's flexibility in supporting diverse workflows.}
Participants appreciated the system’s ability to support various workflows, whether by starting with a detailed outline, a brain dump, or a narrative-first approach. 
Some participants highlighted the system’s flexibility in allowing them to begin with content and then refine it iteratively, while others valued the ability to focus on narrative and then move on to visual elements. 
The system’s adaptability to different working styles, such as sequential versus big-picture thinking, was also noted as beneficial. 

\textbf{Participants appreciated the outline editor for structuring ideas and overview.}
Additionally, the Outline Editor was particularly praised for helping users map out their ideas and providing a clear presentation overview, making it easier to refine individual slides or the entire deck as needed. 
One participant (P06) also preferred a more sequential workflow, suggesting that instead of showing all tag groups at once, they could focus on one at a time.

\begin{figure*}[t]
  \centering
  \includegraphics[width=\linewidth]{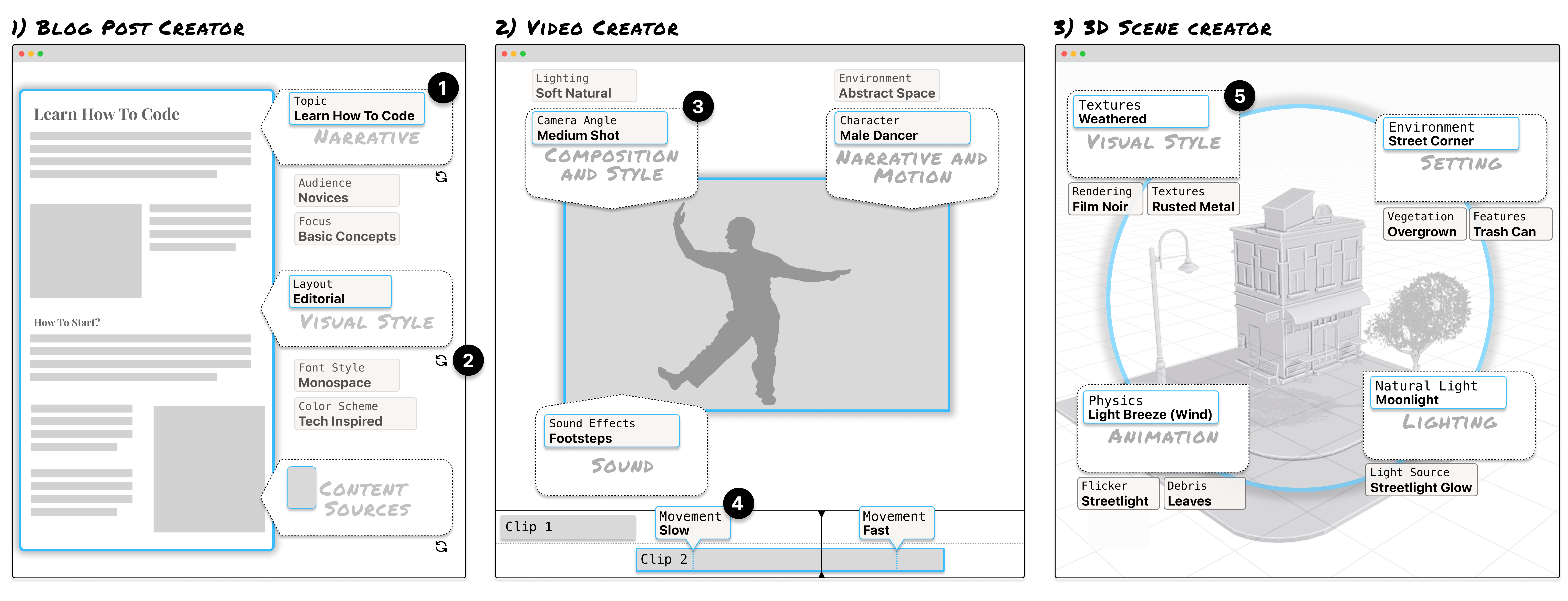}
  \caption{UX interface sketches illustrating alternative use-cases for Intent Tagging-based steering of GenAI content creation; (Left) Blog post creator: (1) Steering of a post generation through intent tags in a side panel; (2) new tag suggestion can be requested per group; (middle) Video creator: Steering of a selected video clip through intent tags; (4) additionally, Keyframe Tags allow to control temporal aspects of the clip generation in the timeline view; (Right) 3D scene creator: (5) Steering of a 3D scene generation through intent tags.  }
  \Description{ The figure shows three interface sketches demonstrating different use cases for Intent Tagging-based content creation: (1) Blog Post Creator: On the left, the system is steering the generation of a blog post titled “Learn How To Code” with tags such as “Audience: Novices” and “Layout: Editorial.” New tags can be requested for each group. (2) Video Creator: In the middle, a video clip featuring a male dancer is being steered through tags like “Camera Angle: Medium Shot” and “Sound Effects: Footsteps.” Keyframe Tags at the bottom allow control over the movement speed in the timeline, from slow to fast. (3) 3D Scene Creator: On the right, a 3D scene of a street corner is being generated with tags like “Textures: Weathered,” “Environment: Street Corner,” and “Lighting: Moonlight.” Intent Tags control elements of the visual style, animation, and setting of the scene.}
  \label{fig:ux-scenarios}
\end{figure*}

\textbf{Users saw value in tags for tailoring presentations to different purposes.}
Furthermore, some users mentioned the system’s tag-based approach could be useful for creating different versions of presentations tailored to specific audiences or contexts. The ability to easily drag tags in and out allowed for quick adjustments, making it convenient to include or exclude certain content based on the needs of a shorter talk, a quick checkup, or a presentation adapted for a particular audience:
\textit{"Sometimes I need to make different versions of presentations [...] This tag system of dragging things out would be an easy way to include and exclude some things if I needed to."} (P03)

\textbf{Participants appreciated the fluid integration of deck-wide and slide-specific edits.}
Participants valued the system’s ability to manage both the entire deck and individual slides, appreciating the flexibility to make broad changes across the deck or to focus on specific slides for detailed adjustments. This dual approach was seen as more efficient compared to traditional slide authoring workflows, where changes often have to be made slide-by-slide: \textit{"I think that having this create the whole deck in one shot is super time-saving, and I don't expect it to be hit in one shot. And having the option to click into each slide, and if I like what I do here and want to apply it to the rest... I like the ability to go from everything, update everything, zoom in on one, and then experiment on that one. And then again, update all."} (P02)

\textbf{Users demanded more control by locking specific design elements. }
While they appreciated the flexibility and ambiguity offered by the system, participants expressed a desire for more control over certain aspects of their presentations by being able to lock or pin specific slide elements (like background colors, fonts, or other design choices) once they were satisfied with them to prevent them from changing unexpectedly during further generation cycles: \textit{"I do love the slider. I only just don't love that if I slide it one way and I get the color I want that it can subsequently change. I want to be able to anchor it."} (P11)

\subsubsection{\textbf{\rev{Meta-Intent Elicitation: Helping Creators Figure Out What They Want and Need}}}

One of the key benefits highlighted by participants was the system’s ability to help them discover and refine their intentions when creating presentations. Many mentioned that in their daily work, the process of selecting the right content and design elements was often difficult without clear guidance or inspiration. The system’s suggestions helped participants think more critically about their choices and provided new ideas they might not have initially considered. This process of "meta-intent" elicitation allowed users to identify important elements that they may have overlooked or not realized they wanted to include, making it easier to achieve their desired outcome: \textit{"I thought it was cool and interesting because it suggests things that I just didn't think about... And I like these suggestions because I don't always know what are the best style choices for something." (P03)}

\textbf{Participants appreciated the system guidance's gradual granularity}
Participants appreciated the system's converging granularity through its suggestions, allowing them to gradually refine their presentations in more specific and thoughtful ways:\textit{"I feel like [IntentTagger] suggests a lot more things that I can control. Versus with [the chat-based system], there are like four prompts and they're like buckets of things you can do rather than the specific things you can tweak. And so for me, this system of these tags feels like I can get more granular towards some kind of vision that I want because I think inherently I'm making some of those choices."} (P12)   
This guidance made the process more engaging and aligned with their vision while also simplifying the decision-making process by presenting contextually relevant options and ideas to respond to: \textit{"It kind of breaks down those decision points that I'm doing sort of subconsciously in a way that is fairly easy. And, like, it suggests a lot of different things [...] it feels like I can get to a decent outcome fairly quickly."} (P12)

\textbf{Participants imagined Intent Tagging to be beneficial for prototyping presentations and initial brainstorming.}
For some participants who reported frequently struggling to organize their thoughts or create structured presentations, the system could also serve as a useful tool for brainstorming and organizing their thoughts. Through the drag-and-drop interface, participants imagined to quickly include or exclude elements, allowing them to organize and prototype their presentations more intuitively and efficiently: \textit{"I can organize my thoughts down and drag things in I want to include, drag things out that I want to exclude."} (P03)

\textbf{Some users felt overwhelmed by visual clutter and suggestions.}
However, while the majority of participants found the amount and visual arrangement of the system-suggested tags helpful, a few noted that too many suggestions could feel visually cluttered and overwhelming. Some expressed a preference for a more simplified or list-based representation, allowing them to focus more easily on the suggestions without feeling overwhelmed by the options: \textit{"...so a bit more visual hierarchy, maybe less overloaded with suggestions."} (P01)

\textbf{Participants praised the slider for stimulating their critical thinking.}
Additionally, features such as the opposite slider helped participants think more deeply about the intention behind their choices. By offering a way to explore different dimensions of a tag or design element, participants were able to visualize various possibilities and fine-tune their decisions in a way that aligned with their goals and encouraged reflection on the impact of their choices: \textit{"I really like the slider idea of like... 'oh, you have this idea, but then you want to see what the opposite end is,' and you can put a slider on that value [...] it just makes you think more about your intention of what you're trying to do."} (P04)

\section{Intent Tagging Interactions Beyond Slide Deck Creation}

To illustrate how intent tagging could be utilized to facilitate human-GenAI co-creation interactions across diverse rich content creation tasks beyond slide deck creation, we explored alternative \textbf{user experience (UX) interface scenarios} for rich content creation applications. 
Some of these were inspired by participants' speculations during the interviews on how IntentTagger could be helpful in applications like Excel for complex data visualizations, educational materials, and tasks such as creating videos or book reports. 

In a design sprint, we sketched out three UX interface scenarios for intent tag-based steering of GenAI-driven creation of blog posts, videos, and 3D scenes (Figure \ref{fig:ux-scenarios}). 
We defined different tag groups for each application, matching the creation task's high-level categories.  
We also explored how to possibly expand the "tag vocabulary" and their functionalities by, for example, envisioning \textit{Keyframe Tags} to allow control over temporal aspects of clips in the video creator  (Figure \ref{fig:ux-scenarios}-4). 
Overall, while only sketches, these explorations aim to highlight the versatility of intent tagging for steering rich content generation across a range of applications and interfaces to be further explored in future research.

\section{Discussion and Design Considerations}

\rev{In the following sections,} we discuss our explorations' learnings from the user study and UX scenarios, reflecting on \textit{Intent Tagging as an enabler of non-linear GenAI workflows} and its role in \textit{promoting Human-AI Co-creation and reflection}. Further, we will discuss the challenges of \textit{blending steering interactions alongside manual content editing} and, lastly, reflect on \textit{trade-offs between UI customization and user distractions}. 
Each section discusses challenges, design considerations, and opportunities for future work.

\subsection{Intent Tagging Enables Non-Linear GenAI Workflows}

Our findings demonstrate how IntentTagger was able to facilitate flexible, non-linear, and non-destructive workflows with high degrees of freedom: Once users became comfortable with the tool, they were able to seamlessly switch between the Deck Steering Canvas, Slide Overlay, and Outline Editor and steering the GenAI system without having to follow a rigid sequence (see 6.2). 
However, this non-linear steerability comes with a cost: While participants valued the system’s flexibility, they also expressed frustration about a need for greater control by "locking" or "pining" specific content or design elements, such as background colors and fonts, once satisfied to prevent unintended changes during further generation cycles (see 6.5).
While it is a powerful concept to enable multi-modal and multi-directional steering of GenAI content generation between tag, outline, slide, and deck levels, we currently lack interfaces allowing users to gradually lock elements across such dimensions. 
As a starting point, future research in this direction could be partially informed by design space exploration and parametric design systems that support users navigating and gradually reducing the degree of freedom across large option sets \cite{zaman_gemni_2015, matejka_dream_2018}.

In terms of user experience, Intent Tagging presents a novel way of steering GenAI systems and creating slide decks that are new to users. 
Our findings showed that while all participants appreciated the flexibility to specify tags with different levels of specificity, many were initially unsure about the tags' correct terminology and group associations (see 6.2.2).
This uncertainty posed initial challenges, which participants quickly overcame after experiencing the systems' flexibility in interpreting their intent after the first generations. 
To better support user onboarding, we suggest that future intent tagging-based interfaces should more explicitly scaffold that learning process, for example, by partially pre-defining common tag labels (such as \textit{"audience"} or \textit{"length"}) or offering optional "tag templates" for specific content creation scenarios to help users better get started and learn intent tagging.

\subsection{Intent Tagging Promotes Human-AI Co-creation and Reflection-In-Action }

Figuring out one's intentions and requirements is often the hardest part of complex tasks like slide deck creation. 
Generally, \textit{Schön} describes this process as “reflection-in-action” \cite{schon_reflective_1983}, where creators interact with the material at hand while the material "talks back" to them, creating an iterative process of acting and reflection. 
When intent tagging, the tags on the canvas talk back to the user, whether self-generated or system-suggested, helping them (along with the outline and slides) tackle the cognitively demanding task of clarifying their intent. 
Furthermore, the more task-relevant tags a user specifies, the more accurately the system can provide contextually tailored suggestions, better aligning with the user’s objectives. 
As users add more tags, they convey a fuller picture to the GenAI system, which in turn enhances the quality and precision of the generated content. 
The system and user engage in a close feedback loop, allowing users to refine their intentions while the AI system responds dynamically---like connecting a mind map to a GenAI content creation system. 
This process relates closely to previously framed ``enactive'' models of Human-AI co-creation \cite{davis_enactive_2015}, relating to the theory that cognition arises from improvised interactions shaped by feedback from the surrounding environment.

\rev{In this version of IntentTagger, we decided that users must explicitly request tag and image suggestions from the system. 
On the one hand, this approach ensures persistency between requests, allowing users to explore suggested tags without the risk of suggestions disappearing without their consent.
On the other hand, tag suggestions also offer an exciting possibility of being combined with mixed-initiative interactions in the future, where the system can automatically initiate tag suggestions in response to the user's active tags \cite{lin_prompts_2023}.}

\rev{In terms of user-system communication, in this study, we focused on analyzing users’ general interactions with system-generated tags (see 6.2.3). 
Future work could delve into a deeper semantic analysis, exploring how users expressed their intent and what specific tags the system suggested in response. Such insights could further improve the effectiveness of tag-based human-AI co-creation systems.}

Furthermore, while many participants experimented with the \textit{Opposite Slider widget}, no one used it to steer slide generation (see 6.2.3). 
However, several participants voiced the clear value of this feature for prompting reflection on their design choices (see 6.6). 
This suggests that, while not used directly for steering generation, the slider still serves as a valuable tool for encouraging and steering deeper thinking. 
This bears interesting further implications for designing future GenAI steering interfaces, which might consider integrating steering controls alongside "intent exploration" UI to promote reflection independent of content generation.

Our findings also revealed that users often got inspired by system suggestions but then manually overrode the tag's value field to better align with their vision. 
This interaction pattern hints at an effective combination of guided intent exploration through system-generated options alongside the flexibility for users to override suggestions if desired. 
Adding to this observation, a recent study in the context of AI-assisted writing found that participants, despite the many integrated AI suggestions, felt ownership over the content \textit{"because of the numerous authorial decisions"} they made \cite{singh_where_2023}. Together, such findings suggest that it is critical for effective human-AI co-creation to enable users to manually edit and override system suggestions to foster a sense of ownership and control. 

Our research indicates that user interactions with IntentTagger's LLM-generated \textit{tag suggestions}, \textit{tag grounding}, and \textit{pre-generated real-time previews} can help align user expectations with GenAI systems' capabilities and recognize their limitations (see 6.4). While some suggested tags referred to aspects beyond the slide generation capabilities (such as icons or background textures), interestingly, users were not significantly frustrated by that. 
Instead, they found the suggestions helpful for improving their presentations, regardless of whether the system could fully execute them. 
This tolerance was likely influenced by the pre-generated real-time tooltip previews, which helped users quickly grasp the system’s limitations without frustration, as they had not invested much time in these explorations.

Adding to previous work on surfacing affordances and providing feed-forward mechanisms~\cite{vermeulen2013crossing, boy2015suggested, TerryCreativeNeedsUIDesign2002}, we promote dynamically \textbf{pre-}generating slide previews (or other content) on a granular level (such as only alternating a slide's font or text format) to support users’ intent exploration and foster greater tolerance when GenAI falls short by reducing emotional investment in misaligned expectations. 
Additionally, \textit{tag grounding}---such as the grounding from text prompt or slide features---might also improve alignment between user expectations and system capabilities by helping users quickly understand how the system interprets their inputs. 
\rev{Lastly, presenting distinct differences between versions of generated content may further enhance reflection-in-action during the creation process, building on insights from Drucker et al. \cite{drucker_comparing_2006}. }
All these mechanisms show promising potential for enhancing user alignment in GenAI workflows.

\subsection{\textbf{Blending Steering Interactions and Manual Content Editing }}  

Our prototype system is designed to probe the concept of intent tagging-based interactions for steering slide deck generation and, therefore, does not support rich content editing that full-fledged applications such as PowerPoint offer. 
In the study, the system allowed participants to create slide decks with GenAI support with more control and satisfaction than with the chat-based system. 
However, going further, we envision that IntentTagger would also integrate rich content editing capabilities of slides in addition to AI-driven generation. Enabling users to interweave manual and AI-driven editing activities opens up many interesting questions relating to the orchestration of the hand-off between user and system. 
For example, devising mechanisms for identifying situations in which steering GenAI actually becomes more tedious than directly editing content and surfacing cues to users nudging them towards direct editing is an intriguing direction to explore.

\subsection{\textbf{Trade-offs between UI Customization and UI Management }}

Lastly, we reflect on requests we gathered from participants to enable them to customize IntentTagger more. A salient example is the requests from a few participants for enabling customizable groupings of tags rather than the three fixed groups we offered (see 6.4). While relatively straightforward to achieve, such added capability may come as a tradeoff between user control of the interface and its management.

While giving users the ability to create their own tag groups' structure would allow for greater flexibility---especially for catering to users' specific needs and creation process---it would also add an additional cognitive layer by requiring users to reflect on what such process is and how to best articulate it in the interface. 
This leads to the classic question of identifying which interface aspects are best given to the users to shape and which to leave in the hands of design and system architects. 
Probing this delicate balance between supporting flexibility and minimizing effort, enabling rich interactions, and making them simple to use in the context of generative AI paves the way for future research in our field.

\subsection{\rev{\textbf{Limitations of the Evaluation}}}
\rev{We highlight three limitations of this work:
Firstly, all participants were recruited from the same large technology company, and while they represent a diverse set of job titles and backgrounds, it is possible that participants from other companies or from non-corporate environments might provide additional insights not captured in the current study. 
Secondly, the study's lab setting, limited task times, and constrained number of slides allowed us to study Intent Tagging in a controlled and concise format. 
However, future work is needed to study the impacts of Intent Tags on creation processes \textit{in-vivo} and over \textit{longer periods of time} to gain further insights into real-life content creation scenarios. 
Lastly, we deliberately limited our comparison of IntentTagger with the widespread commercial Copilot and Designer features in PowerPoint. 
Future comparisons with other tools and interfaces may surface a broader range of benefits and tradeoffs.}

\section{Conclusion}

GenAI offers immense potential for augmenting content creation, but challenges around intent elicitation and alignment, prompt formulation, and workflow flexibility impede effective human-AI co-creation. To address these issues, we propose intent tagging: graphical micro-prompting interactions for supporting granular and non-linear workflows with GenAI systems in the context of slide deck creation. To explore its benefits and challenges, we developed IntentTagger, an intent tagging-based LLM-driven system allowing users to iteratively create and modify slide deck presentations, and conducted a user study. Our findings revealed that users preferred intent tag-based interactions over chat- and gallery-based systems, valuing the system’s ability to support non-linear workflows, flexible intent expression, and integrated suggestions that helped them clarify their goals and think through the slide creation tasks. Based on these findings, we discuss design considerations for integrating intent tagging into GenAI-assisted slide authoring. Although our system and study primarily focus on slide creation, intent tags seem to present a design pattern that researchers could explore further in other AI-assisted workflows and scenarios, thereby offering designers a new interaction technique for HCI+AI applications.

\begin{acks}
We thank all study participants,  as well as Anna Offenwanger and Zheng Ning, for their support and feedback throughout the project and the reviewers for their constructive feedback on the paper. 
This work was supported in part by the National Science Foundation under Grant No. 2118924.
\end{acks}

\bibliographystyle{ACM-Reference-Format}
\bibliography{References}

\appendix

\newpage
\onecolumn

\section{Additional Materials}

\begin{table*}[h]
\caption{Overview of study participants.}
\begin{tabular}{lclll}
\toprule
\textbf{ID} & \textbf{Age Range (Years)} & \textbf{Gender}     & \textbf{Job Title}  & \textbf{Frequency of PowerPoint Usage} \\
\midrule
P01            & 18-29     & Female     & Senior Product Designer       & Multiple times per week                      \\
P02            & 30-39     & Female     & Senior Product Manager        & Every day                                    \\
P03            & 18-29     & Female     & Research Intern (PhD)             & Multiple times per month                     \\
P04            & 18-29     & Male       & Senior Human Factors Engineer & Every day                                    \\
P05            & 30-39     & Female     & Senior Designer               & Multiple times per week                      \\
P06            & 40-49     & Female     & Senior User Researcher        & Multiple times per week                      \\
P07            & 30-39     & Male       & Senior Product Designer       & Every day                                    \\
P08            & 18-29     & Female     & Content Designer              & Multiple times per month                     \\
P09            & 30-39     & Male       & Research Intern (PhD)         & Multiple times per month                     \\
P10            & 18-29     & Non-binary & Research Intern (PhD)         & Multiple times per year                      \\
P11            & 30-39     & Female     & Design Program Manager        & Multiple times per week                      \\
P12            & 40-49     & Female     & UX Researcher                 & Multiple times per month      \\                                       
\bottomrule
\end{tabular}
\label{tab:participants}
\Description{Study participant demographics. (Table is machine readable). Study participant demographics. (Table is machine readable). The table provides demographic information for 12 participants (P01 to P12) involved in the study. Columns include ID, Age Range (Years), Gender, Job Title, and Frequency of PowerPoint Usage. The age ranges span from 18-29 to 40-49 years, with job titles such as Senior Product Designer, Research Intern (PhD), and UX Researcher. Participants’ frequency of PowerPoint usage varies from “Multiple times per year” to “Every day.” Two participants, P10 and P04, are noted as non-binary and male, respectively.}
\end{table*}

\rev{\subsection{Further System Implementation Details }}

\subsubsection{\rev{\textbf{Tag Suggestion Mechanism}}}

\rev{To generate intent tag suggestions based on a user's existing tags on the slide deck steering canvas, we construct a call to GPT comprising the static \texttt{SYSTEM\_PROMPT} and the dynamic \texttt{USER\_CONTEXT} below. 
\texttt{USER\_CONTEXT} is a string dynamically constructed from all active intent tags \texttt{[attribute:value]} for each tag group \texttt{(Narrative, Visual Style, Content Sources)}. }

\begin{lstlisting}
SYSTEM_PROMPT (static):  
You are an assistant to help users author slide presentations in PowerPoint. 
You are helping the user develop the narrative and visual style and extracting content from source documents relevant to the presentation. 
You will receive instructions to help with one of these three buckets: Narrative, Visual Style, and Content Sources.  
The user will already have descriptions, tags, and media elements that describe different aspects of the presentation.  
Your task is to suggest additional keywords and concepts for each of these buckets.  
The user will prompt you in the format:  
**Bucket** attribute1:value1, attribute2:value2, ...  
A bucket might be empty, in which case you should suggest concepts based on the other buckets.  
A value might not be associated with an attribute.  
For each user prompt, suggest 7 concepts for each of the requested buckets.  
A concept is a keyword or phrase that is relevant to the user's presentation. A concept should be a single word or a short phrase. Each concept should have an attribute and value.  
Return your response in JSON format. Only return JSON format without any additional Markdown wrapper code blocks.  
 
Here is an example of the JSON return format:  
{
    "Narrative": ["attribue1:value1", "attribue1:value1", ...],  
    "Visual Style": ["attribue1:value1", "attribue1:value1", ...],  
    "Content Sources": ["attribue1:value1", "attribue1:value1", ...]  
}

USER_CONTEXT (dynamic): 
**Narrative** attribute1:value1, attribute2:value2, ... 
**Visual Style** attribute1:value1, attribute2:value2, ... 
**Content Sources** attribute1:value1, attribute2:value2, ... 
\end{lstlisting}

\subsubsection{\rev{\textbf{Outline Generation From Intent Tags}}}

\rev{To generate an outline based on a user's existing tags on the slide deck steering canvas, we construct a call to GPT comprising the static \texttt{SYSTEM\_PROMPT} and the dynamic \texttt{USER\_CONTEXT} below.
\texttt{USER\_CONTEXT} is a string dynamically constructed from all active intent tags \texttt{[attribute:value]} for each tag group \texttt{(Narrative, Visual Style, Content Sources)}.
GPT is instructed to generate an outline in Markdown format, including the URLs of image reference tags if present. }

\begin{lstlisting}
SYSTEM_PROMPT (static): 
You are an assistant to help users author slide presentations in PowerPoint. 
You are creating an outline for the user's presentation. 
The user will have descriptions and keywords that describe different aspects of the presentation. 
The user will prompt you in the format: attribute1:value1, attribute2:value2, ... 
Ignore all the attributes related to the visual style.  
If the content sources include ImageUrl attributes, you should include these images in the outline as images in the markdown. 
Respond with an outline of the presentation in the format:  
section title - section content bullet points. 
Return the outline in markdown format without ```markdown at the start and end. Only return markdown content for the outline. 

USER_CONTEXT (dynamic): 
**Narrative** attribute1:value1, attribute2:value2, ... 
**Visual Style** attribute1:value1, attribute2:value2, ... 
**Content Sources** attribute1:value1, attribute2:value2, ... 
\end{lstlisting}

\subsubsection{\rev{\textbf{Slide generation mechanism}}}

\rev{ To generate the JSON structure for a slide deck, we construct a call to GPT comprising the static \texttt{SYSTEM\_PROMPT} and the dynamic \texttt{USER\_CONTEXT}. 
\texttt{USER\_CONTEXT} is a string containing the \texttt{PRESENTATION OUTLINE} in markdown, \texttt{META INFORMATION}, which is a string containing all active intent tags \texttt{[attribute:value]}, and optionally the \texttt{REFERENCE SLIDE DECK TEMPLATE}, which is a JSON string of a reference slide deck (if provided).  
The general slide deck template schema and slide generation mechanism are based on the Spectacle\footnote{https://github.com/FormidableLabs/spectacle} library. }

\begin{lstlisting}
SYSTEM_PROMPT (static): 
Your task is to generate a slide presentation deck from an outline and meta information.  
You will receive the presentation OUTLINE in markdown format from the user. 
You will also receive additional META INFORMATION about the desired presentation's Narrative, Visual Style, and Content Sources.  
The META INFORMATION will describe different aspects of the presentation.  
You will receive the META INFORMATION in the format:  
**Bucket** attribute1:value1 (optional description), attribute2:value2 (optinal description), ...  
A bucket might be empty, in which case you should suggest concepts based on the other buckets.  
A value might not be associated with an attribute.  
The META INFORMATION might also contain USER DOCUMENTS with content that has to be included in the presentation. 

GENERATING SLIDES  
Your task is to generate a slide deck. Each slide has to consist of a LAYOUT, CONTENT, and THEME.  

LAYOUT:  
The layout determines the arrangement of the content on the slide.  
There are five different layouts to choose from:  
"title", "listOrParagraph", "verticalImage", "fullImage". 

CONTENT:  
Each layout takes different content parameters depending on the layout type. Optionally, each slide can take an image url as background image in the content.  
The background image will replace the background color of the slide.  
The "listOrParagraph" and "verticalImage" layouts take either a list or a paragraph as content besides the title and image.  
For bullet lists, never use more than 3 bullets. 

THEME:  
The theme determines the fonts, colors, and font sizes used in the slide.  
 
fonts: { header: fontname, text: fontname }  
"header" is the font used for titles and "text" is the font used for body, lists, paragraphs.  
There are six fonts to choose from: "Quicksand", "Playfair Display", "Montserrat", "Merriweather", "Roboto" and "Roboto Condensed". 

colors: { primary: color, secondary: color, tertiary: color }  
"primary" is the main color used for text, "secondary" is the color used for accents, and "tertiary" is the color used for backgrounds. 

fontSizes: { h1: size, text: size }  
"h1" is the font size used for titles and "text" is the font size used for body text. 

REFERENCE SLIDE DECK TEMPLATE  
The user might also provide a REFERENCE SLIDE DECK TEMPLATE with a slide deck template they want to use for the visual style.  
Important: If a REFERENCE SLIDE DECK TEMPLATE is provided, use only this template for the visual style (font, backgroundImage, sizes, colors) and IGNORE all visual style information from the generic example slide deck below.  
ALWAYS use the backgroundImge URLs from the REFERENCE SLIDE DECK TEMPLATE on each slide! 

REWORK CONTENT FROM THE OUTLINE  
You should rework the content from the OUTLINE to fit the slide deck based on the desired format specified in the META INFORMATION.  
For example, if the META INFORMATION specifies that the presentation should have a formal tone, you should rework the content to fit this tone.  
If the META INFORMATION specifies that the presentation should have a more verbose text, you should rework the content to fit this style. 

YOUR TASK:  
Your task is to compose a slide deck based on the outline and to adapt the content to the layout and theme.  
Your task is to define the theme and layout of the slides based on the OUTLINE and META INFORMATION provided by the user.  
Generate the number of slides based on the outline and meta information provided by the user.  
Keep the visual style consistent across all slides.  
Especially the fonts, colors, background color and font sizes.  
You should return the slide deck in JSON format. Only return JSON format without any additional Markdown wrapper code blocks.  

Here is a generic example of a slide deck in JSON format: 
[{ 
    slideNumber: 1, 
    layout: "title", 
    content: { 
      title: "A presentation about something", 
      subtitle: "by someone", 
      backgroundImage:  "url(<image url placeholder>)" 
    }, 
    theme: { 
      fonts: { 
        header: '"Playfair Display", serif', 
        text: '"Quicksand", sans-serif', 
      }, 
      colors: { 
        primary: "#000", 
        secondary: "#000", 
        tertiary: "#fff", 
      }, 
      fontSizes: { 
        h1: "100px", 
        text: "44px", 
      }, 
      space: [16, 24, 32], 
    }, 
  },{ 
    slideNumber: 2, 
    layout: "listOrParagraph", 
    ...
    (PARTS OMITTED) 
    ...
}] 
 

USER_CONTEXT (dynamic): 
PRESENTATION OUTLINE: 
$CURRENT_OUTLINE_IN_MARKDOWN_STRING 

META INFORMATION:  
$CURRENT_ACTIVE_TAGS_AS_ATTRIBUTE:VALUE_PAIRS_CLUSTERED_BY_GROUP 

(optional) REFERENCE SLIDE DECK TEMPLATE: 
$REFERENCE_SLIDE_DECK_JSON 
\end{lstlisting}

\subsubsection{\rev{\textbf{Mechanism to create tags for an existing slide (tag grounding act)}}}

\rev{To generate a collection of intent tags based on a provided image of the current slide, we construct a call to GPT comprising the static \texttt{SYSTEM\_PROMPT} and the dynamic \texttt{USER\_CONTEXT} below.
\texttt{USER\_CONTEXT} is a string of the current slide in a base64 image format.
In addition to the slide image, we experimented with including the active steering board tags in the prompt but refrained from including these in this version of IntentTagger for simplicity. 
Future explorations outside the scope of this paper should investigate how "global" steering board-level intent tags can meaningfully propagate to individual slides and how changes on single slides can propagate back to the global steering board level while meaningfully resolving conflicts between global and slide-level tags. 
}

\begin{lstlisting}
SYSTEM_PROMPT (static): 
You are an assistant to help users author slide presentations in PowerPoint. 
You will receive an image of a slide from their slide presentation deck.  
Your task is to analyze the slide deck and return descriptive attributes that best describe the slide.  
You should return these attributes related to three categories: Narrative, Visual Style, and Content Sources.     
Each attribute should consist of a label and a value, like Tonality:Formal or Typography:Modern. 
For each bucket, return between 2 to 6 attributes.  
Return your response in JSON format.  
Only return JSON format without any additional Markdown wrapper code blocks. 

Here is an example of the JSON format: 
{ 
    "Narrative": ["attribue1:value1", "attribue1:value1", ...], 
    "Visual Style": ["attribue1:value1", "attribue1:value1", ...], 
    "Content Sources": ["attribue1:value1", "attribue1:value1", ...] 
} 

USER_CONTEXT (dynamic): 
$IMAGE_OF_CURRENT_SLIDE_AS_BASE64_STRING 
\end{lstlisting}

\subsection{Semi-structured Task Outcomes }
This section contains screenshots of IntentTagger (deck steering board and slide panel) taken at the end of each participant's semi-structured slide creation task from our user study (see study phase 3 at section \ref{sec:study-procedure}).

\begin{figure}[H]
    \centering
  \includegraphics[width=0.99\linewidth]{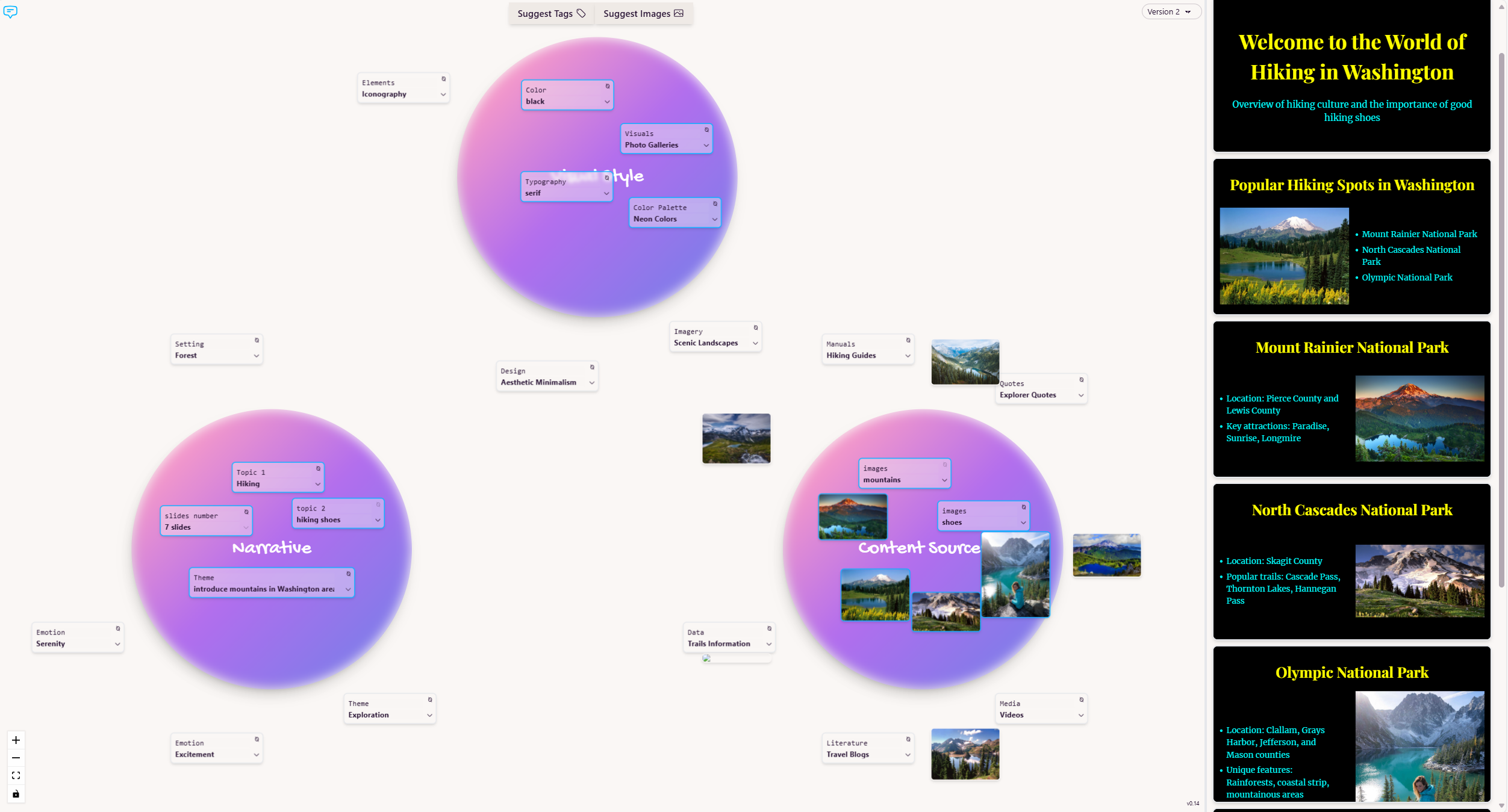}
  \caption{Semi-structured task outcome of P01}
  \Description{The figure displays a screenshot of the IntentTagger interface for participant P01’s open-ended slide creation task. The left side shows the Deck Steering Board, where tags related to Narrative, Visual Style, and Content Sources are organized. Tags include topics such as “Hiking,” “Hiking Shoes,” and imagery of “Mountains.” The right side shows the Slide Panel, featuring a generated slide deck about hiking in Washington, including information on popular hiking spots like Mount Rainier and the Olympic National Park.} 
  \label{fig:appendix_task4_P01}
\end{figure}

\begin{figure}[H]
    \centering
  \includegraphics[width=0.99\linewidth]{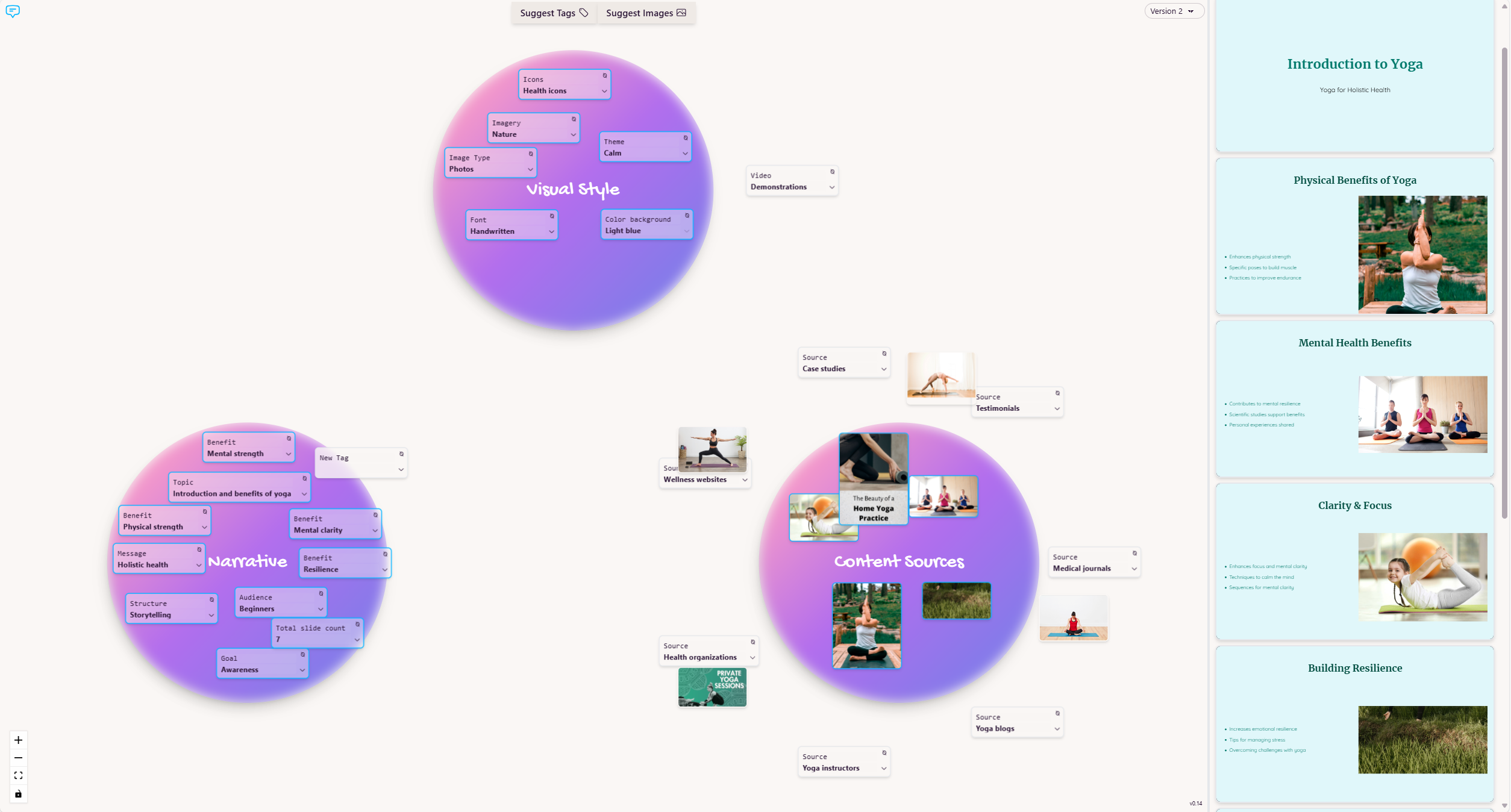}
  \caption{Semi-structured task outcome of P02}
  \Description{ The figure shows the IntentTagger interface for participant P02’s semi-structured task. The Deck Steering Board on the left features tags related to Narrative, Visual Style, and Content Sources. The tags include topics like “Introduction and benefits of yoga,” “Physical strength,” and “Mental clarity” under Narrative, while Visual Style contains tags such as “Font: Handwritten” and “Color Background: Light blue.” The Content Sources include wellness websites, case studies, and testimonials with relevant images. On the right, the Slide Panel shows a slide deck on the benefits of yoga, including sections on physical and mental health benefits, clarity, and focus.} 
  \label{fig:appendix_task4_P02}
\end{figure}

\begin{figure}[H]
    \centering
  \includegraphics[width=0.99\linewidth]{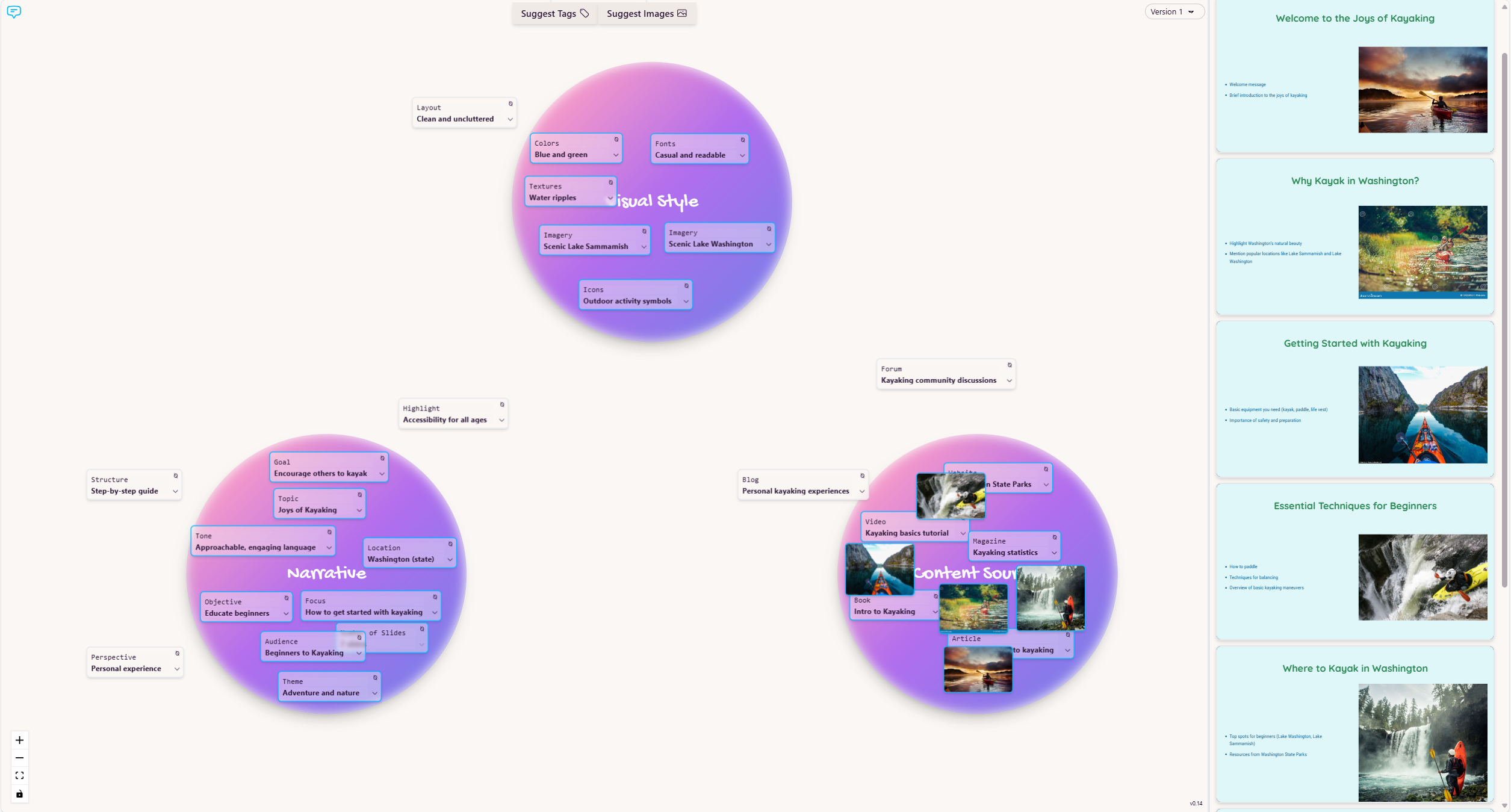}
  \caption{Semi-structured task outcome of P03}
  \Description{The figure shows the IntentTagger interface for participant P03’s semi-structured task. The Deck Steering Board on the left contains tags related to Narrative, Visual Style, and Content Sources. The Narrative tags include “Topic: Joys of Kayaking,” “Goal: Encourage others to kayak,” and “Structure: Step-by-step guide.” The Visual Style tags include “Colors: Blue and Green” and “Textures: Water ripples.” The Content Sources section includes videos, blogs, and articles related to kayaking. On the right, the Slide Panel shows a generated slide deck on kayaking, with slides covering topics such as “Why Kayak in Washington” and “Essential Techniques for Beginners.” } 
  \label{fig:appendix_task4_P03}
\end{figure}

\begin{figure}[H]
    \centering
  \includegraphics[width=0.99\linewidth]{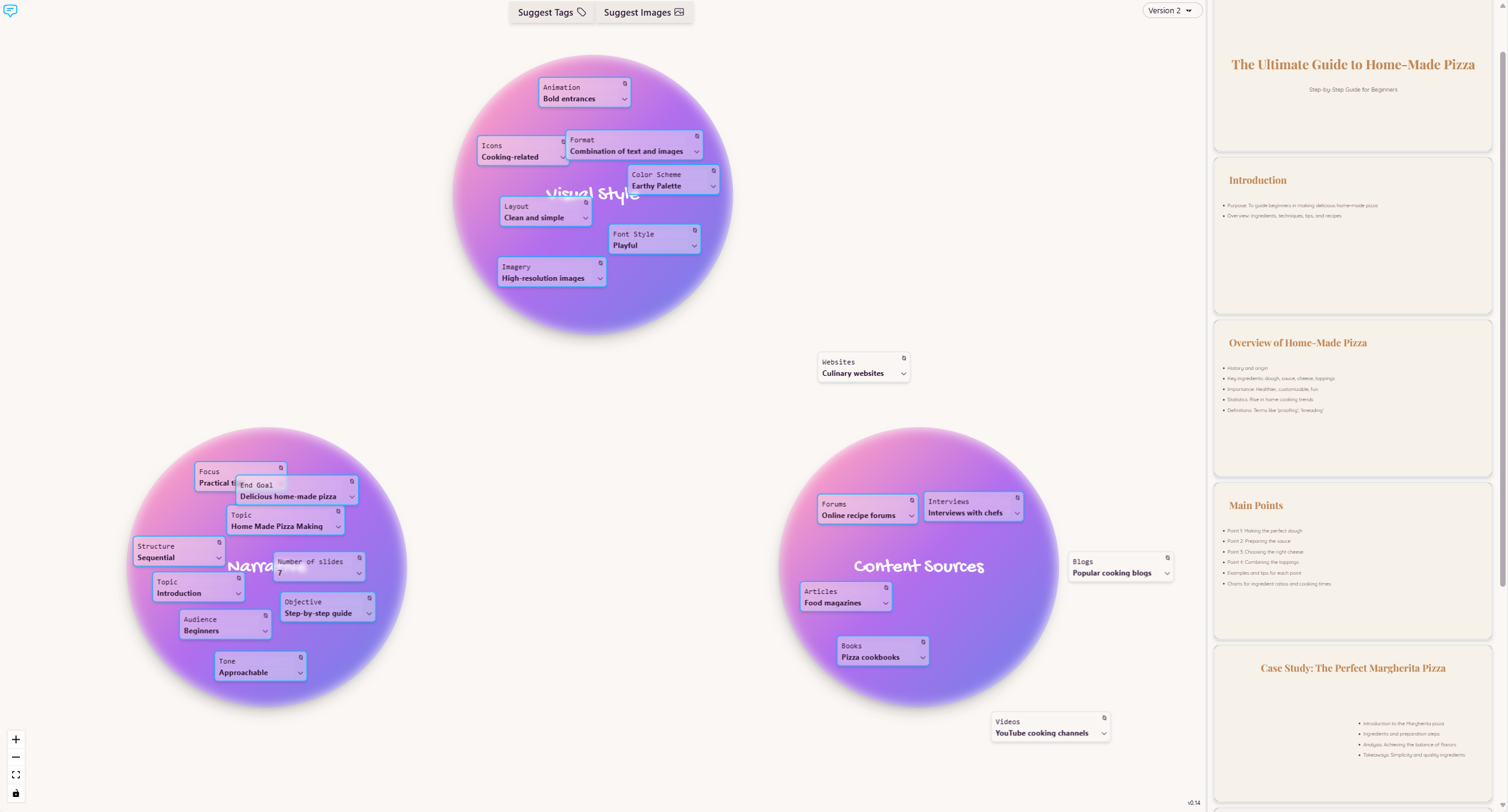}
  \caption{Semi-structured task outcome of P04 (Note: The user was unable to add images to their presentation, due to a temporary outage of the image search API.)}
  \Description{The figure shows the IntentTagger interface for participant P04’s semi-structured task. The Deck Steering Board on the left contains tags related to Narrative, Visual Style, and Content Sources. The Narrative tags include “Topic: Home Made Pizza Making,” “Focus: Practical,” and “Objective: Step-by-step guide.” The Visual Style tags include “Color Scheme: Earthy Palette” and “Layout: Clean and simple.” The Content Sources include online recipe forums, interviews with chefs, and food magazines. On the right, the Slide Panel shows a slide deck on pizza-making, covering topics such as “Overview of Home-Made Pizza” and “Case Study: The Perfect Margherita Pizza.” Note: The user was unable to add images to the presentation due to an API outage. } 
  \label{fig:appendix_task4_P04}
\end{figure}

\begin{figure}[H]
    \centering
  \includegraphics[width=0.99\linewidth]{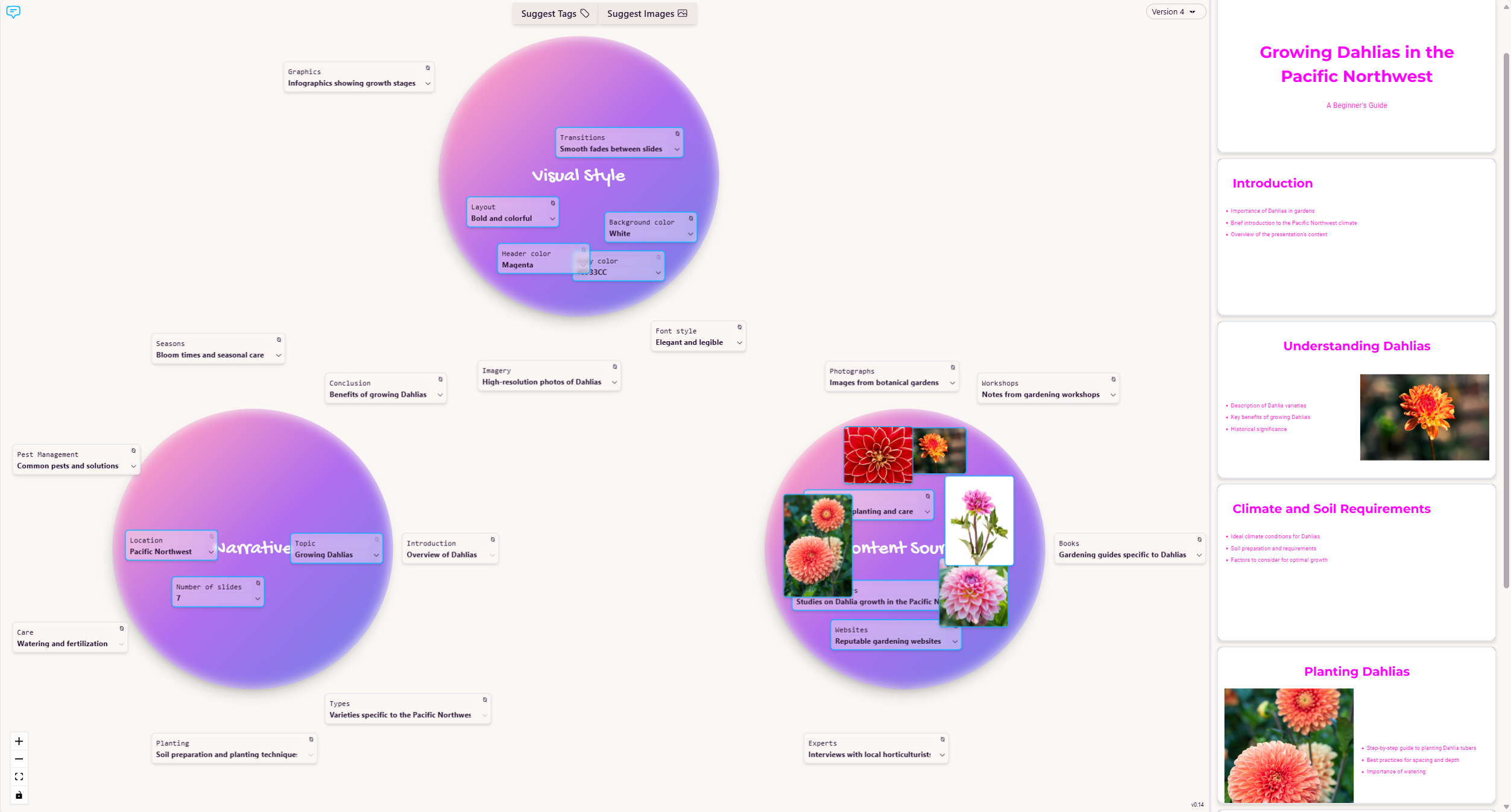}
  \caption{Semi-structured task outcome of P05}
  \Description{The figure shows the IntentTagger interface for participant P05’s semi-structured task. The Deck Steering Board on the left contains tags for Narrative, Visual Style, and Content Sources. The Narrative section includes tags like “Topic: Growing Dahlias,” “Location: Pacific Northwest,” and “Introduction: Overview of Dahlias.” The Visual Style section includes tags such as “Layout: Bold and colorful,” “Background: White,” and “Font Style: Elegant and legible.” The Content Sources include gardening guides, images from botanical gardens, and notes from gardening workshops. On the right, the Slide Panel displays a slide deck on growing dahlias, covering topics such as “Understanding Dahlias,” “Climate and Soil Requirements,” and “Planting Dahlias.”} 
  \label{fig:appendix_task4_P05}
\end{figure}

\begin{figure}[H]
    \centering
  \includegraphics[width=0.99\linewidth]{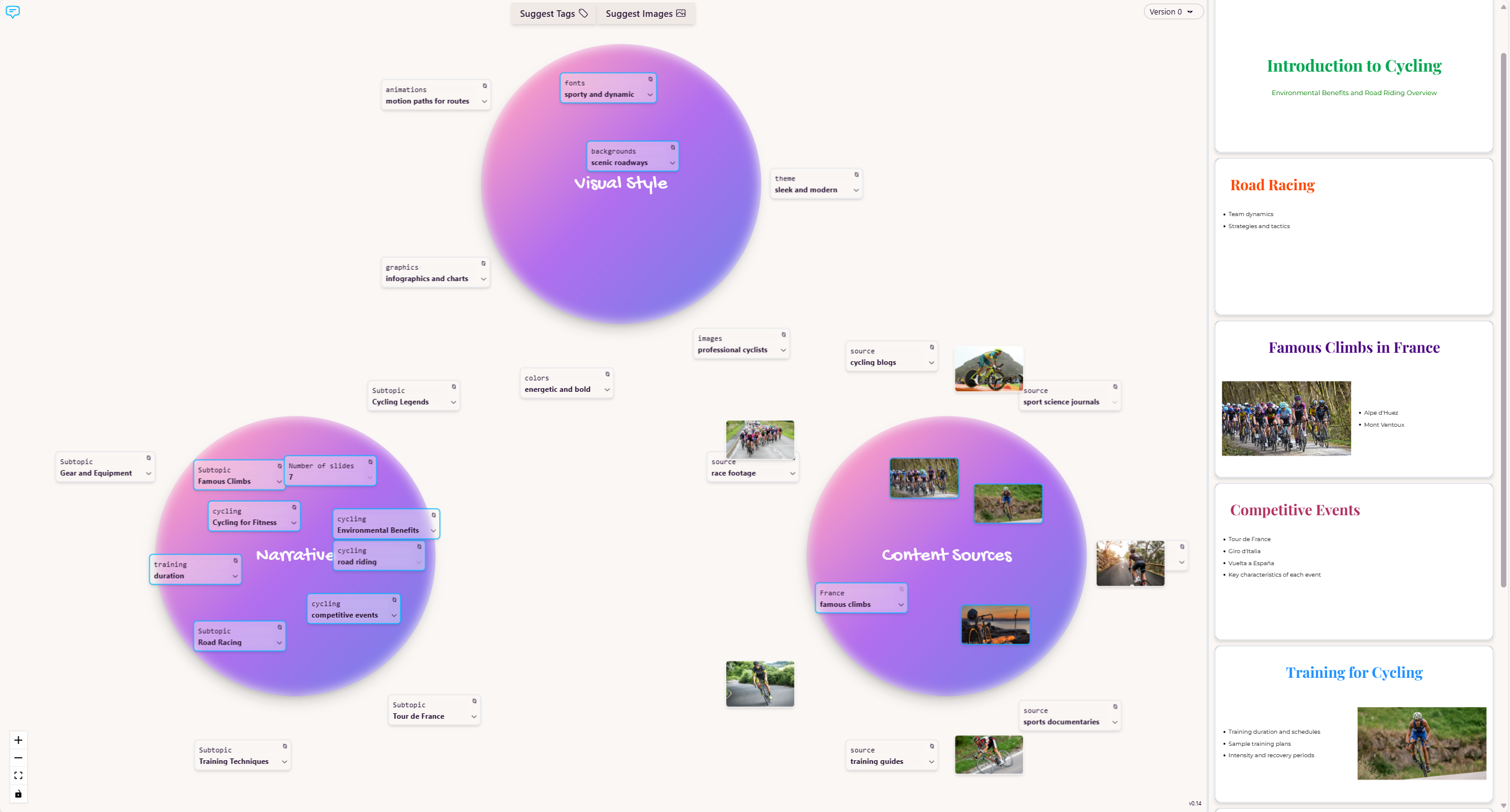}
  \caption{Semi-structured task outcome of P06}
  \Description{The figure shows the IntentTagger interface for participant P06’s semi-structured task. The Deck Steering Board on the left contains tags for Narrative, Visual Style, and Content Sources. The Narrative section includes tags like “Cycling for Fitness,” “Competitive Events,” and “Famous Climbs in France.” The Visual Style section includes tags such as “Fonts: Sporty and dynamic” and “Backgrounds: Scenic roadways.” The Content Sources include images of professional cyclists, race footage, and sports documentaries. On the right, the Slide Panel displays a slide deck on cycling, covering topics such as “Famous Climbs in France,” “Competitive Events,” and “Training for Cycling.”} 
  \label{fig:appendix_task4_P06}
\end{figure}

\begin{figure}[H]
    \centering
  \includegraphics[width=0.99\linewidth]{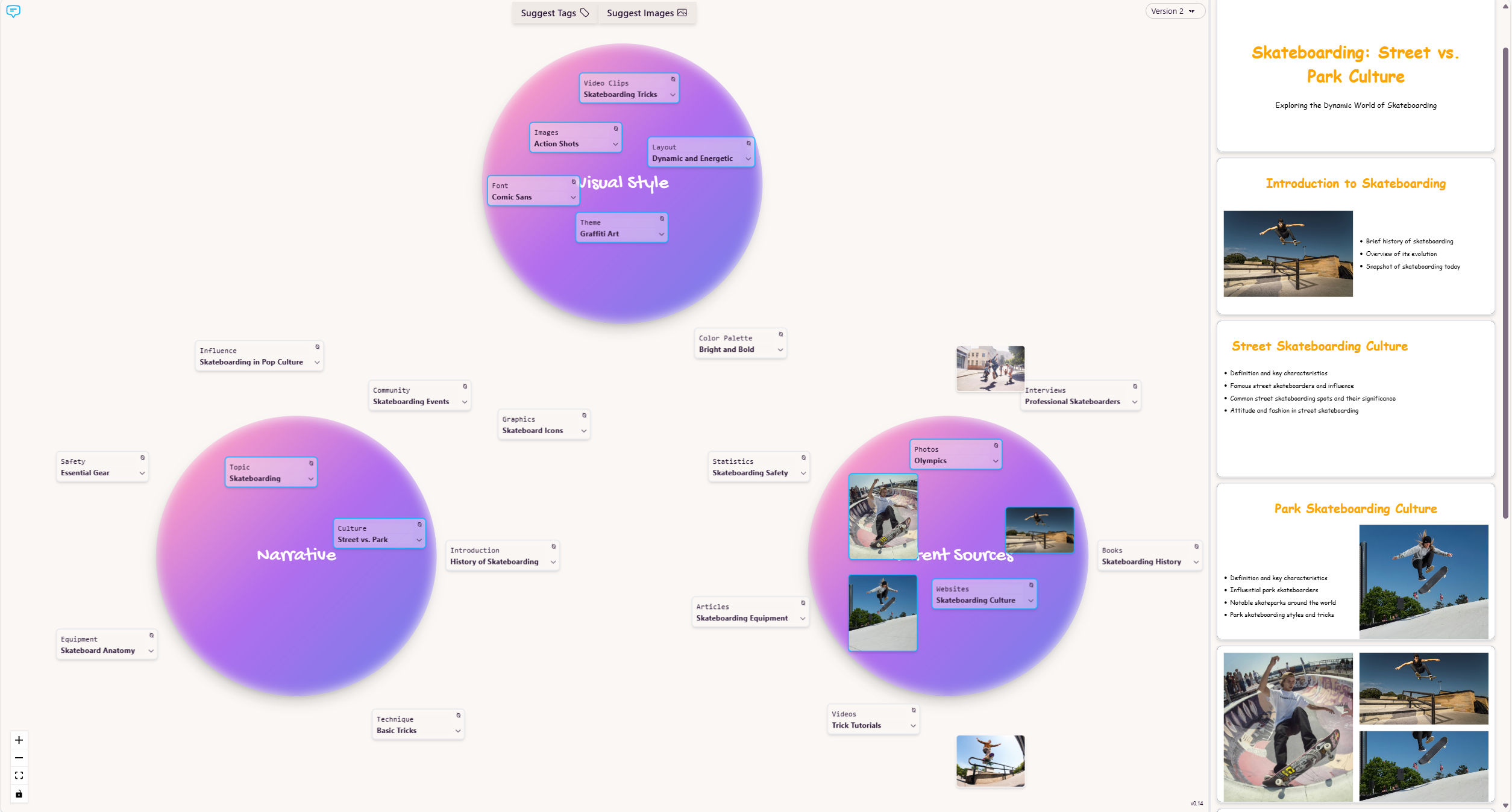}
  \caption{Semi-structured task outcome of P07}
  \Description{The figure shows the IntentTagger interface for participant P07’s semi-structured task. The Deck Steering Board on the left contains tags for Narrative, Visual Style, and Content Sources. The Narrative section includes tags such as “Topic: Skateboarding,” “Culture: Street vs. Park,” and “Introduction: History of Skateboarding.” The Visual Style section includes tags like “Font: Comic Sans,” “Theme: Graffiti Art,” and “Layout: Dynamic and Energetic.” The Content Sources include videos, articles, interviews with professional skateboarders, and skateboarding culture websites. On the right, the Slide Panel shows a slide deck on skateboarding culture, with sections on “Introduction to Skateboarding,” “Street Skateboarding Culture,” and “Park Skateboarding Culture.” } 
  \label{fig:appendix_task4_P07}
\end{figure}

\begin{figure}[H]
    \centering
  \includegraphics[width=0.99\linewidth]{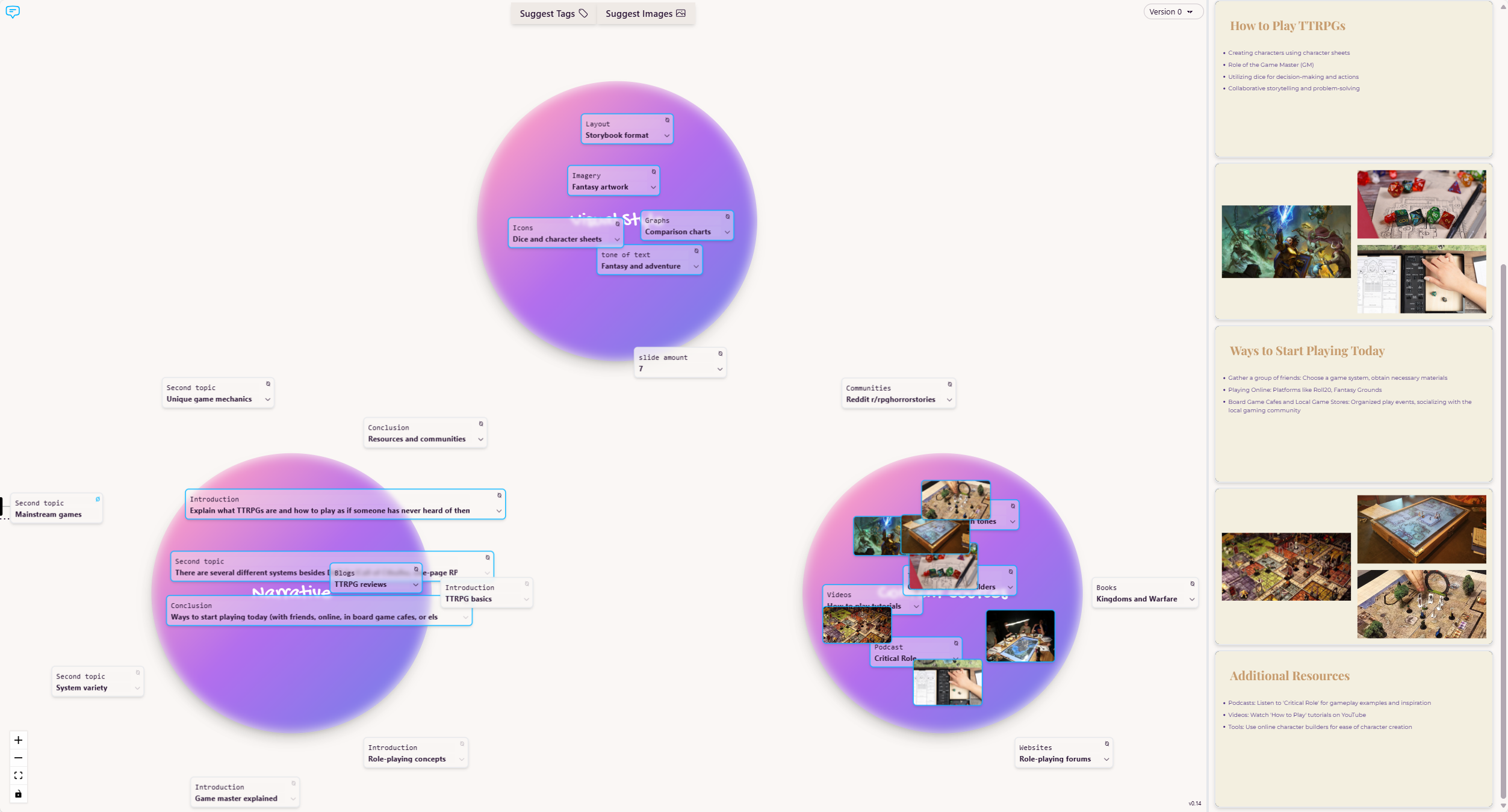}
  \caption{Semi-structured task outcome of P08}
  \Description{The figure shows the IntentTagger interface for participant P08’s semi-structured task. The Deck Steering Board on the left contains tags related to Narrative, Visual Style, and Content Sources. The Narrative section includes tags such as “Topic: How to play TTRPGs,” “Second topic: Unique game mechanics,” and “Conclusion: Resources and communities.” The Visual Style section includes tags like “Layout: Storybook format” and “Imagery: Fantasy artwork.” The Content Sources include videos, books, and role-playing forums related to tabletop RPGs. On the right, the Slide Panel shows a slide deck on how to play TTRPGs, covering topics like “How to Play TTRPGs,” “Ways to Start Playing Today,” and “Additional Resources.” } 
  \label{fig:appendix_task4_P08}
\end{figure}

\begin{figure}[H]
    \centering
  \includegraphics[width=0.99\linewidth]{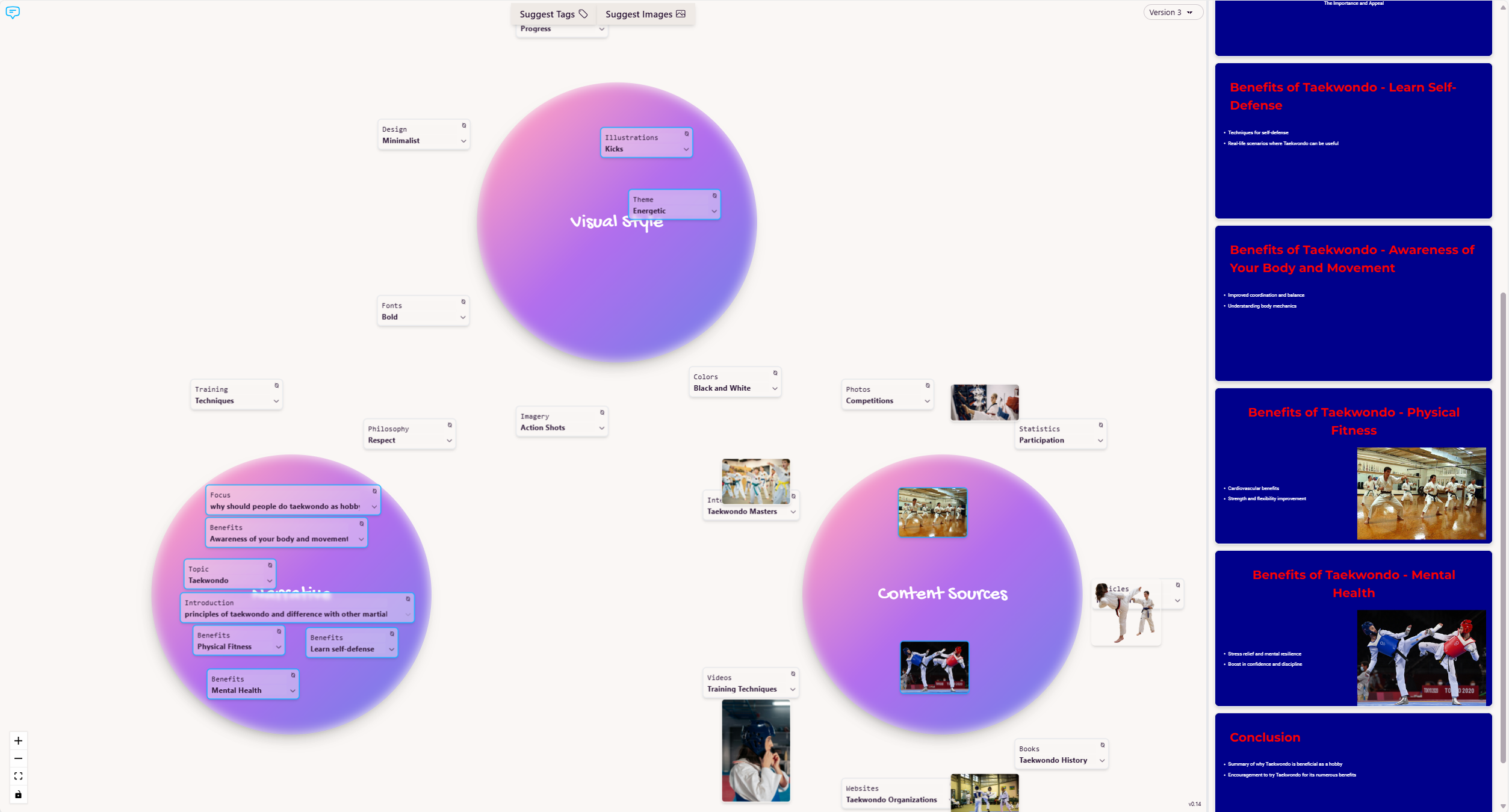}
  \caption{Semi-structured task outcome of P09}
  \Description{The figure shows the IntentTagger interface for participant P09’s semi-structured task. The Deck Steering Board on the left contains tags related to Narrative, Visual Style, and Content Sources. The Narrative section includes tags like “Topic: Taekwondo,” “Benefits: Physical Fitness,” “Benefits: Learn Self-Defense,” and “Introduction: Principles of Taekwondo.” The Visual Style section includes tags such as “Theme: Energetic,” “Colors: Black and White,” and “Illustrations: Kicks.” The Content Sources include statistics on participation, photos from competitions, and training videos. On the right, the Slide Panel displays a slide deck on the benefits of Taekwondo, covering topics such as self-defense, physical fitness, mental health, and body awareness.} 
  \label{fig:appendix_task4_P09}
\end{figure}

\begin{figure}[H]
    \centering
  \includegraphics[width=0.99\linewidth]{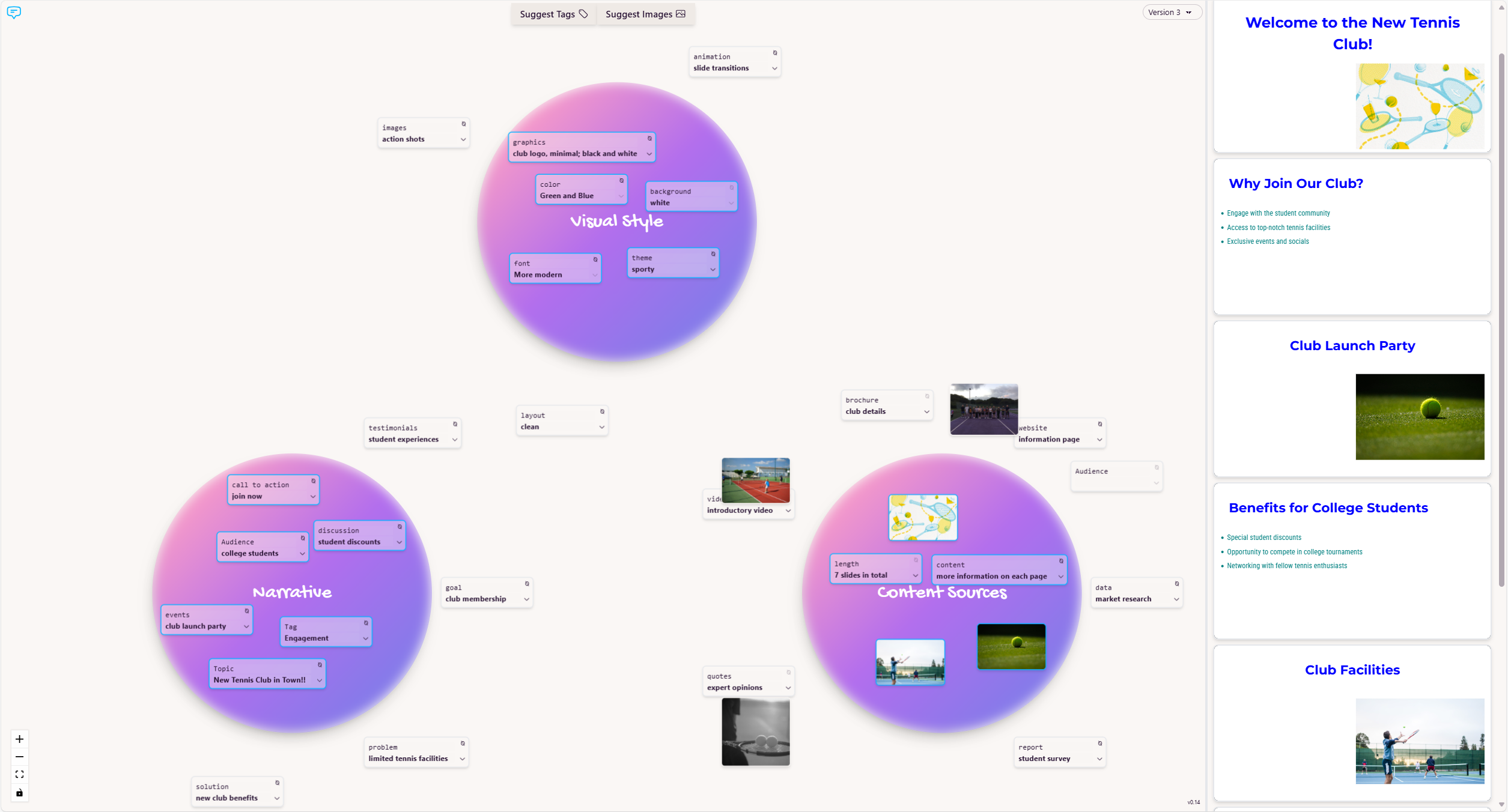}
  \caption{Semi-structured task outcome of P10}
  \Description{ The figure shows the IntentTagger interface for participant P10’s semi-structured task. The Deck Steering Board on the left contains tags related to Narrative, Visual Style, and Content Sources. The Narrative section includes tags like “Topic: New Tennis Club in Town,” “Call to Action: Join Now,” and “Audience: College Students.” The Visual Style section includes tags such as “Font: Neo Modern,” “Theme: Sporty,” and “Color: Green and Blue.” The Content Sources include introductory videos, student surveys, brochures with club details, and expert opinions. On the right, the Slide Panel displays a slide deck introducing the tennis club, covering sections like “Why Join Our Club?” “Club Launch Party,” and “Benefits for College Students.” } 
  \label{fig:appendix_task4_P10}
\end{figure}

\begin{figure}[H]
    \centering
  \includegraphics[width=0.99\linewidth]{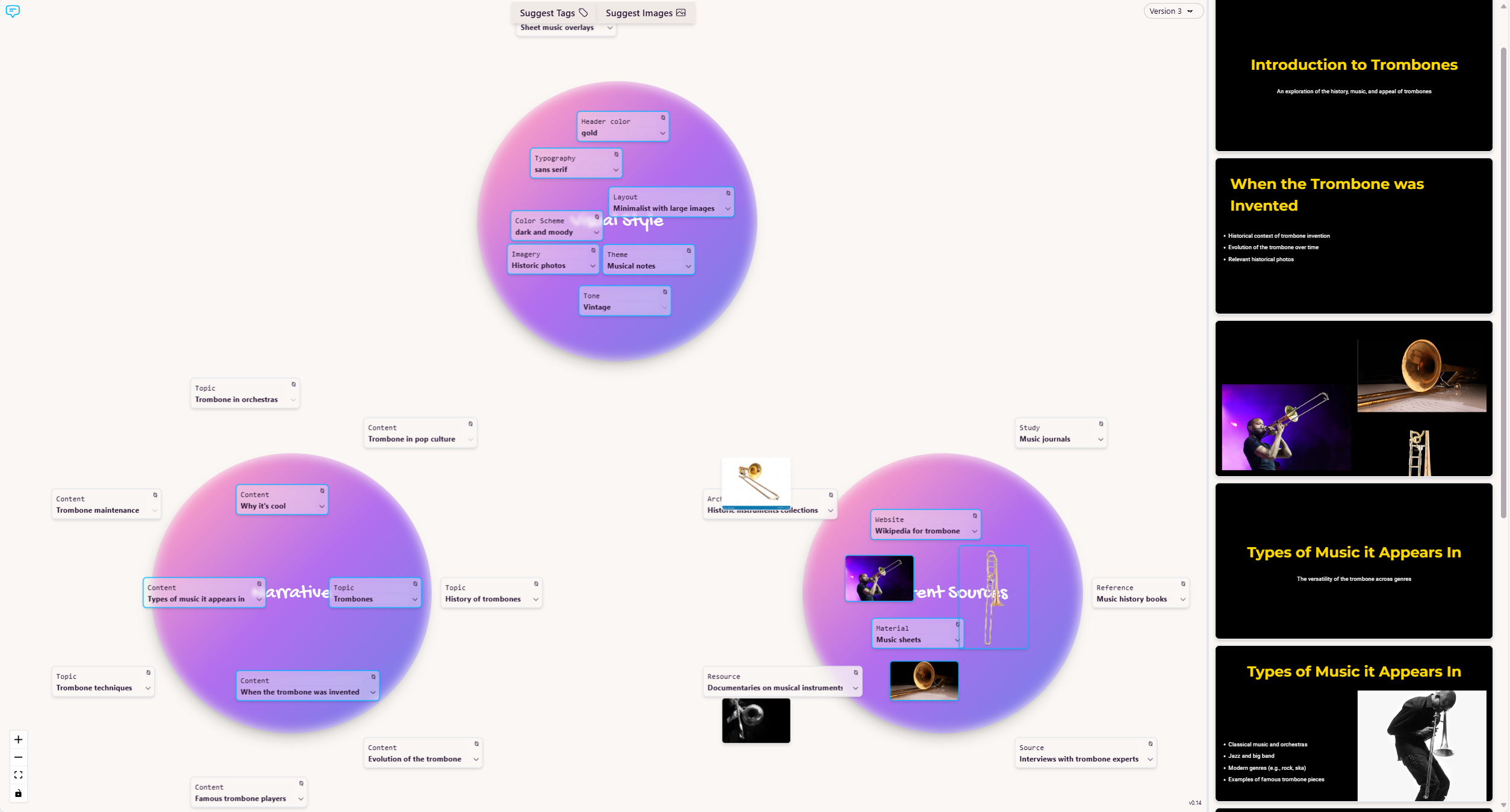}
  \caption{Semi-structured task outcome of P11}
  \Description{The figure shows the IntentTagger interface for participant P11’s semi-structured task. The Deck Steering Board on the left contains tags related to Narrative, Visual Style, and Content Sources. The Narrative section includes tags like “Topic: Trombones,” “Content: History of Trombones,” and “Content: Types of Music It Appears In.” The Visual Style section includes tags such as “Color Scheme: Dark and Moody,” “Typography: Sans Serif,” and “Imagery: Historic Photos.” The Content Sources include music journals, music history books, and interviews with trombone experts. On the right, the Slide Panel displays a slide deck on the trombone, covering sections like “Introduction to Trombones,” “When the Trombone Was Invented,” and “Types of Music It Appears In.” } 
  \label{fig:appendix_task4_P11}
\end{figure}

\begin{figure}[H]
    \centering
  \includegraphics[width=0.99\linewidth]{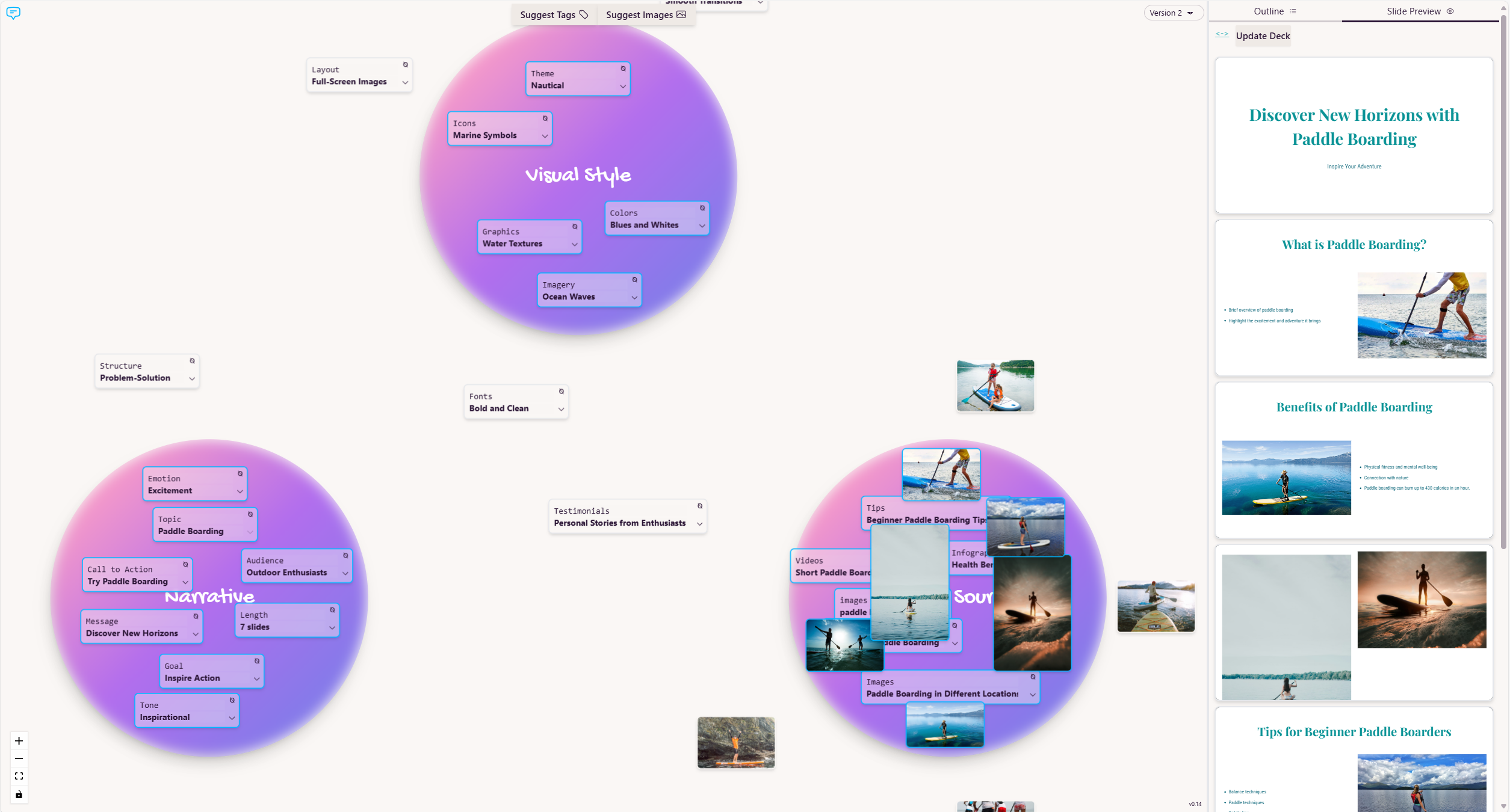}
  \caption{Semi-structured task outcome of P12}
  \Description{The figure shows the IntentTagger interface for participant P12’s semi-structured task. The Deck Steering Board on the left contains tags related to Narrative, Visual Style, and Content Sources. The Narrative section includes tags such as “Topic: Paddle Boarding,” “Call to Action: Try Paddle Boarding,” and “Audience: Outdoor Enthusiasts.” The Visual Style section includes tags like “Theme: Nautical,” “Colors: Blue and White,” and “Imagery: Ocean Waves.” The Content Sources include videos, tips for beginners, and personal stories from paddle boarding enthusiasts. On the right, the Slide Panel displays a slide deck on paddle boarding, covering sections like “What is Paddle Boarding?” “Benefits of Paddle Boarding,” and “Tips for Beginner Paddle Boarders.” } 
  \label{fig:appendix_task4_P12}
\end{figure}

\end{document}